\documentclass[twocolumn]{aastex63}
\usepackage{amsmath}
\usepackage{textcomp}
\usepackage{amssymb}
\usepackage{float}
\usepackage{color}
\usepackage{graphicx}

\newcommand{\Msun}{\ensuremath{M_{\odot}}}
\newcommand{\lum}{erg\,s$^{-1}$}
\newcommand{\fermi}{{\it Fermi}}
\newcommand{\nustar}{{\it NuSTAR}}
\newcommand{\xmm}{{\it XMM-Newton}}
\newcommand{\swift}{{\it Swift}}
\newcommand{\chandra}{{\it Chandra}}

\newcommand{\ergflux}{\mbox{${\rm \, erg \,\, cm^{-2} \, s^{-1}}$}}
\newcommand{\phflux}{\mbox{${\rm \, ph \,\, cm^{-2} \, s^{-1}}$}}
\newcommand{\gm}{$\gamma$}
\newcommand{\ld}{$L_{\rm disk}$}
\newcommand{\mbh}{$M_{\rm BH}$}

 \font\sevenrm=cmr7 scaled 1000

\received{Jan 1, 2020}
\revised{January 1, 2020}
\accepted{\today}
\submitjournal{ApJ}

\shorttitle{High-$z$ blazars}
\shortauthors{Paliya et al.}

\begin{document}

\title{Blazars at the Cosmic Dawn}

\correspondingauthor{Vaidehi S. Paliya}
\email{vaidehi.s.paliya@gmail.com}

\author[0000-0001-7774-5308]{Vaidehi S. Paliya}
\affiliation{Deutsches Elektronen Synchrotron DESY, Platanenallee 6, 15738 Zeuthen, Germany}

\author[0000-0002-6584-1703]{M. Ajello}
\affiliation{Department of Physics and Astronomy, Clemson University, Kinard Lab of Physics, Clemson, SC 29634-0978, USA}

\author[0000-0003-1514-881X]{H. -M. Cao}
\affiliation{INAF Istituto di Radioastronomia, via Gobetti 101, 40129 Bologna, Italy}

\author[0000-0002-8657-8852]{M. Giroletti}
\affiliation{INAF Istituto di Radioastronomia, via Gobetti 101, 40129 Bologna, Italy}

\author[0000-0002-0878-1193]{Amanpreet Kaur}
\affiliation{The Pennsylvania State University, 525 Davey Lab, University Park, PA 16802, USA}

\author{Greg Madejski}
\affiliation{Kavli Institute for Particle Astrophysics and Cosmology, Department of Physics and SLAC National Accelerator Laboratory, Stanford University, Stanford, CA 94305, USA}

\author[0000-0003-2186-9242]{Benoit Lott}
\affiliation{Universit\'e Bordeaux 1, CNRS/IN2P3, Centre d'\'Etudes Nucl\'eaires de Bordeaux Gradignan, 33175 Gradignan, France}

\author[0000-0002-8028-0991]{D. Hartmann}
\affiliation{Department of Physics and Astronomy, Clemson University, Kinard Lab of Physics, Clemson, SC 29634-0978, USA}

\begin{abstract}
The uncharted territory of the high-redshift ($z\gtrsim3$) Universe holds the key to understand the evolution of quasars. In an attempt to identify the most extreme members of the quasar population, i.e., blazars, we have carried out a multi-wavelength study of a large sample of radio-loud quasars beyond $z=3$. Our sample consists of 9 \gm-ray detected blazars and 133 candidate blazars selected based on the flatness of their soft X-ray spectra (0.3$-$10 keV photon index $\leq1.75$), including 15 with \nustar~observations. The application of the likelihood profile stacking technique reveals that the high-redshift blazars are faint \gm-ray emitters with steep spectra. The high-redshift blazars host massive black holes ($\langle \log~M_{\rm BH,~M{\odot}} \rangle>9$) and luminous accretion disks ($\langle L_{\rm disk} \rangle>10^{46}$ \lum). Their broadband spectral energy distributions are found to be dominated by high-energy radiation indicating their jets to be among the most luminous ones. Focusing on the sources exhibiting resolved X-ray jets (as observed with the \chandra~satellite), we find the bulk Lorentz factor to be larger with respect to other $z>3$ blazars, indicating faster moving jets. We conclude that the presented list of the high-redshift blazars may act as a reservoir for follow-up observations, e.g., with \nustar, to understand the evolution of relativistic jets at the dawn of the Universe.

\end{abstract}


\section{Introduction} \label{sec:intro}
Relativistic jets are the manifestation of the extreme processes that occur within the central regions of galaxies \citep[cf.][for a review]{2019ARA&A..57..467B}. Active galactic nuclei (AGN) hosting relativistic jets closely aligned to the line of sight are called blazars. Due to their peculiar orientation, the relativistic amplification of the non-thermal jetted radiation \citep[Doppler boosting, see, e.g.,][]{1979rpa..book.....R} leads to the observation of a number of interesting phenomena. A few examples are detection at all accessible frequencies \citep[e.g.,][]{2011ApJ...736..131A}, observation of temporal and spectral variability \citep[][]{1996Natur.383..319G,2011ApJ...729....2A,2014MNRAS.441.1899F,2017ApJ...844...32P}, superluminal motion and high brightness temperature \citep[][]{1979Natur.277..182S,2019ApJ...874...43L}. The optical and radio emissions detected from blazars are found to be significantly polarized \citep[e.g.,][]{2008PASJ...60..707F,2016ApJ...833...77I}. The flux enhancement also makes blazars a dominating class of \gm-ray emitters in the extragalactic high-energy sky \citep[][]{2020ApJ...892..105A} and one of the very few astrophysical source classes detected at cosmic distances \citep[e.g.,][]{2004ApJ...610L...9R,2013ApJ...777..147S}. Blazars are classified as flat spectrum radio quasars (FSRQs) and BL Lac objects based on their optical spectroscopic properties. FSRQs are characterized by broad emission lines (rest-frame equivalent width $>$5\AA), whereas, BL Lac sources exhibit weak or no emission lines in their optical spectra thereby making it challenging to detect their redshift \citep[][]{1991ApJ...374..431S}. BL Lac objects are known to exhibit a negative or mildly positive evolution compared to strong positive evolution noticed in FSRQs \citep[][]{2012ApJ...751..108A,2014ApJ...780...73A}. Altogether, FSRQs dominate the known population of the high-redshift ($z\gtrsim3$) blazars and are found to be much more luminous with respect to the BL Lac population \citep[e.g.,][]{2009ApJ...699..603A,2017ApJ...837L...5A,2019ApJ...881..154P}. 

The broadband spectral energy distribution (SED) of a blazar is dominated by non-thermal emission from the jet and shows a characteristic double hump structure. The low-frequency hump is associated with synchrotron radiation emitted by relativistic electrons in the presence of a magnetic field. On the other hand, in the leptonic radiative scenario, the high-energy X-ray-to-\gm-ray emission from blazars is attributed to inverse Compton up-scattering of low-energy photons by the jet electrons. The reservoir of the seed photons for the inverse Compton emission could be the synchrotron photons originated within the jet \citep[so-called synchrotron self Compton or SSC;][]{1985ApJ...298..114M}.  Alternatively, thermal IR-to-UV radiation emitted by various AGN components such as the accretion disk, broad line region (BLR), and dusty torus can also get up-scattered to X-ray-to-\gm-ray energies, a process termed external Compton or EC mechanism \citep[see, e.g.,][]{1994ApJ...421..153S,2001ApJ...561..111G} since the seed photons originate externally to the jet. The high-energy radiation from BL Lac sources is primarily explained via SSC process peaking at MeV-to-TeV energies \citep[e.g.,][]{2010MNRAS.401.1570T}, thereby making them bright in this energy range. The X-ray-to \gm-ray emission observed from FSRQs, on the other hand, peaks at relatively low frequencies ($\sim$MeV energies) and is found to be well explained by the EC mechanism \citep[e.g.,][]{2016ApJ...826...76A}.

Based on the location of the synchrotron peak, blazars have also been classified as low-synchrotron peaked (LSP, $\nu^{\rm syn}_{\rm peak,~Hz}<10^{14}$), intermediate-synchrotron peaked ($10^{14}\leqslant\nu^{\rm syn}_{\rm peak,~Hz}\leqslant10^{15}$), and high-synchrotron peaked (HSP, $\nu^{\rm syn}_{\rm peak,~Hz}>10^{15}$) objects \citep[][]{2010ApJ...716...30A}. BL Lac objects display a wide range of synchrotron peak location, i.e., from LSP-to-HSP, whereas, FSRQs are mostly LSP type blazars \citep[e.g.,][]{2020ApJ...892..105A}. Since the synchrotron peak in FSRQs is located in the sub-milimeter-to-infrared (IR) band, the emission from the accretion disk, so-called big blue bump, has been observed in many FSRQs, especially the high-redshift ones, at optical-ultraviolet frequencies \citep[cf.][]{2010MNRAS.405..387G,2016ApJ...825...74P}. The inverse Compton peak in the high-redshift blazars, on the other hand, is usually located at hard X-ray-to-MeV energy band, as revealed by the observation of flat hard X-ray and steep falling \gm-ray spectra \citep[see, e.g.,][for recent multi-frequency campaigns]{2017ApJ...837L...5A,2017ApJ...839...96M,2019ApJ...871..211P,2019A&A...627A..72G}. It has been noticed that as the bolometric luminosity of blazars increases, the SED peaks shift to lower frequencies and the inverse Compton peak dominates the SED\footnote{The prevalence of the inverse Compton peak over the synchrotron one can be quantified with the term `Compton dominance' which is defined as the ratio of the inverse Compton to synchrotron peak luminosities \citep[see, e.g.,][]{2013ApJ...763..134F}.} \citep[][]{1998MNRAS.299..433F,2010ApJ...710...24S}. Since at high redshifts, only the most luminous sources are expected to be detected, a major fraction of the bolometric output of the high-redshift blazars is found to be radiated in the form of high-energy X-ray-to-\gm-ray emission, leading to the observation of the Compton dominated SEDs.

High-redshift blazars are crucial to study relativistic jets and their connection with the central engine (i.e., the black hole and the accretion disk) at the early epoch of the evolution of the Universe. Supermassive black holes are reported to evolve quicker in jetted quasars compared to radio-quiet AGNs \citep[][]{2015MNRAS.446.2483S}, thus indicating a connection between the jet and the black hole growth \citep[e.g.,][]{2014MNRAS.442L..81F,2017ApJ...836L...1T}. The detection of a few sources in a given redshift bin and determination of their physical properties enable us to constrain the behavior of the whole jetted population in that redshift bin. This is because the identification of a single blazar with jet velocity or bulk Lorentz factor $\Gamma$ implies the existence of 2$\Gamma^2$ sources with similar intrinsic properties but having a jet pointed elsewhere \citep[e.g.,][]{2011MNRAS.416..216V}. Therefore, it is important to identify and study the high-redshift blazars to understand the evolution of jetted AGNs and massive black holes at the cosmic dawn.

Only a handful of the high-redshift blazars are known so far \citep[see, e.g.,][]{2018ApJS..235....4O,2020ApJ...892..105A,2019MNRAS.484..204C} and even fewer have been studied in detail \citep[e.g.,][]{2005A&A...443L..33D,2010A&A...509A..69B,2012MNRAS.421..390L,2015MNRAS.446.2921F,2015MNRAS.450L..57G,2016MNRAS.462.1542S,2017MNRAS.467..950C,2017MNRAS.468...69Z,2018ApJ...856..105A,2019ApJ...879L...9L,2019A&A...629A..68B,2019MNRAS.489.2732I}. This is likely due to their faintness, intrinsic rareness, and/or difficulty in identifying blazars among the high-redshift radio-loud quasars. A \gm-ray detection with the \fermi-Large Area Telescope (LAT) could be a definitive signature for the presence of a closely aligned relativistic jet \citep[][]{2017ApJ...837L...5A}, however, the energy shift of the SED peaks to low frequencies, along with \gm-ray attenuation due to extragalactic background absorption \citep[cf.][]{2011MNRAS.410.2556D,2019ApJ...874L...7D}, makes high-redshift blazars fainter and steepen their \gm-ray spectrum in the \fermi-LAT energy range. Most importantly, the current \fermi-LAT  sensitivity is likely to be too low to detect a large number of $z>3$ blazars due to their great distances, hence low flux. A large radio-loudness along with the observation of a flat radio spectrum provide evidence supporting the beamed nature of the observed radiation. However, most of the known radio-loud, high-redshift quasars only have single frequency radio flux density measurements from the NRAO VLA Sky Survey \citep[NVSS;][]{1998AJ....115.1693C}, Faint Images of the Radio Sky at Twenty-centimeters \citep[FIRST;][]{1997ApJ...475..479W,2015ApJ...801...26H} or Sydney University Molonglo Sky Survey \citep[SUMSS;][]{2003MNRAS.342.1117M}. The fact that many well-studied, high-redshift blazars exhibit Gigahertz peaked spectra \citep[e.g., QSO J0906+6930 at $z=5.47$;][]{2017MNRAS.467.2039C}, indicates that a flat radio spectrum alone cannot be a definitive feature. A large brightness temperature ($\gtrsim10^{11}$ K) can also give some hints about the relativistic beaming \citep[cf.][]{2016MNRAS.463.3260C}. Other methods, such as the observation of superluminal motion, requires multi-epoch monitoring covering long time periods and thus are limited to study only the brightest radio sources \citep[e.g.,][]{2019arXiv191212597Z}.

Since the high-redshift blazars are usually LSP type objects, they are expected to exhibit a flat or rising X-ray spectrum (in the $\nu F_{\nu}$ versus $\nu$ plane), especially in the hard X-ray band. This, along with the radio-loudness, can be used to ascertain the blazar nature of a high-redshift, radio-loud quasar. Again, due to limited sensitivity of the hard X-ray surveying instrument \swift~Burst Alert Telescope \citep[BAT, 14$-$195 keV;][]{2005SSRv..120..143B}, only a few ($<$10), extremely bright, $z>3$ quasars are confirmed as beamed AGNs using this approach \citep[][]{2018ApJS..235....4O}. The  Nuclear Spectroscopic Telescope Array \citep[\nustar;][]{2013ApJ...770..103H}, on the other hand, has a considerably improved sensitivity which has led to the confirmation of relatively faint radio-loud quasars as blazars \citep[][]{2013ApJ...777..147S}. However, due to limited field of view of \nustar, only one source can be observed in a single pointing. In this regard, a useful strategy could be to explore the soft X-ray spectral behavior of the high-redshift, radio-loud quasars and identify `candidate' blazars among them \citep[see, e.g.,][for a similar approach]{2015MNRAS.450L..34G}. This is because hundreds of the high-redshift quasars are observed with soft X-ray instruments either as target of interest or lying as background objects in the field of other observations), e.g., \chandra~X-ray observatory, \xmm~and \swift-X-ray Telescope (XRT), and hence a meaningful population study can be done. The best candidates can then be followed up, e.g, with \nustar~and Very Large Array \citep[VLA, see, e.g.,][]{2014arXiv1406.4797G}, to confirm their blazar identity and study the physical properties of relativistic jets at the beginning of the Universe. This is the primary objective of the work discussed in this article.

Here we present the results of an exhaustive investigation to explore the multi-frequency behavior of 142 $z>3$ radio-loud quasars that are likely to be blazars, using all of the publicly available data. Other than studying the physical properties, our goal is also to prepare a list of the most promising high-redshift, radio-loud quasars that have a high probability of hosting closely aligned relativistic jets. This list would serve as the reservoir from which sources can be picked to follow with \nustar~and other multi-wavelength observing facilities. We discuss the criteria to define the sample in Section~\ref{sec:sample}. The data reduction techniques are described in Section~\ref{sec:analysis} and the adopted leptonic radiative model is elaborated in Section~\ref{sec:model}. We present the derived results in Section~\ref{sec:obs_prop} and \ref{sec:sed_prop}. Section~\ref{sec:x-ray_jets} is devoted to our findings on extended X-ray jets and we summarize in Section~\ref{sec:summary}. We adopt a cosmology of $H_0=67.8$~km~s$^{-1}$~Mpc$^{-1}$, $\Omega_m = 0.308$, and $\Omega_\Lambda = 0.692$ \citep[][]{2016A&A...594A..13P}.

\section{The Sample} \label{sec:sample}
We started with the Million Quasar Catalog \citep[MQC v6.4;][]{2019arXiv191205614F} and considered all sources with $z\geq3$. This catalog is a regularly updated compendium of 757991 type 1 quasars/AGNs and $\sim$1.1 million quasar candidates with high-confidence ($\geq$80\% likelihood). It is primarily based on SDSS and AllWISE catalogs and also covers the southern hemisphere using 2 degree-field quasar redshift Survey and 6 degree-field galaxy survey \citep[][]{2000MNRAS.317.1014B,2009MNRAS.399..683J}, along with $>$1000 individual publications. Both spectroscopically confirmed quasars and sources with photometric redshifts have been considered in MQC.

The 75940 $z>3$ objects selected from MQC were then cross-matched with NVSS, SUMSS, and FIRST radio catalogs using a 3$^{\prime\prime}$ search radius to identify radio detected high-redshift quasars. Using the $R$ band magnitude from MQC and flux density information from the matches in the radio catalogs, we computed the radio-loudness parameter \citep[$R$;][]{1989AJ.....98.1195K} for the selected quasars. To determine the rest-frame 5 GHz and optical $B$-band flux densities, we extrapolated the measured radio fluxes assuming a flat radio spectrum ($\alpha=0, F_{\nu}\propto \nu^{\alpha}$) and considered an optical spectral index of $\alpha=-0.5$ \citep[][]{2007AJ....133..313A}. At this stage, we only retained radio-loud ($R>10$) quasars leading to a total of 2226 sources. We also cross-matched this sub-sample with the 5$^{\rm th}$ ROMA-BZCAT catalog \citep[][]{2015Ap&SS.357...75M} and found that all but one $z\geq3$ BZCAT sources are already included in our sample. We included the missing object BZQ J0941$-$8615 \citep[$z=3.697$;][]{2013AJ....146...10T} to ensure that all BZCAT blazars are considered in our work. Then, we searched for the availability of the X-ray data in \chandra, \xmm, and \swift-XRT data archives and kept 156 objects with existing X-ray observations\footnote{Some  of the observations were carried out as a part of our own proposals in \nustar~(proposal id: 3279, PI: Paliya) and \xmm~guest investigator cycles (proposal id: 80200, PI: Paliya). Along with this, we also acquired data for a few sources lacking any previous X-ray measurements by \swift-XRT target of opportunity observations.}. These high-redshift, radio-loud, X-ray detected quasars were subjected to X-ray spectral analysis as described in the next section.

The average X-ray spectral shape of radio-quiet quasars is found to be softer \citep[X-ray photon index $\Gamma_{\rm X} \gtrsim 1.9$,][]{2005ApJ...630..729S} than for relativistically beamed, radio-loud quasars\footnote{Note that Compton thick AGNs can have a flat X-ray spectrum owing to severe absorption at soft X-rays \citep[e.g.,][]{2007A&A...466..823G,2018ApJ...854...49M}. However, since they are primarily radio-quiet, our sample is free from such objects.} \citep[e.g.,][]{2013ApJ...763..109W}. To identify the best blazar candidates, we, therefore, considered only those objects that have $\Gamma_{\rm X}\lesssim 1.75$ \citep[see also][for a similar approach]{2019MNRAS.489.2732I}. This exercise led to a final sample of 142 high-redshift, radio-loud, candidate blazars. For the sake of brevity, we simply call them high-redshift blazars in the rest of the paper.

The recently released fourth catalog of the \fermi-LAT detected AGNs \citep[4LAC;][]{2020ApJ...892..105A} has listed 10 \gm-ray emitting $z\geq3$ blazars. All but one, 4FGL J1219.0+3653  \citep[$z=3.52$;][]{2017A&A...597A..79P}, are present in our sample. The source 4FGL J1219.0+3653 had no existing X-ray data and our two \swift~target of opportunity observations (target id: 12058 and 13082, summed exposure $\sim$4 ksec) failed to determine the spectral parameters of the source. Therefore, it is not considered in this work. Altogether, our sample consists of 9 \gm-ray detected and 133 \fermi-LAT undetected blazars. The basic properties of these 142 sources are presented in Table~\ref{tab:basic_info}.

\begin{deluxetable*}{lccccc}[t!]
\tabletypesize{\small}
\tablecaption{Basic properties of 142 high-redshift blazars studied in this work.\label{tab:basic_info}}
\tablewidth{0pt}
\tablehead{
\colhead{Name} & \colhead{R. A.} & \colhead{Decl.} & \colhead{redshift} & \colhead{$R_{\rm mag}$} & \colhead{$F_{\rm radio}$} \\
\colhead{} & \colhead{degrees} & \colhead{degrees} & \colhead{} & \colhead{} & \colhead{(mJy)}}
\startdata
\colhead{} & \multicolumn{3}{c}{\gm-ray detected blazars} & \colhead{} & \colhead{}\\
NVSS J033755$-$120404 &   54.48104 & $-$12.06793& 3.442  &  20.19& 475.3 \\
NVSS J053954$-$283956 &   84.97617 & $-$28.66554& 3.104  &  18.97& 862.2 \\
NVSS J073357+045614 &   113.48941& 4.93736  & 3.01   &  18.76& 218.8 \\
NVSS J080518+614423 &   121.32575& 61.73992 & 3.033  &  19.81& 828.2 \\
NVSS J083318$-$045458 &   128.32704& $-$4.9165  & 3.5    &  18.68& 356.5 \\
NVSS J135406$-$020603 &   208.52873& $-$2.10089 & 3.716  &  19.64& 733.4 \\
NVSS J142921+540611 &   217.34116& 54.10309 & 3.03   &  19.84& 1028.3\\
NVSS J151002+570243 &   227.51216& 57.04538 & 4.313  &  19.89& 202.0 \\
NVSS J163547+362930 &   248.94681& 36.49164 & 3.615  &  20.55& 151.8 \\
\hline
\colhead{} & \multicolumn{3}{c}{\gm-ray undetected blazars} & \colhead{} & \colhead{}\\
NVSS J000108+191434   &  0.28589  & 19.24269 & 3.1    &  20.5 & 265.1 \\
NVSS J000657+141546   &  1.73971  & 14.26299 & 3.2    &  18.86& 183.4 \\
NVSS J001708+813508   &  4.28531  & 81.58559 & 3.387  &  16.61& 692.5 \\
NVSS J012100$-$280623   &  20.25309 & $-$28.10616& 3.119  &  18.82& 122.0 \\
NVSS J012201+031002   &  20.50794 & 3.16733  & 4.0    &  19.78& 98.4  \\
\enddata
\tablecomments{The positional coordinates (R.A. and Decl., in J2000), redshift, and $R$-band magnitudes are taken from MQC. The name and radio flux density values are adopted from  NVSS, SUMSS, or FIRST catalogs depending in which catalog the radio counterpart was identified. For the source SUMSS J094156$-$861502 (or BZQ J0941$-$8615), we provide the relevant information from the BZCAT catalog.\\
(This table is available in its entirety in a machine-readable form in the online journal. A portion is shown here for guidance regarding its form and content.)}

\end{deluxetable*}

\section{Data Reduction Methods} \label{sec:analysis}
\subsection{Gamma-ray analysis}
We analyzed the \fermi-LAT data for all sources present in the sample including blazars from 4LAC. The goals are to: (i) update the spectral parameters of the known \gm-ray emitters, (ii) identify new \gm-ray emitting blazars, (iii) determine the flux sensitivity limits for undetected objects and stack their likelihood profiles to derive the cumulative \gm-ray detection significance. The data cover the period of almost 11 years of the \fermi-LAT operation (2008 August 5 to 2019 July 14). We defined a region of interest (ROI) of 15$^{\circ}$ centered at the target quasar and selected P8R3 SOURCE class events ({\tt evclass=128} and {\tt evtype=3}) in the energy range of 0.1$-$300 GeV. A filter ``DATA\_QUAL$>$0 \&\& LAT\_CONFIG==1" was also applied to determine the good time intervals. Additionally, a zenith angle cut of $z_{\rm max}=90^{\circ}$ was used to limit the contamination from the Earth limb \gm-rays. To generate the \gm-ray sky model, we adopted the sources present in the recently released \fermi~Large Area Telescope Fourth Source Catalog \citep[4FGL;][]{2019arXiv190210045T} and lying within 25$^{\circ}$ of the target position. The latest diffuse background models\footnote{https://fermi.gsfc.nasa.gov/ssc/data/access/lat/BackgroundModels.html}. i.e., {\tt gll\_iem\_v07.fits} and {\tt iso\_P8R3\_SOURCE\_V2\_v1.txt} were also adopted in the analysis. We computed the maximum likelihood test statistic as TS =  $2\log(\mathcal{L}_1-\mathcal{L}_0$), where $\mathcal{L}_0$ and $\mathcal{L}_1$ denote the likelihood values without and with a point source at the position of interest, respectively \citep[][]{1996ApJ...461..396M}. We first optimized the ROI to get a crude estimation of the TS for each source and then allowed the spectral parameters of all the sources with TS$>$25 to vary during the likelihood fit. Since the time period considered in this work is longer than that covered in the 4FGL catalog, TS maps were generated to search for \gm-ray emitting objects present in the data but not in the catalog. Whenever an excess emission with TS$>$25 was identified, we modeled it with a power law and insert in the sky model. Once all excess emissions were found and included in the sky model, we performed a final likelihood fit to optimize the spectral parameters left free to vary and to determine the parameters and detection significance for the target quasar. In this work, a source is considered to be \gm-ray detected if the derived TS is larger than 25. The entire data analysis was performed using the publicly available package {\tt fermiPy} \citep[][]{2017arXiv170709551W} and fermitools\footnote{https://github.com/fermi-lat/Fermitools-conda/wiki}. The uncertainties were computed at 1$\sigma$ confidence level.

We stacked the likelihood profiles of all the \gm-ray undetected sources to calculate the overall detection significance of the sample. This was done by computing the likelihood values for each object over a grid of photon flux and photon index. Such likelihood profiles were generated for all high-redshift blazars and then stacked to estimate the combined TS and spectral parameters associated with the TS peak. Further details of this technique can be found in \citet[][]{2019ApJ...882L...3P}.

\subsection{Hard X-ray analysis}
There are 15 sources in our sample that have existing \nustar~observations. We adopted the tool {\tt nupipeline} to reduce the raw \nustar~data and calibrate the event files. To extract the source and background spectra, circular regions of  30$^{\prime\prime}$ and 70$^{\prime\prime}$ radii, respectively, were considered from the same chip. We used the pipeline {\tt nuproducts} to extract the spectra and response matrix and ancillary files. The spectra of bright sources were grouped to have 20 counts per bin, whereas, we adopted a binning of 1 count per bin for faint objects using the tool {\tt grppha}. We performed the spectral fitting in XSPEC \citep[v 12.10.1;][]{1996ASPC..101...17A} with a power law model. The uncertainties are estimated at the 90\% confidence level.

We used publicly available 14$-$195 keV spectra of 9 high-redshift blazars present in the 105-month \swift-BAT catalog\footnote{https://swift.gsfc.nasa.gov/results/bs105mon/} \citep[][]{2018ApJS..235....4O} and applied a power law model in XSPEC to extract the spectral data points.

\subsection{Soft X-ray analysis}
\chandra: The observations from Advanced CCD Imaging Spectrometer (ACIS, 0.5$-$7 keV) onboard \chandra~X-ray observatory were reduced using Chandra Interactive Analysis of Observations (CIAO, version 4.11) software package and the CALDB version 4.8.2. For sources with more than one \chandra~pointings, we considered the observations which has the longest exposure. We first ran the tool {\tt chandra\_repro} to generate the cleaned and calibrated event files and then used the tool {\tt specextract} to extract the source and background spectra. For this purpose, we adopted a source region of 3$^{\prime\prime}$ centered at the target quasar and a 10$^{\prime\prime}$ circle was considered from a nearby source-free region to represent the background. In 7 out of 54 sources, we have found evidence for the presence of extended X-ray jets. We selected a source region as a circle of 1.5$^{\prime\prime}$-2$^{\prime\prime}$ excluding the extended X-ray emission in these objects. For the spectral analysis, the generated source spectra were binned to have at least 1 count per bin and the fitting was performed in XSPEC following the C-statistics \citep[][]{1979ApJ...228..939C}. We considered an absorbed power law model and adopt the Galactic neutral hydrogen column density from \citet[][]{2005AA...440..775K}.

In order to ascertain the detection of extended X-ray jets, we generated exposure-corrected 0.5$-$7 keV images using the tool {\tt fluximage} and adjusted the X-ray core position to match with the VLA position using the task {\tt wcs\_update}.

\xmm: The \xmm~data were analyzed following the standard procedure\footnote{http://www.cosmos.esa.int/web/xmm-newton/sas-threads} using the package Science Analysis Software 15.0.0. In particular, we adopted the task {\tt epproc} to create EPIC-PN event files and then used {\tt evselect} to remove the high flaring background periods. We considered the source region as a circle of 40$^{\prime\prime}$ radius centered at the source of interest and the background region was selected as a circle of the same size from the same chip$-$but free from source contamination. The tool {\tt evselect} was also used to extract the source and background spectra. The pipelines {\tt rmfgen} and {\tt arfgen} were used to generate the response and ancillary files. Finally, we bin the source spectra using {\tt specgroup} with 20 counts per bin and performed the fitting in XSPEC.

\swift-XRT: We used the online \swift-XRT data product facility\footnote{http://www.swift.ac.uk/user\_objects/} \citep[][]{2009MNRAS.397.1177E} to generate the source, background, and ancillary response files. This tool automatically determines the sizes of the source and background regions based on the count rate of the source \citep[see also][]{2009MNRAS.397.1177E}. We rebinned the source spectra with 1 or 20 counts per bin, depending on the source brightness, and performed the fitting in XSPEC keeping the neutral hydrogen column density fixed to the Galactic value. We derived the uncertainties in the parameters at 90\% confidence level.

\subsection{Optical spectral analysis}
\begin{figure}[t!]
\hbox{\hspace{0cm}
\includegraphics[width=\linewidth]{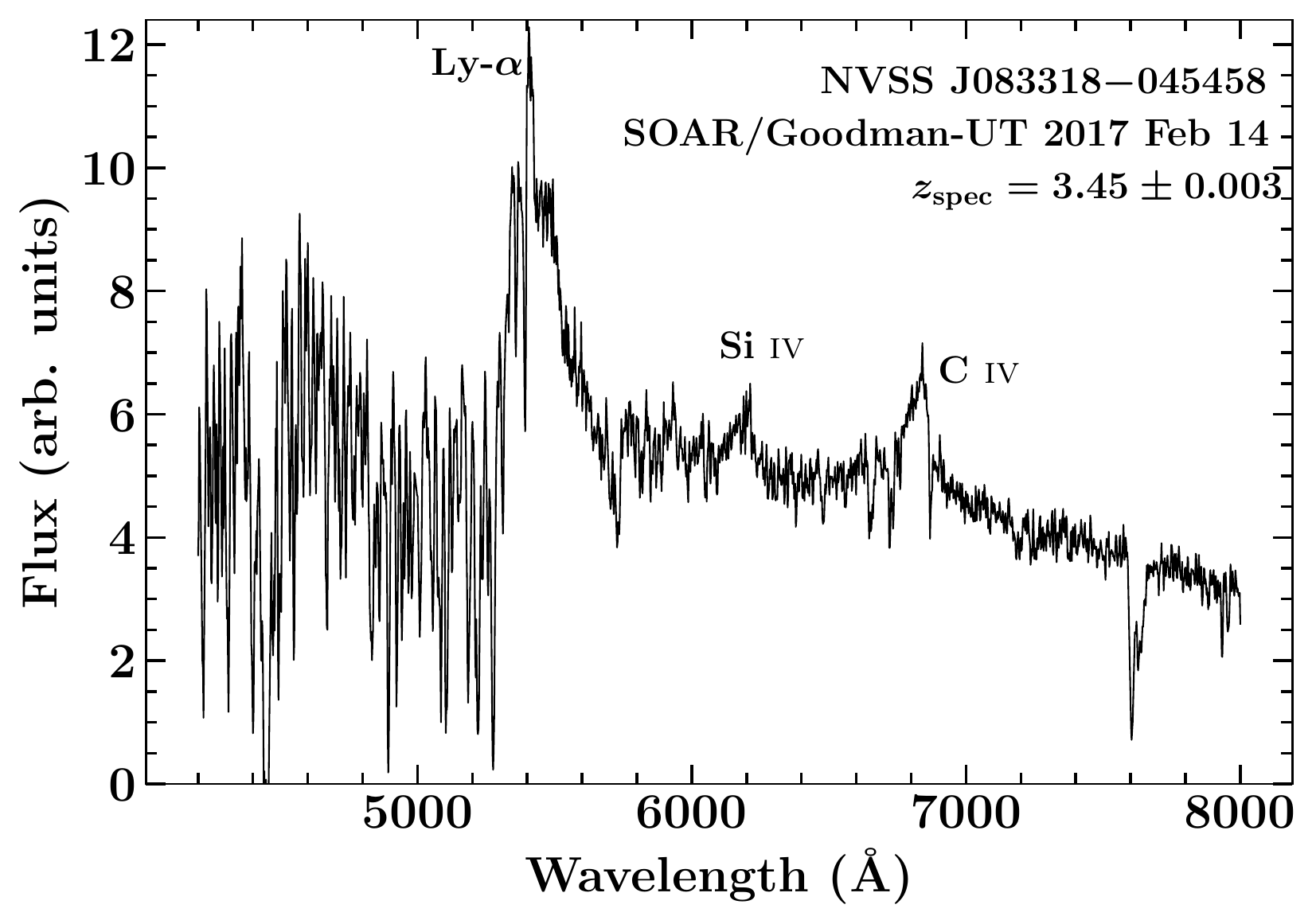}
}
\caption{Optical spectrum of NVSS J083318$-$045458 taken with the Goodman spectrograph mounted at 4 m SOAR telescope. A few prominent emission lines are labeled which enabled the spectroscopic redshift measurement and confirmed the high-redshift nature of the source with $z_{\rm spec}=3.45$.} \label{fig:J0833_spec}
\end{figure}

One of the \gm-ray emitting sources present in our sample, 4FGL J0833.4$-$0458 or NVSSJ083318$-$045458, had only photometric redshift information in MQC \citep[$z_{\rm phot}=3.5$;][]{2015ApJS..219...39R}. We observed this object with the Goodman Spectrograph mounted on the 4.1 m SOAR (Southern Astrophysical Research Telescope) on 2017 February 14. The data were obtained with a 400 l/mm grating in conjunction with a 1.07 arcsec slit. Three spectra were obtained for a total exposure of 3600 sec (1200 sec$\times$3) and then combined in order to remove any artificial features due to cosmic ray or instrumental effects. The standard optical spectroscopic reduction procedure was utilized using the IRAF \citep[][]{1986SPIE..627..733T} pipeline. The obtained spectra were first cleaned by subtracting bias and applying flat field normalization. These cleaned data were then wavelength calibrated using Fe-Ar lamp spectra, which were obtained after every source observation. All the spectra were flux calibrated using a spectrophotometric standard obtained during the night of observation. Finally, each spectra were corrected for Galactic extinction, using the E(B-V) values obtained from \citet[][]{2011ApJ...737..103S}. The resultant optical spectrum of J083318$-$045458 is shown in Figure~\ref{fig:J0833_spec}. Various broad emission lines, e.g., Ly-$\alpha$ and C{\sevenrm IV}, are observed leading to a spectroscopic redshift of $z_{\rm spec}=3.45\pm0.003$.

\subsection{Radio analysis}
We analyzed VLA data of the 7 high-redshift blazars that have exhibited traces of extended X-ray emission. In order to find the radio counterparts for these X-ray jets, we reprocessed the raw data downloaded from the VLA Archive\footnote{https://archive.nrao.edu/archive/advquery.jsp}. The data reduction was conducted in the NRAO Astronomical Image Processing System \citep[AIPS;][]{2003ASSL..285..109G}. The sources were first calibrated, and then the amplitude and phase solutions were transferred to the targets. The calibrated data were imaged in Difmap \citep[][]{1997ASPC..125...77S}. We prefer to use the data acquired with the VLA at L Band and in A configuration, so as to obtain better resolution and better sensitivity for resolving and detecting the extended radio emission which usually has a steep spectrum. For the source NVSS J090915+035443, the C-band data was used. The observing and image information are summarized in Table~\ref{tab:radio}.

\begin{table*}
\caption{The VLA observing and imaging information for 7 targets showing traces of X-ray jets.\label{tab:radio}}
\begin{center}
\begin{tabular}{lcccccc}
\hline
Name       & Obs. date        &  Freq.    &   Beam size    &     PA         &  Peak br.  & RMS  \\
                &                  &   GHz     &    arcsec      &     Degr.      &  mJy/beam  & mJy/beam \\
~[1] & [2] & [3]  & [4] & [5] & [6]  & [7] \\ 
\hline
J090915+035443  & 1984-12-17       &  4.8    & 0.5$\times$0.3   & $-$41.0         & 188.9    & 0.3\\
J140501+041536  & 1987-08-16       &  1.4    & 1.3$\times$1.2   &  12.4         & 590.3    & 0.3\\
J142107$-$064355 & 2004-12-22       &  1.4    & 1.4$\times$1.1   &  2.0          & 342.8    & 0.2\\
J143023+420436  & 2004-12-06       &  1.4    & 1.4$\times$1.0   &  53.0         & 155.0    & 0.1\\
J151002+570243  & 1995-07-14       &  1.4    & 1.6$\times$1.1   & $-$6.8          & 227.0    & 0.1\\
J161005+181143  & 1987-08-16       &  1.4    & 1.2$\times$1.1   & $-$6.8          & 203.9    & 0.3\\
J174614+622654  & 1991-09-08       &  1.4    & 1.7$\times$1.0   & $-$69.7         & 477.1    & 0.4\\
\hline
\end{tabular}
\end{center}
\tablecomments{All the experiments were carried out in VLA A-configuration. Col.[2]: observing date; Col.[3]: observing frequency; Col.[4]: restoring beam size at full width of half maximum; Col.[5]: position angle of the restoring beam major axis, measured north through east; Col.[6]: peak brightness of the CLEAN images; Col.[7]: off-source image noise.}
\end{table*}
\subsection{Other archival observations}
To cover the radio-to-UV part of the SED, we relied on the archival spectral measurements from Space Science Data Center SED Builder\footnote{https://tools.ssdc.asi.it/}. These measurements primarily come from NVSS, SUMSS, FIRST, {\it Planck}, {\it Wide-field Infrared Survey Explorer}, and Sloan Digital Sky Survey quasar catalogs and allowed us to determine the level of the synchrotron emission and also constrain the accretion disk spectrum at optical-UV energies.

\section{The Leptonic Radiative Model}\label{sec:model}
We used the conventional synchrotron, inverse Compton emission model \citep[see, e.g.,][]{2009herb.book.....D} to reproduce the broadband SEDs of the high-redshift blazars and explain it here in brief. We assume a spherical emission region of radius $R_{\rm blob}$ covering the whole cross-section of the jet and moving along with the bulk Lorentz factor $\Gamma$. The jet is considered to be of conical shape with semi-opening angle 0.1 radian and this connects $R_{\rm blob}$ with the distance of the emission region ($R_{\rm diss}$) from the central engine. The energy distribution of the relativistic electrons present in the emission region is adopted to follow a smooth broken power law. In the presence of a uniform but tangled magnetic field, these relativistic electrons radiate via synchrotron, SSC and EC processes. For the latter, we compute the comoving-frame radiative energy densities of the BLR, dusty torus, and the accretion disk following the prescriptions of \citet[][]{2009MNRAS.397..985G}. The radiative profile of the standard optically thick, geometrically thin accretion disk \citep[][]{1973A&A....24..337S} is assumed to follow a multi-color blackbody \citep[][]{2002apa..book.....F}. Both the BLR and dusty torus are considered as thin spherical shells whose radii depend on the luminosity of the accretion disk as $R_{\rm BLR} = 10^{17} L^{1/2}_{\rm disk,45}$ and $R_{\rm torus} = 2.5\times10^{18} L^{1/2}_{\rm disk,45}$ cm, respectively, where $L_{\rm disk,45}$ is the accretion disk luminosity ($L_{\rm disk}$) in units of 10$^{45}$ \lum. We assume that 10\% and 30\% of $L_{\rm disk}$ is reprocessed by the BLR and the torus, respectively. Various jet powers are computed following \citet[][]{2008MNRAS.385..283C} and we assume no pairs in the jet, i.e., equal number density of electrons and cold protons, while deriving the kinetic jet power.

{\it SED Modeling Guidelines:} Our model does not perform any statistical fit and we merely reproduce the observed SED following a fit-by-eye approach. The uniqueness of the SED parameters mainly depends on the availability of the simultaneous observations covering all accessible bands as much as possible. There is a clear dearth of multi-wavelength data for the high-redshift blazars. Most of them are undetected in the \gm-ray band and only a few have existing hard X-ray observations. Due to their great distances and hence faintness, the measured uncertainties are also large at soft X-rays. Furthermore, most of the X-ray observations were carried out with different science objectives, e.g. to search for soft X-ray flattening \citep[cf.][]{2001MNRAS.323..373F} and extended X-ray jets \citep[e.g.,][]{2018ApJ...856...66M}, and thus, they do not represent any particular high/low activity state of sources. Therefore, we collected all available, `non-simultaneous' data sets and treated them as a representation of the average behavior of the blazars under consideration. The motivation here is to study the overall physical properties of the high-redshift jetted population and determine interesting objects that can be followed up for deeper studies. While doing so, we were driven by our current understanding of blazar radiative processes based on previous works reported in the literature and we tried to constrain the SED parameters as described below.

Two crucial parameters in the modeling of FSRQs are \ld~and the mass of the central black hole (\mbh). Since these sources exhibit strong emission lines in their optical spectra, one can reliably derive the luminosity of the BLR and \ld~from the emission line information \citep[e.g., using scaling relations of][]{1991ApJ...373..465F} and \mbh~assuming the virial relations to hold valid \citep[e.g.,][]{2006ApJ...641..689V,2012ApJ...748...49S}. Above redshift 3, only the C{\sevenrm IV} line remains in the wavelength range covered by the optical spectroscopic facilities, e.g., SDSS. However, as demonstrated in various studies \citep[e.g.,][]{2011AJ....141..167R,2014ApJS..215...12C}, C{\sevenrm IV} is likely not suitable to derive \mbh~due to blueshifts and/or absorption troughs, indicating strong outflows. An alternative approach to determine \ld~and \mbh~is by modeling the optical-UV spectrum with the accretion disk model, provided the big blue bump is visible  \citep[e.g.,][]{2013MNRAS.431..210C}. In this technique, there are two free parameters, the mass accretion rate and \mbh. The level of the optical-UV spectrum constrains the former, hence \ld~for a certain accretion efficiency, leaving only \mbh~as a free parameter. A small mass refers to a smaller accretion disk surface and for a given \ld, it implies a hotter disk, thus the accretion disk radiation peaking at higher frequencies. A few studies have recently shown that \ld~and \mbh~derived from this method agree well with that computed from optical spectroscopy \citep[][]{2015MNRAS.448.1060G,2017ApJ...851...33P,2019ApJ...881..154P}. Therefore, we derive the two central engine parameters by adopting the accretion disk modeling approach.

The accuracy of the above mentioned technique depends on the visibility of the peak of the big blue bump. For a source with \ld~$\sim$10$^{47}$ \lum, the peak lies at far-UV (i.e., $>10^{15}$ Hz, in the rest frame) if the mass of the central black hole is $<$10$^{9}$ \Msun. Constraining \mbh~for such objects with disk modeling approach may not be possible since the emission bluer to the Lyman-$\alpha$ frequency is severely absorbed by the intervening clouds. To overcome this problem, we determined \ld~from the C{\sevenrm III}, C{\sevenrm IV}, and/or Lyman-$\alpha$ line luminosity information taken from literature \citep[e.g.,][]{1994ApJ...436..678O,2003ApJ...596L..39S,2011ApJS..194...45S,2012RMxAA..48....9T,2012ApJ...748...49S} by using the flux scaling of \citet[][]{1991ApJ...373..465F} and \citet[][]{1997MNRAS.286..415C} to calculate BLR luminosity and assuming 10\% of the disk emission is reprocessed by BLR. Assuming an uncertainty of 0.3 dex, this additional piece of information provided a range of \ld~values that can be used to estimate the peak of the disk emission \citep[e.g.,][]{2015MNRAS.450L..34G}. Finally, for a good IR-optical data coverage, both \mbh~and \ld~can be reasonably constrained within a factor of 2. Even for objects with poorer data availability, the uncertainty  is of the order of that associated with virial estimations, i.e., $\sim$0.3 dex. This has been demonstrated in the appendix (Section~\ref{sec:disk_uncer}).

 \begin{figure*}[t!]
\hbox{\hspace{0cm}
\includegraphics[scale=0.45]{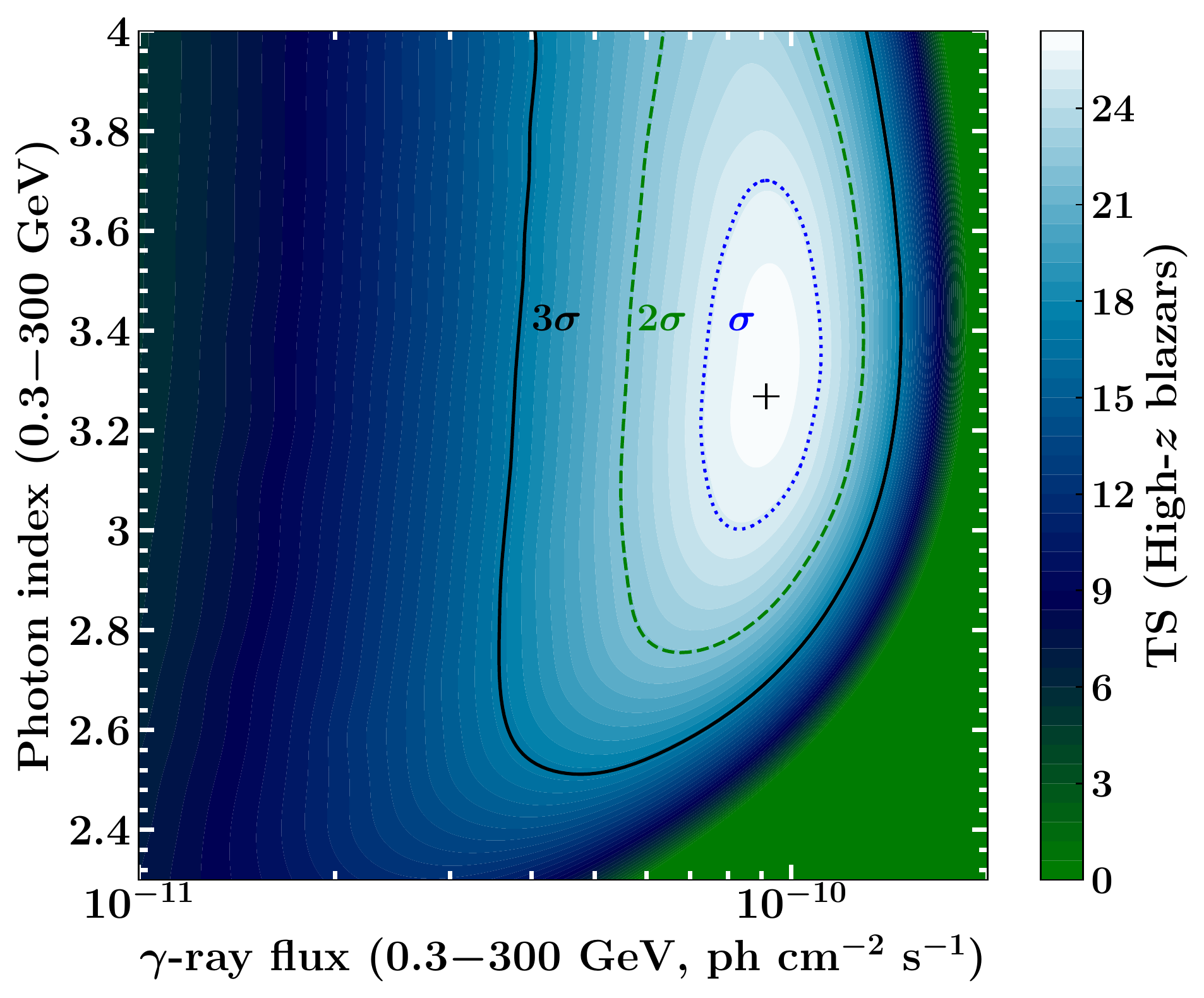}
\includegraphics[scale=0.45]{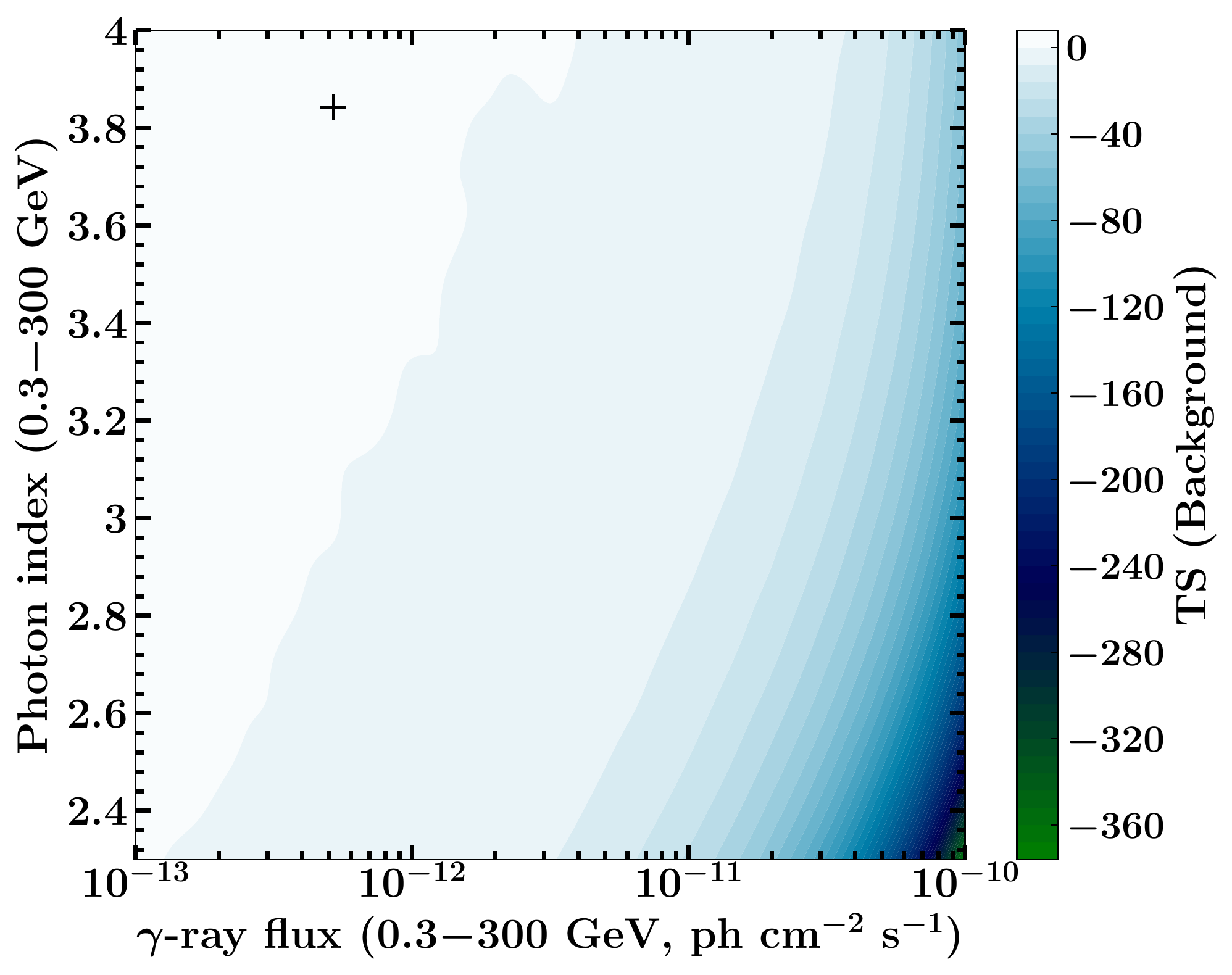}
}
\caption{The stacked TS profile of \gm-ray undetected high-redshift blazars (left) and empty \gm-ray sky positions representing the background (right). The confidence contours are at $\sigma$, 2$\sigma$, and 3$\sigma$ level as labeled and `+' mark shows the peak of the TS profile. In the left plot, we masked the negative TS values to highlight the positive \gm-ray signal. A negative TS indicates that the alternative hypothesis for the presence of a point source characterized by a given flux and photon index is strongly rejected with respect to the null hypothesis of no source.} \label{fig:stacking}
\end{figure*}

The high-energy index of the particle energy distribution can be constrained from the optical-UV data provided it is dominated by the falling part of the synchrotron radiation. However, all the sources studied here are LSP FSRQs with synchrotron emission peaking in the unobserved far-IR to sub-mm wavelengths leaving the accretion disk emission naked at optical-UV frequencies. We also cannot use the \gm-ray spectral shape to constrain the high-energy index, as is usually done in LSP FSRQs \citep[e.g.,][]{2019ApJ...871..211P,2019ApJ...874...47V}, since most of the high-redshift blazars are not detected with \fermi-LAT. Therefore, we froze it to a value of 5.4 derived from the \gm-ray photon index estimated using the stacking technique. To reduce the number of free parameters, we fixed the maximum value of the random Lorentz factor of the electron population (\gm$_{\rm max}$) to 1500. The viewing angle  ($\theta_{\rm v}$) was also frozen to 3$^{\circ}$ which is typically adopted in the blazar SED modeling and consistent with that inferred from radio studies of blazars \citep[][]{2005AJ....130.1418J}. Note that, since blazar jets are viewed within a maximum $\theta_{\rm v}$ of 1/$\Gamma$, one can get a meaningful constraint on the average viewing angle directly from $\Gamma$ also.

In all high-redshift blazars, the synchrotron emission peak was found to be located at self-absorbed frequencies ($<$10$^{12}$ Hz). Therefore, we considered the observed radio emission to get an idea about the typical flux level of the synchrotron radiation which was found to be low. Accordingly, the computed SSC emission remained well below the observed X-ray spectrum allowing us to constrain the size of the emission region, hence $R_{\rm diss}$, and the magnetic field. By reproducing the X-ray SED with the EC process, we were able to determine the low-energy index of the electron energy distribution from the observed X-ray spectral shape. The level of the X-ray flux constrained the bulk Lorentz factor and also controlled $R_{\rm diss}$. This is because the radiative energy densities of various AGN components used to estimate the EC flux vary as a function of $R_{\rm diss}$ \citep[][]{2009MNRAS.397..985G}. Note that due to lack of hard X-ray data and \gm-ray non-detection, the high energy peak is not well constrained. Therefore, we use the soft X-ray spectrum and the \fermi-LAT sensitivity limits (shown with black stars in Figure~\ref{fig:sed}) to get an idea about the approximate position of the inverse Compton peak. Similarly, the level of the synchrotron emission is not well constrained, especially for those that have a single point radio detection. In such cases, we are driven by our current understanding about jet physics. We know that FSRQ SEDs are Compton dominated, however, the Compton dominance (CD) cannot be very large ($>$1000). Since it is the ratio of inverse Compton to synchrotron peak luminosities, neither synchrotron peak can have an extremely low flux value nor inverse Compton peak can have very large flux. The former should be, on average, of the order of the observed radio emission or probably larger, keeping in mind the synchrotron self absorption. The high-energy peak cannot have very large flux, as constrained from the \fermi-LAT sensitivity limits. Also, an extremely bright peak demands a large bulk Lorentz factor ($>$20-30), which is likely to be unrealistic based on previous blazar population studies \citep[e.g.,][]{2015MNRAS.448.1060G,2017ApJ...851...33P}. Altogether, this leaves a limited allowed range for both SED peaks. Further details about the adopted methodology can be found in \citet[][]{2017ApJ...851...33P}.

\section{Observed Properties}\label{sec:obs_prop}
\subsection{Gamma-rays}\label{subsec:gamma}
The analysis of $\sim$11 years of the \fermi-LAT data has not revealed any new \gm-ray emitting blazars beyond $z=3$, other than those present in the 4FGL catalog. The faintness of the high-redshift blazars in the \gm-ray band is not only due to their large distances but also probably has a physical origin. Since the same electron population is expected to radiate both the low- and high-energy peaks, the LSP nature of these sources, in turn, suggests the high-energy SED bump is located at relatively lower ($\sim$MeV) frequencies. Due to the $k$-correction effect \citep[][]{2002astro.ph.10394H}, the SED peak shifts towards hard X-rays, making the \gm-ray spectrum steeper in the \fermi-LAT energy range. The enhancement in the luminosity as redshift increases also contributes to this effect, causing high-redshift blazars to become fainter in \gm-rays and brighter in the hard X-ray-to-MeV band.

\begin{deluxetable*}{lccccccccccll}[t!]
\tablecaption{The results of the spectral analysis of the analyzed X-ray data obtained with \swift-XRT, \xmm, and/or \chandra~satellites. \label{tab:XRT}}
\tabletypesize{\small}
\tablewidth{0pt}
\tablehead{
\colhead{Name} & \colhead{N$_{\rm H}$} & \colhead{Exp.} &  \multicolumn{3}{c}{soft X-ray flux} &  \multicolumn{3}{c}{Photon index} & \colhead{$\chi^2$/C-stat.} & \colhead{dof}& \colhead{Stat.}& \colhead{mission}\\
\colhead{} & \colhead{} & \colhead{} & \colhead{F$_{\rm X}$} & \colhead{F$_{\rm X,~low}$} & \colhead{F$_{\rm X,~high}$} & \colhead{$\Gamma_{\rm X}$}  & \colhead{$\Gamma_{\rm X,~low}$} & \colhead{$\Gamma_{\rm X,~high}$} & \colhead{} & \colhead{}& \colhead{}& \colhead{}\\
\colhead{[1]} & \colhead{[2]} & \colhead{[3]} & \colhead{[4]} & \colhead{[5]} & \colhead{[6]} & \colhead{[7]}  & \colhead{[8]} & \colhead{[9]} & \colhead{[10]} & \colhead{[11]}& \colhead{[12]}& \colhead{[13]}
}
\startdata
J000108+191434		&3.16	&5.56	&0.91	&0.00	&2.55	&1.38	&0.23	&2.74	&7.55	&6	&c-stat	&Swift\\
J000657+141546		&4.62	&11.68	&6.04	&4.64	&8.08	&1.37	&1.11	&1.63	&91.65	&83	&c-stat	&Swift\\
J001708+813508		&13.50	&13.41	&45.30	&50.50	&52.20	&1.40	&1.38	&1.42	&521.18	&494	&chi	&XMM\\
                                 		&13.50	&32.40	&49.50	&47.05	&51.96	&1.33	&1.28	&1.37	&128.01	&112	&chi	&Swift\\
J012100$-$280623		&1.60	&5.70	&2.01	&0.78	&6.18	&0.93	&$-$0.16	&1.98	&8.38	&9	&c-stat	&Swift\\
\enddata
\tablecomments{The column information are as follows. Col.[1]: source name (for brevity, we do not use the prefix NVSS, SUMSS, or FIRST); Col.[2]: the Galactic neutral Hydrogen column density, in 10$^{20}$ cm$^{-2}$; Col.[3]: observing exposure, in ksec; Col.[4], [5], and [6]: observed 0.3$-$10 keV (0.5$-$7 keV for \chandra) flux and its lower and upper limits, respectively, in units of 10$^{-13}$ \ergflux; Col.[7], [8], and [9]: power-law photon index and its lower and upper limits, respectively; Col.[10]: the $\chi^2$ or C-statistics value derived from the model fitting; Col.[11]: degrees of freedom; Col.[12]: adopted statistics, c-stat: C-statistics \citep[][]{1979ApJ...228..939C}, and chi: $\chi^2$ fitting; and Col.[13]: name of the satellite.\\
(This table is available in its entirety in a machine-readable form in the online journal. A portion is shown here for guidance regarding its form and content.)}

\end{deluxetable*}

\begin{deluxetable*}{lcccccccccl}[t!]
\tablecaption{The results of the spectral analysis of the analyzed hard X-ray data obtained with \nustar. \label{tab:nustar}}
\tabletypesize{\small}
\tablewidth{0pt}
\tablehead{
\colhead{Name} & \colhead{Exp.} &  \multicolumn{3}{c}{hard X-ray flux} &  \multicolumn{3}{c}{Photon index} & \colhead{$\chi^2$/C-stat.} & \colhead{dof}& \colhead{Stat.}\\
\colhead{} & \colhead{} & \colhead{F$_{\rm X}$} & \colhead{F$_{\rm X,~low}$} & \colhead{F$_{\rm X,~high}$} & \colhead{$\Gamma_{\rm X}$}  & \colhead{$\Gamma_{\rm X,~low}$} & \colhead{$\Gamma_{\rm X,~high}$} & \colhead{} & \colhead{}& \colhead{}\\
\colhead{[1]} & \colhead{[2]} & \colhead{[3]} & \colhead{[4]} & \colhead{[5]} & \colhead{[6]} & \colhead{[7]}  & \colhead{[8]} & \colhead{[9]} & \colhead{[10]} & \colhead{[11]}}
\startdata
J001708+813508	&31.00	&10.73	&9.98	&11.48	&1.77	&1.71	&1.84	&101.39	&101	&chi\\
J012201+031002	&30.83	&2.14	&1.65	&2.60	&1.61	&1.41	&1.83	&275.73	&276	&c-stat\\
J013126$-$100931	&29.91	&11.19	&9.97	&12.14	&1.43	&1.35	&1.52	&451.19	&498	&c-stat\\
J020346+113445	&31.66	&2.62	&2.15	&3.01	&1.77	&1.59	&1.95	&279.16	&299	&c-stat\\
J052506$-$233810	&20.93	&11.70	&10.07	&12.93	&1.38	&1.29	&1.48	&367.44	&429	&c-stat\\
J064632+445116	&32.16	&2.02	&1.64	&2.36	&1.76	&1.56	&1.96	&242.26	&255	&c-stat\\
J090630+693031	&79.33	&0.20	&0.05	&0.28	&1.93	&1.41	&2.51	&241.89	&244	&c-stat\\
J102623+254259	&59.39	&0.21	&0.00	&0.32	&1.43	&0.61	&2.50	&153.43	&165	&c-stat\\
J102838$-$084438	&30.69	&3.00	&2.42	&3.49	&1.63	&1.47	&1.80	&247.59	&310	&c-stat\\
J135406$-$020603	&53.11	&2.22	&1.69	&2.66	&1.31	&1.14	&1.49	&279.18	&362	&c-stat\\
J143023+420436	&49.19	&5.32	&4.62	&5.88	&1.52	&1.43	&1.62	&407.41	&466	&c-stat\\
J151002+570243	&36.86	&2.91	&2.19	&3.48	&1.19	&1.00	&1.40	&242.29	&284	&c-stat\\
J155930+030447	&53.39	&0.41	&0.23	&0.53	&1.91	&1.52	&2.32	&234.83	&223	&c-stat\\
J193957$-$100240	&39.26	&2.30	&1.95	&2.60	&2.02	&1.86	&2.19	&290.13	&341	&c-stat\\
J212912$-$153841	&33.32	&28.45	&26.71	&29.99	&1.56	&1.51	&1.60	&129.94	&158	&chi\\
\enddata
\tablecomments{The column details are as follows. Col.[1]: name of the source (for brevity, we do not use the prefix NVSS, SUMSS, or FIRST); Col.[2]: net exposure, in ksec; Col.[3], [4], and [5]: observed 3$-$79 keV flux and its lower and upper limits, respectively, in units of 10$^{-12}$ \ergflux; Col.[6], [7], and [8]: power-law photon index and its lower and upper limits, respectively; Col.[9]: the $\chi^2$ or C-statistics value derived from the model fitting; Col.[10]: degrees of freedom; and Col.[11]: adopted statistics, c-stat: C-statistics and chi: $\chi^2$ fitting.}

\end{deluxetable*}
We search for the cumulative \gm-ray signal from the 133 \fermi-LAT undetected sources by stacking their likelihood profiles \citep[][]{2019ApJ...882L...3P}. The derived results are shown in Figure~\ref{fig:stacking} where we also show the stacked TS profile of 133 empty \gm-ray sky positions representing the cumulative background emission. This exercise was done in 0.3$-$300 GeV energy range. The motivation behind using the minimum energy as 300 MeV instead of 100 MeV is to avoid the bright background emission embedded in the data (see appendix A for details). We estimate a combined TS of TS$_{\rm peak}=26.1$ and average photon flux $F_{\rm 0.3-300~GeV}=9.2^{+1.8}_{-1.9}\times 10^{-11}$ \phflux~and photon index $\Gamma_{\rm 0.3-300~GeV}=3.3^{+0.4}_{-0.2}$. This observation suggests that the population of the high-redshift blazars is a \gm-ray emitter, though individual objects are too faint to detect with \fermi-LAT. Furthermore, the steep \gm-ray spectrum is expected from the high-redshift blazar population. The computed photon flux is also about an order of magnitude lower than the detection threshold of the \fermi-LAT revealing the capabilities of the stacking technique in extracting the signal from the \gm-ray undetected population. Also note that above 300 MeV, the signal-to-noise ratio is better, mostly because of the narrowing of the LAT point spread function.

\begin{figure*}[t!]
\hbox{\hspace{0cm}
\includegraphics[scale=0.4]{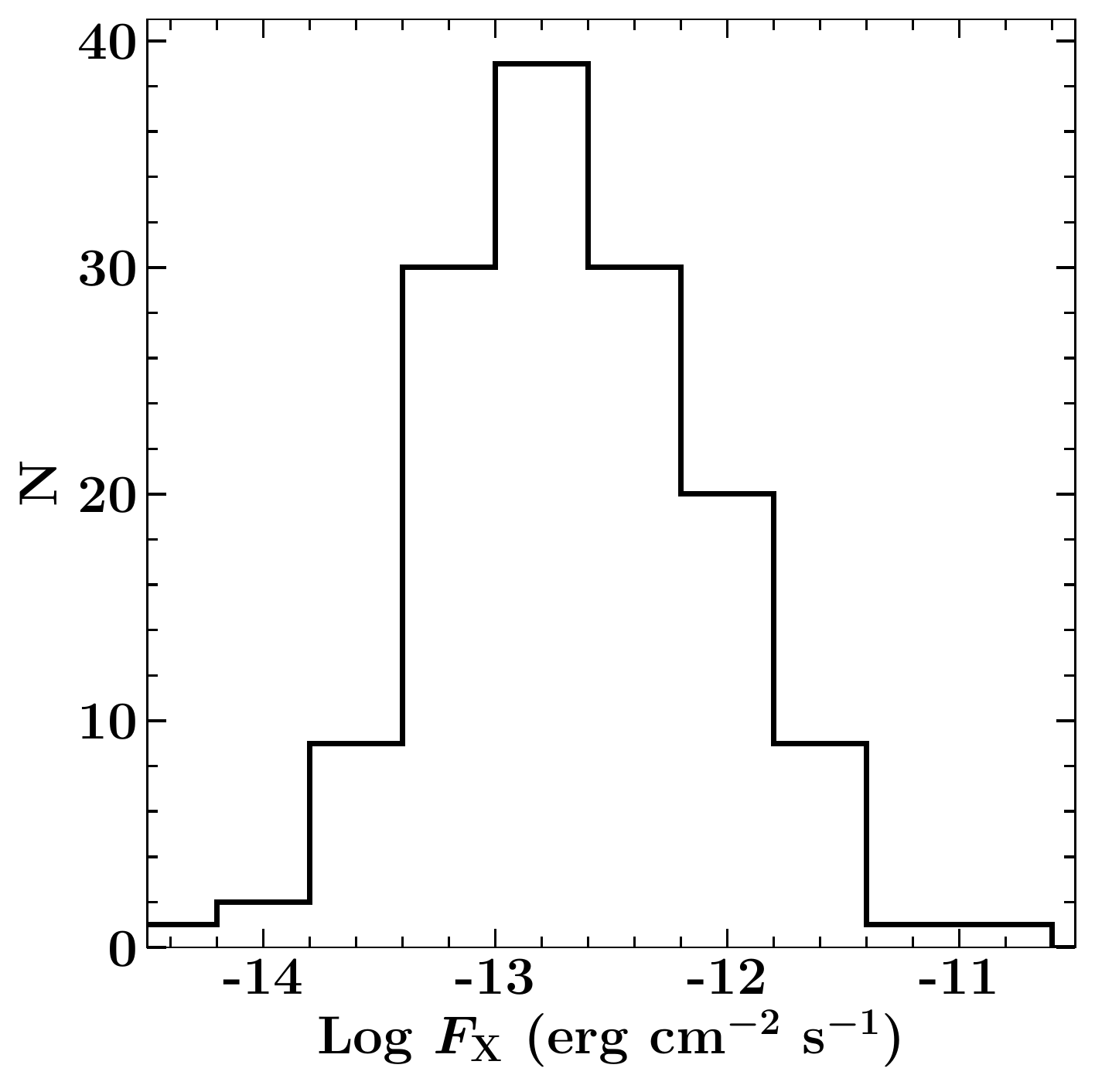}
\includegraphics[scale=0.4]{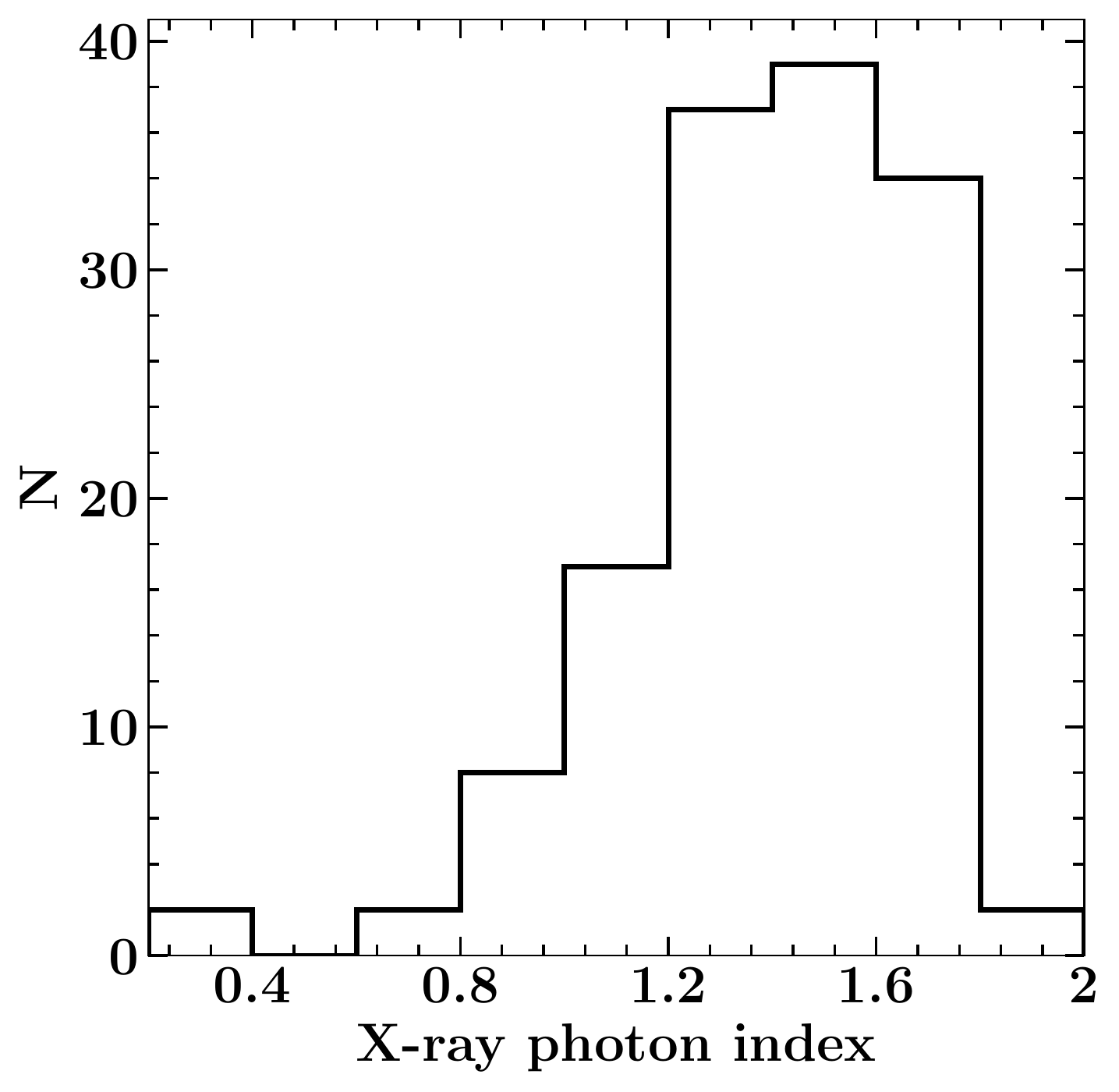}
\includegraphics[scale=0.4]{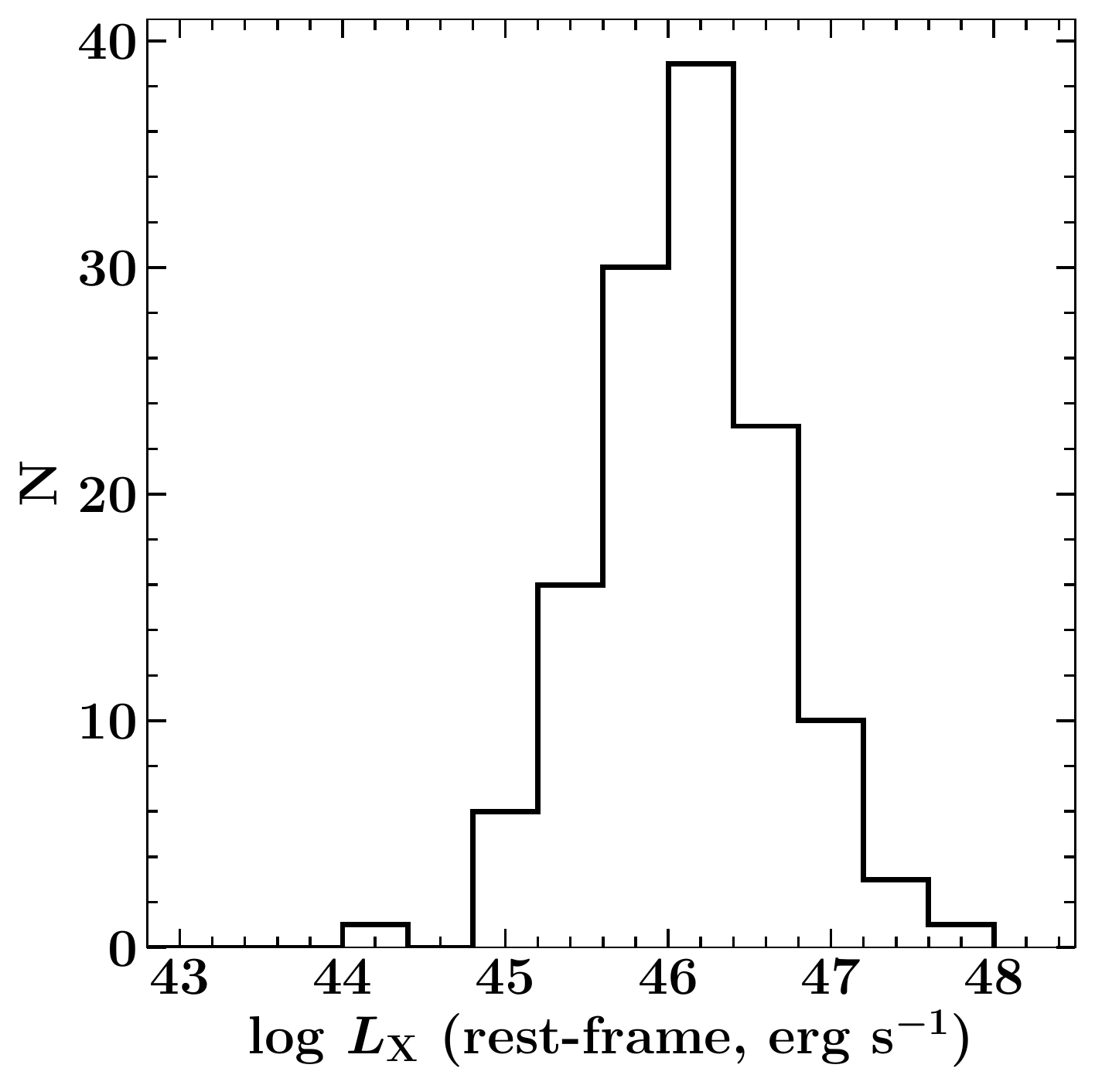}
}
\caption{The histograms of the observed X-ray flux (left), photon index (middle), and luminosity (right) for the high-redshift blazars. Note that for sources with X-ray observations taken with more than one satellite, we consider the one with the smallest uncertainty in the X-ray photon index. \label{fig:x-ray}}
\end{figure*}

\subsection{X-rays}
There are a total of 104 \swift-XRT, 54 \chandra, and 18 \xmm~observations of the high-redshift blazars present in the sample. The X-ray spectral parameters derived by fitting a simple absorbed power law model for all sources are provided in Tables~\ref{tab:XRT} and \ref{tab:nustar} and shown in Figure~\ref{fig:x-ray}.

High-redshift blazars are faint X-ray sources with average X-ray flux $\langle \log~F_{\rm X} \rangle=-12.71$ (in logarithmic scale of \ergflux). Their average X-ray spectral shape is hard with $\langle \Gamma_{\rm X} \rangle=1.42$. This might be due to our criterion of considering only the hardest spectrum objects. The estimated $k$-corrected, rest-frame X-ray luminosity reveals that the high-redshift sources are luminous (Figure~\ref{fig:x-ray}) with $\langle \log~L_{\rm X} \rangle=46.09$  (in logarithmic scale of \lum), likely due to Malmquist bias.

 \begin{figure*}[t!]
\hbox{
\includegraphics[scale=0.75]{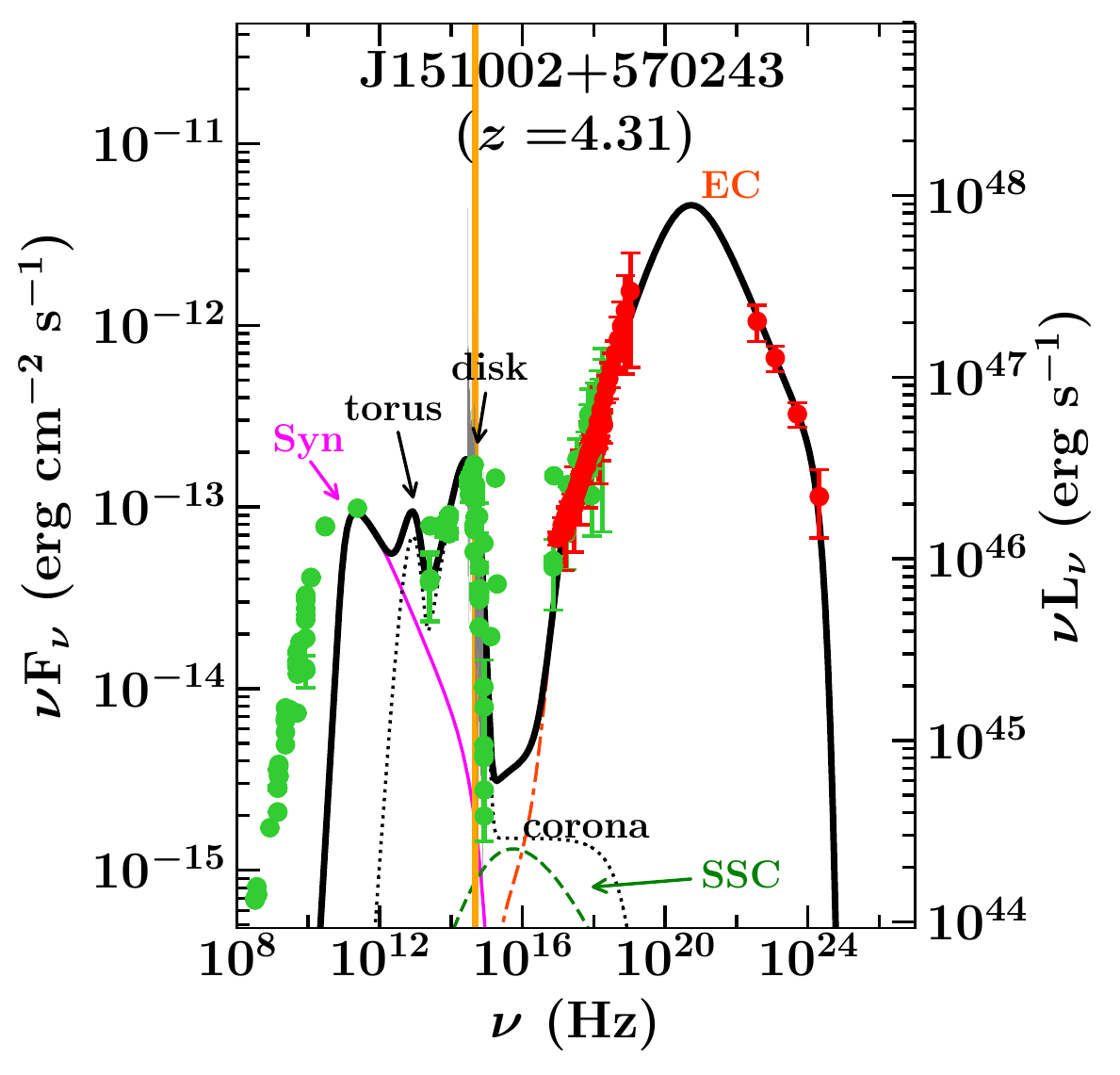}
\includegraphics[scale=0.75]{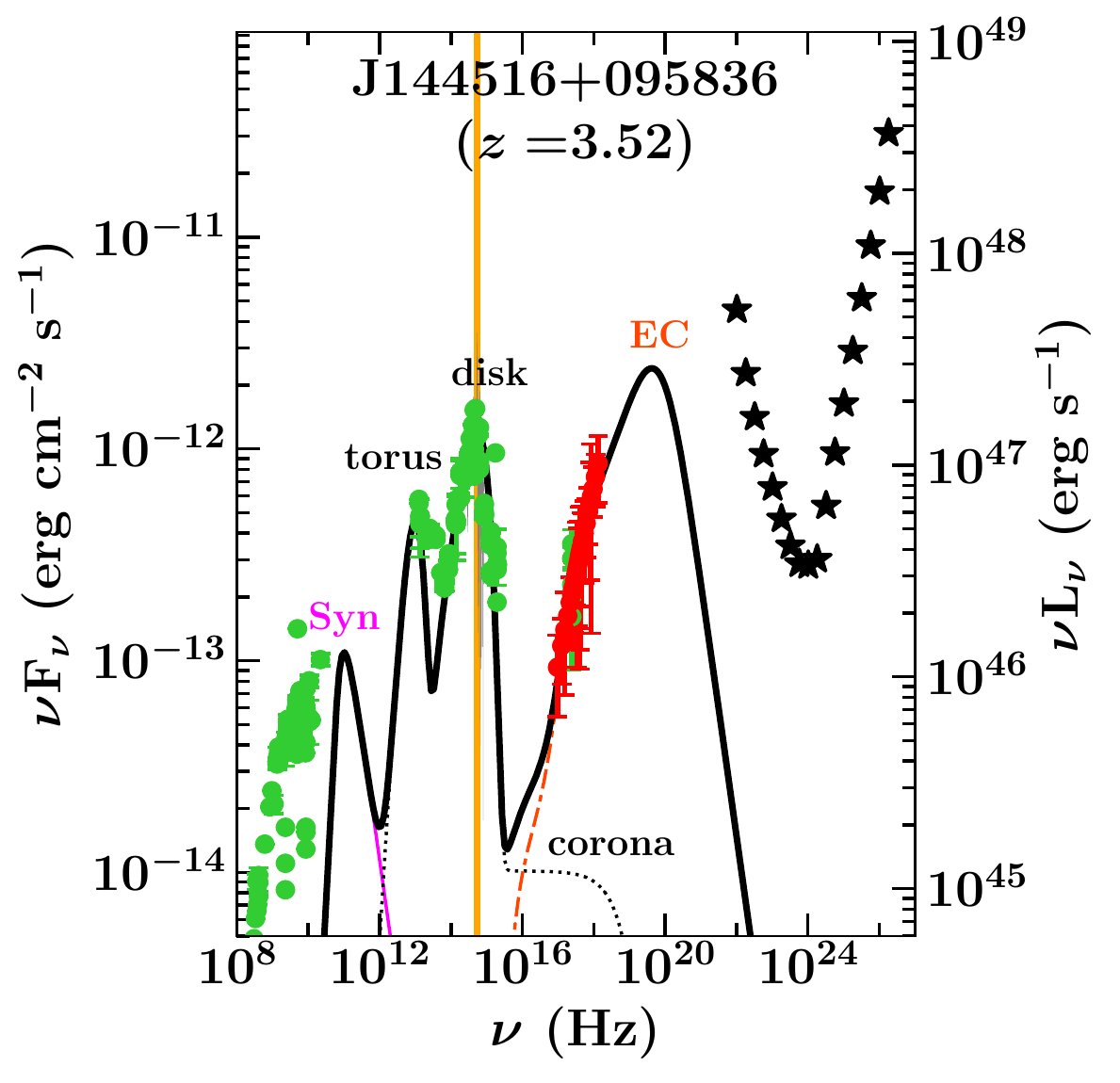}
}
\hbox{
\includegraphics[scale=0.75]{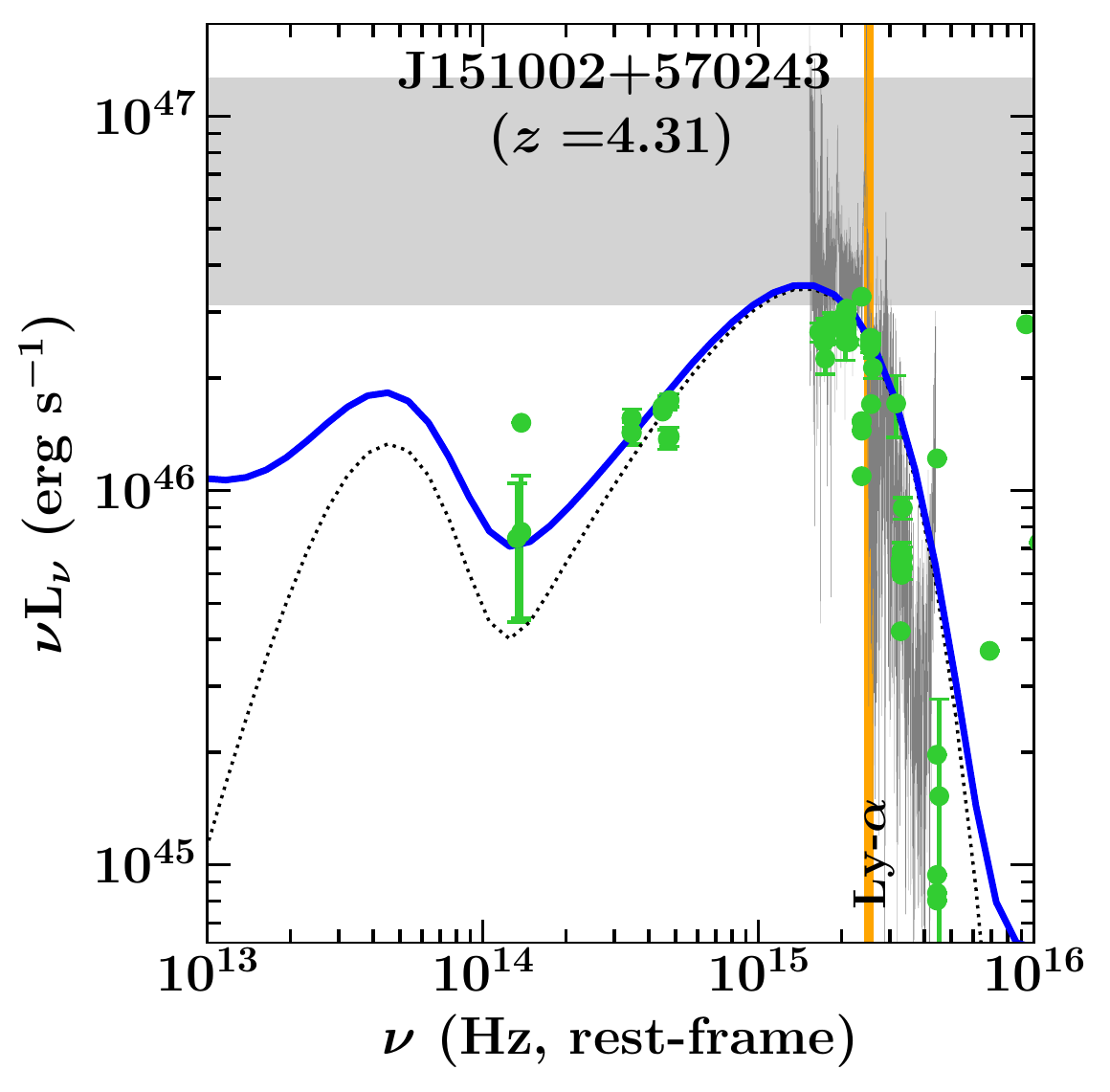}
\includegraphics[scale=0.75]{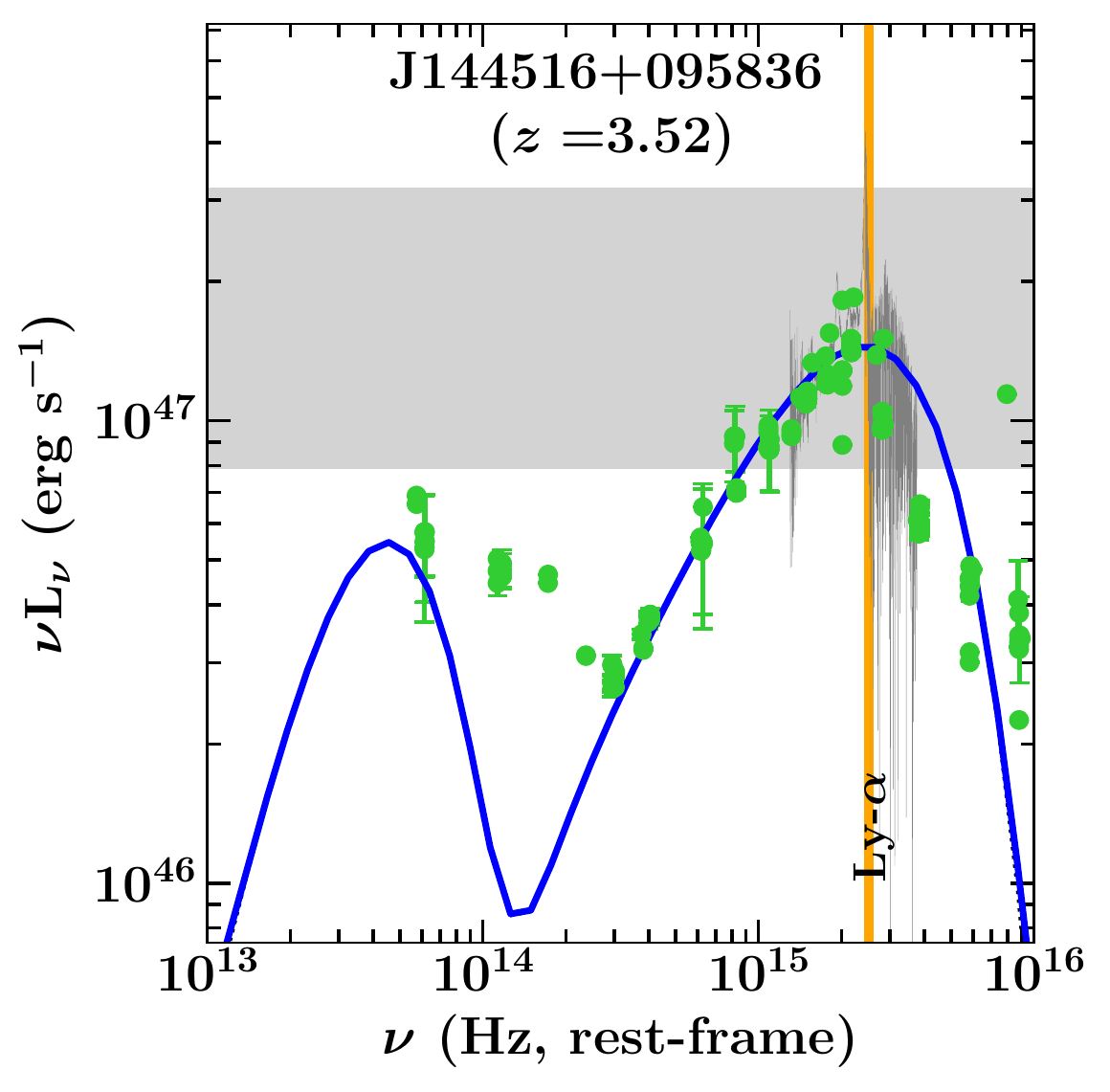}
}\caption{Top: the spectral energy distributions of the farthest \gm-ray detected blazar (left) and a \fermi-LAT undetected object present in our sample. The data analyzed by us (X- and \gm-rays) are shown with red circles, whereas those taken from the SSDC archive are represented with light green circles. Vertical orange line refers to the Lyman-$\alpha$ frequency. Various radiation mechanisms associated with the used leptonic model are labeled. In particular, pink thin solid, green dashed, and orange dash-dash-dot lines correspond to synchrotron, SSC, and EC processes, respectively. Black dotted line represent the thermal emission from the dusty torus, accretion disk, and the X-ray corona. The black thick solid line is the sum of all of the radiative components. At \gm-ray energies, black stars denote the \fermi-LAT sensitivity for the period covered in this work. Bottom: Same as above but zooming on the IR-UV part of the SED to highlight the measurement of the \ld~and \mbh~following  the disk modeling approach. The gray band refers to the \ld~estimated from the C{\sevenrm III},  C{\sevenrm IV}, or Lyman-$\alpha$ line luminosities, assuming an uncertainty of 0.3 dex. We also show the optical spectrum, whenever available, with grey line. We have not used data points bluer to the Lyman-$\alpha$ frequency in the modeling and show them only for completeness.\\
All the modeled SED plots for the other blazars are shown in the figure set.} \label{fig:sed}
\end{figure*}

\section{Physical Properties Inferred from the SED Modeling}\label{sec:sed_prop}

\begin{table*}
\caption{The parameters used/derived from the SED modeling of the high-redshift blazars.\label{tab:sed_param}}
\begin{center}
\begin{tabular}{lccccccccccccc}
\hline
Name & $z$ & $M_{\rm BH}$ & $L_{\rm disk}$ & $R_{\rm diss}$ & $R_{\rm BLR}$ & $\delta$ & $\Gamma$ & $B$ & $p$ & $\gamma_{\rm min}$ & $\gamma_{\rm b}$ & $U_{\rm e}$ & CD \\
~[1] & [2] & [3]  & [4] & [5] & [6]  & [7] & [8] & [9] & [10] & [11] & [12] & [13] & [14] \\ 
\hline
J000108+191434	& 3.10	& 9.30	& 46.23	& 0.165	& 0.133	& 12.3	& 7	& 1.2	& 1.7	    & 1	& 99	 & $-$1.39	& 4.3\\
J000657+141546	& 3.20	& 9.18	& 47.00	& 0.144	& 0.323	& 14.7	& 9	& 1.0	& 1.9	& 1	& 71	 & $-$1.37	& 89.8\\
J001708+813508	& 3.37	& 10.00& 	48.00	& 0.383	& 1.020	& 12.2	& 8	& 2.2	& 1.9	& 1	& 41	 & $-$1.55	& 25.3\\
J012100$-$280623	& 3.12	& 9.00	& 46.76	& 0.191    & 0.244	& 12.3	& 7	& 1.5	& 1.7    & 	1	& 89	 & $-$1.89	& 18.0\\
J012201+031002	& 4.00	& 9.43	& 46.08	& 0.116    & 0.112	    & 15.7	&10	& 0.8	& 2.0	& 1	& 65 &	$-$0.90	& 159.9\\

\hline
\end{tabular}
\end{center}
\tablecomments{The column information are as follows: Col.[1] source name; Col.[2]: redshift of the blazar; Col.[3]: log-scale black hole mass, in units of the solar mass; Col.[4]: log-scale luminosity of the accretion disk, in \lum; Col.[5]: distance of the emission region from the central black hole, in parsec; Col.[6]: radius of the spherical BLR, in parsec; Col.[7] and [8]: the Doppler factor and the bulk Lorentz factor, respectively; Col.[9]: magnetic field, in Gauss; Col.[10]: slopes of the broken power law electron energy distribution before the peak; Col.[11], and [12]: the minimum and the break Lorentz factors of the radiating electrons; Col.[13]: the log-scale electron energy density, in erg cm$^{-3}$; and Col.[14]: the Compton dominance.\\
(This table is available in its entirety in a machine-readable form in the online journal. A portion is shown here for guidance regarding its form and content.)}
\end{table*}

\begin{table}
\caption{Statistical summary of the SED parameters derived for the high-redshift blazars studied in this work. \label{tab:stat}}
\begin{center}
\begin{tabular}{lll}
\hline
Parameters & Mean &  Range\\
\hline
Disk luminosity   (log-scale, in \lum) & 46.7$\pm$0.4 & 45.9--48.0 \\
Black hole mass (log-scale, in \Msun) & 9.5$\pm$0.3 & 8.7--10.3 \\
Electron spectral index ($p$) & 1.8$\pm$0.2 & 1.1--2.2 \\
Break Lorentz factor (log-scale) & 1.8$\pm$0.2 & 1.3--2.3 \\
Magnetic field (in Gauss) & 1.0$\pm$0.5 & 0.2--3.2 \\
Dissipation distance (log-scale, in cm) & 17.8$\pm$0.2 & 17.3--18.5 \\
Compton dominance (log-scale) & 1.6$\pm$0.5 & 0.5--2.8 \\
Bulk Lorentz factor & 7.0$\pm$1.9 & 5.0--14.0 \\
Doppler factor & 12.3$\pm$2.3 & 9.3--22.6 \\
Electron jet power (log-scale, in \lum) & 45.0$\pm$0.5 & 43.4--46.1 \\
Magnetic jet power (log-scale, in \lum) & 45.2$\pm$0.6 & 43.4--46.6 \\
Radiative jet power (log-scale, in \lum) & 46.1$\pm$0.6 & 44.4--47.7 \\
Kinetic jet power (log-scale, in \lum) & 47.5$\pm$0.5 & 45.8--48.7 \\
\hline
\end{tabular}
\end{center}
\tablecomments{Quoted uncertainties in the mean values are the 1$\sigma$ standard deviation.}
\end{table}

We generate the broadband SEDs of all sources considered in this work using the methodology described in Section~\ref{sec:analysis} and reproduce them with a single-zone leptonic emission model as explained in Section~\ref{sec:model}. The modeled SEDs are shown in Figure~\ref{fig:sed} and we provide the associated SED parameters in Table~\ref{tab:sed_param}. In Table~\ref{tab:stat}, we provide the mean and 1$\sigma$ standard deviation for all the SED parameters. 

\subsection{Central Engine}\label{subsec:bh}
\begin{figure*}[t!]
\gridline{\fig{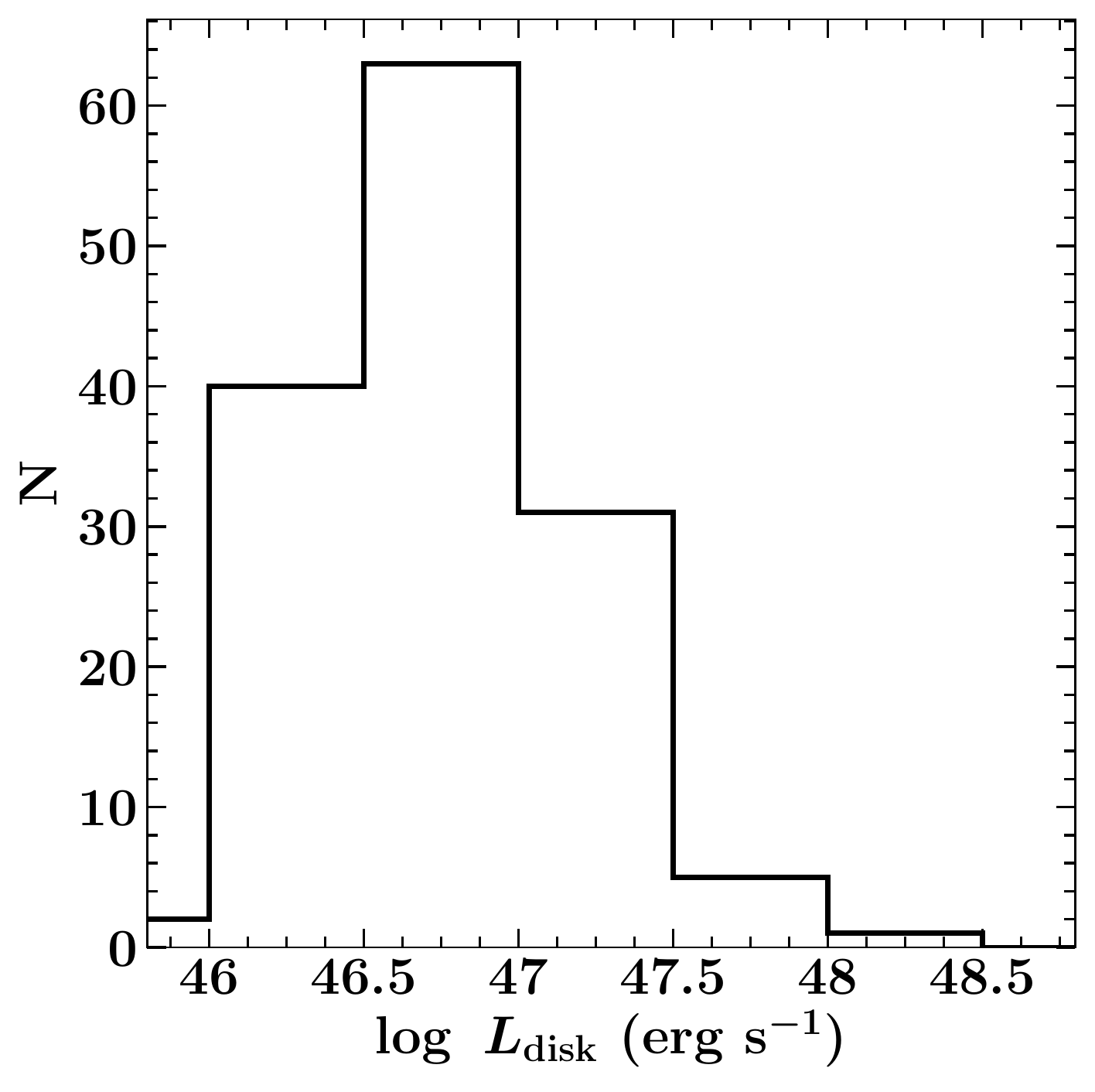}{0.24\textwidth}{(a)}
          \fig{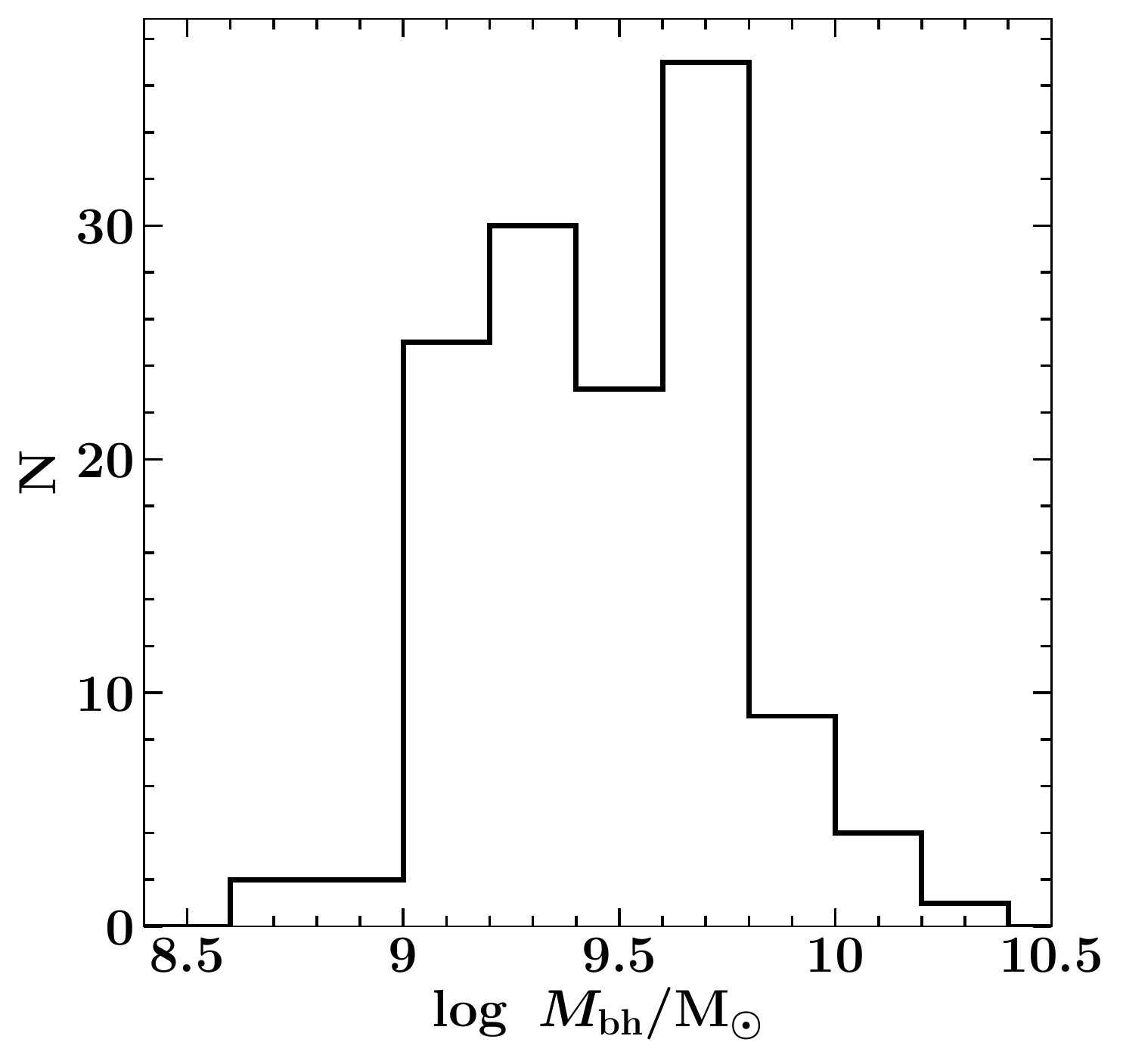}{0.24\textwidth}{(b)}
          }
\gridline{\fig{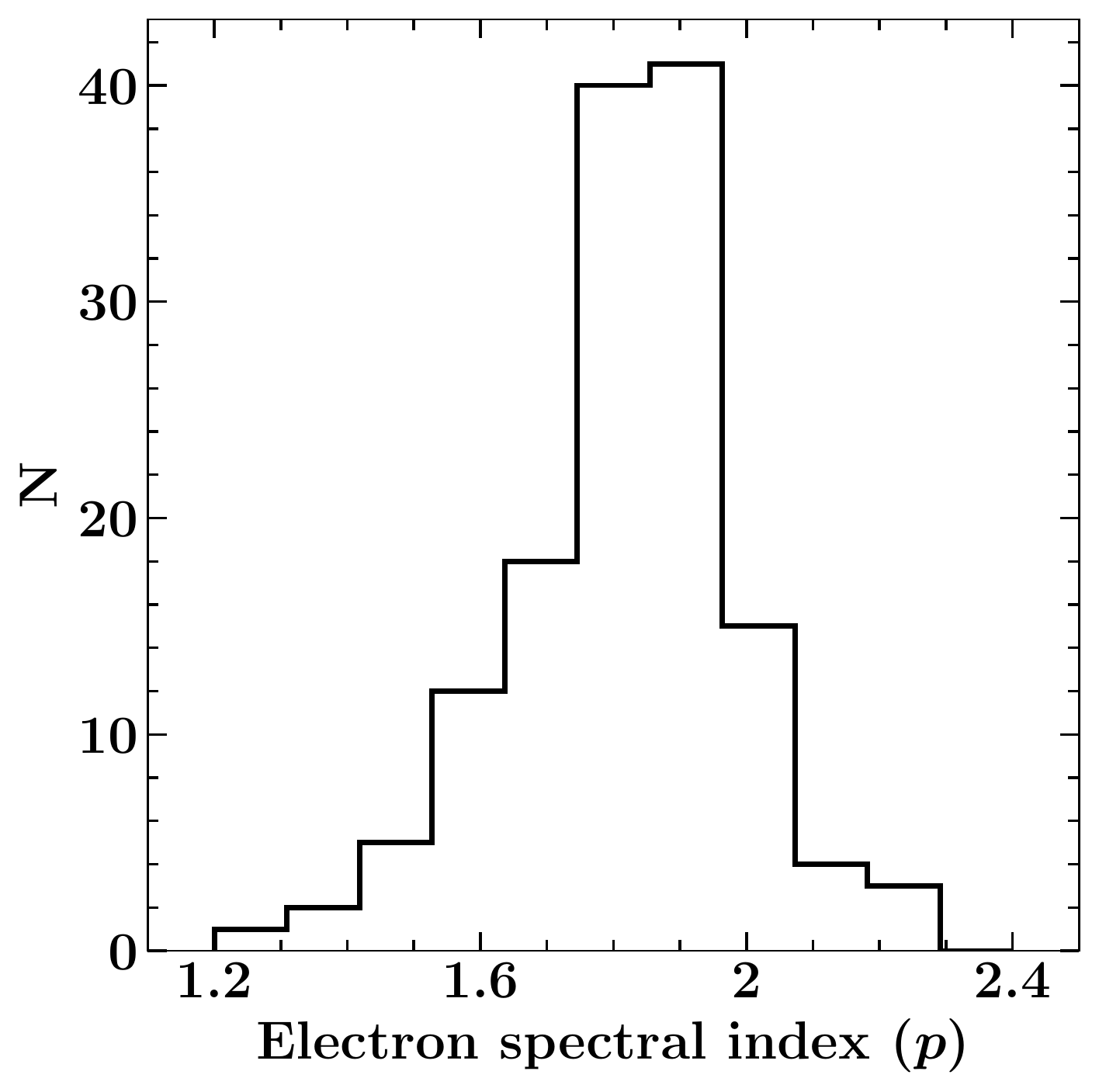}{0.24\textwidth}{(c)}
          \fig{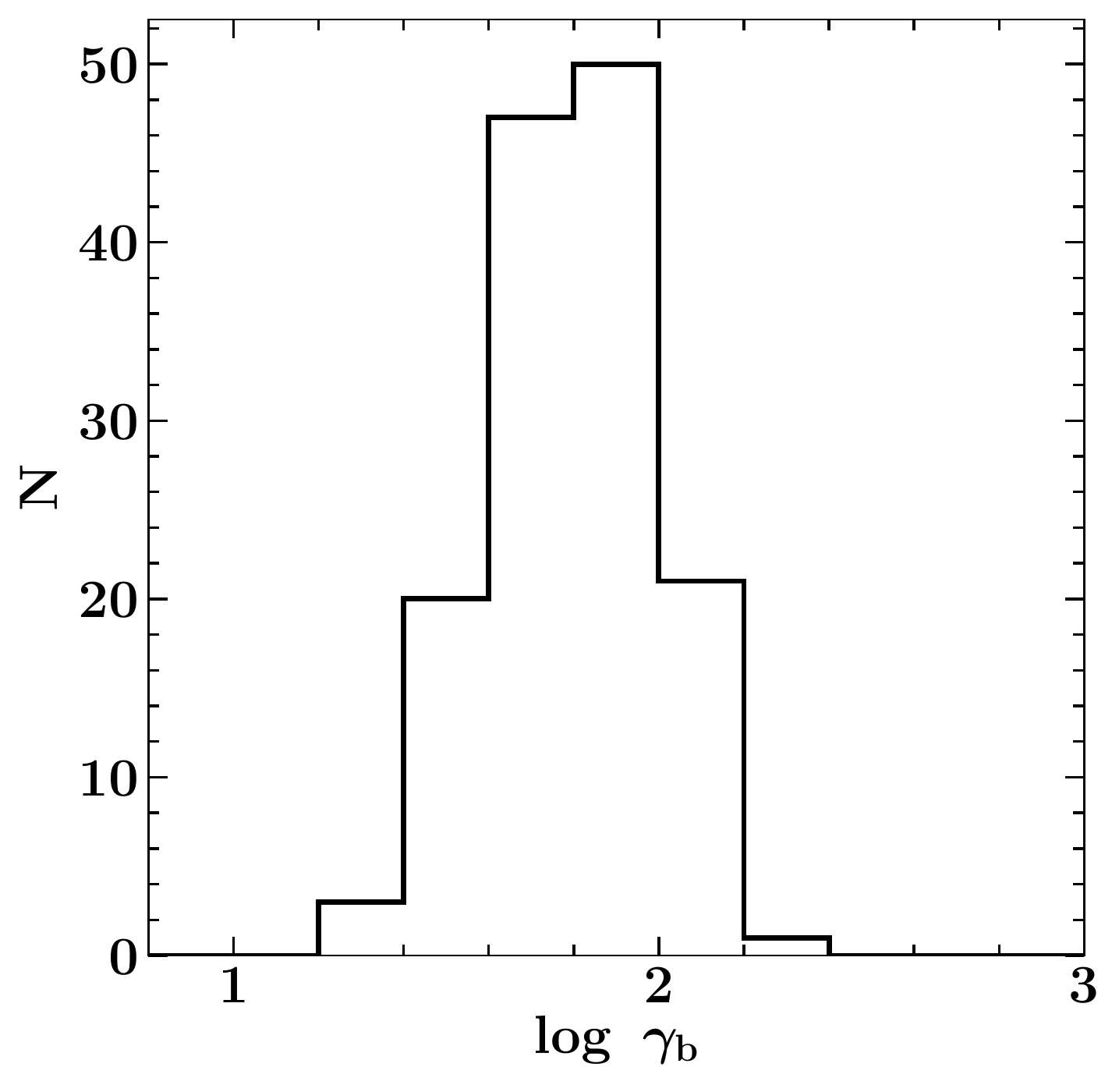}{0.24\textwidth}{(d)}
          \fig{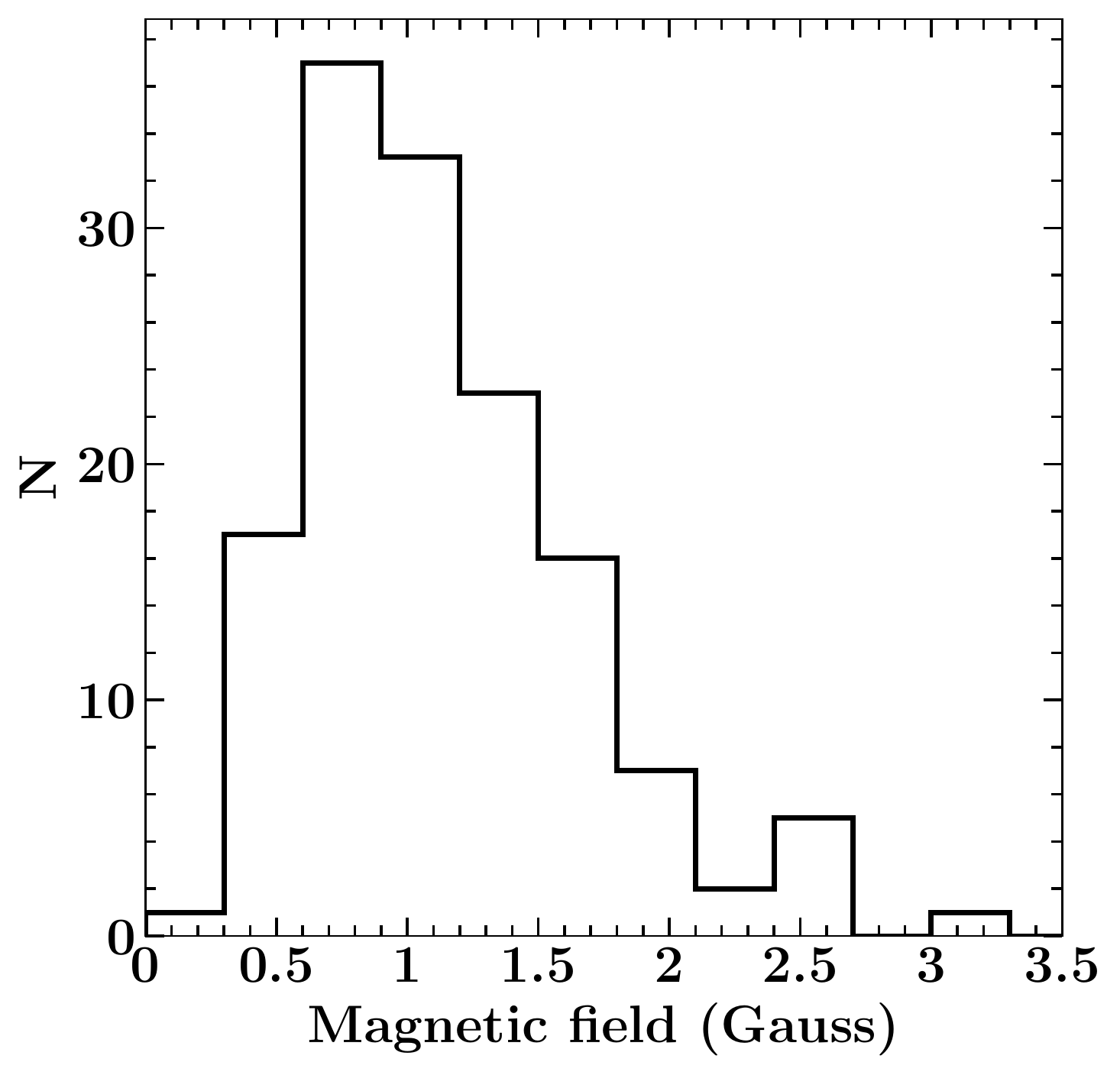}{0.24\textwidth}{(e)}
          \fig{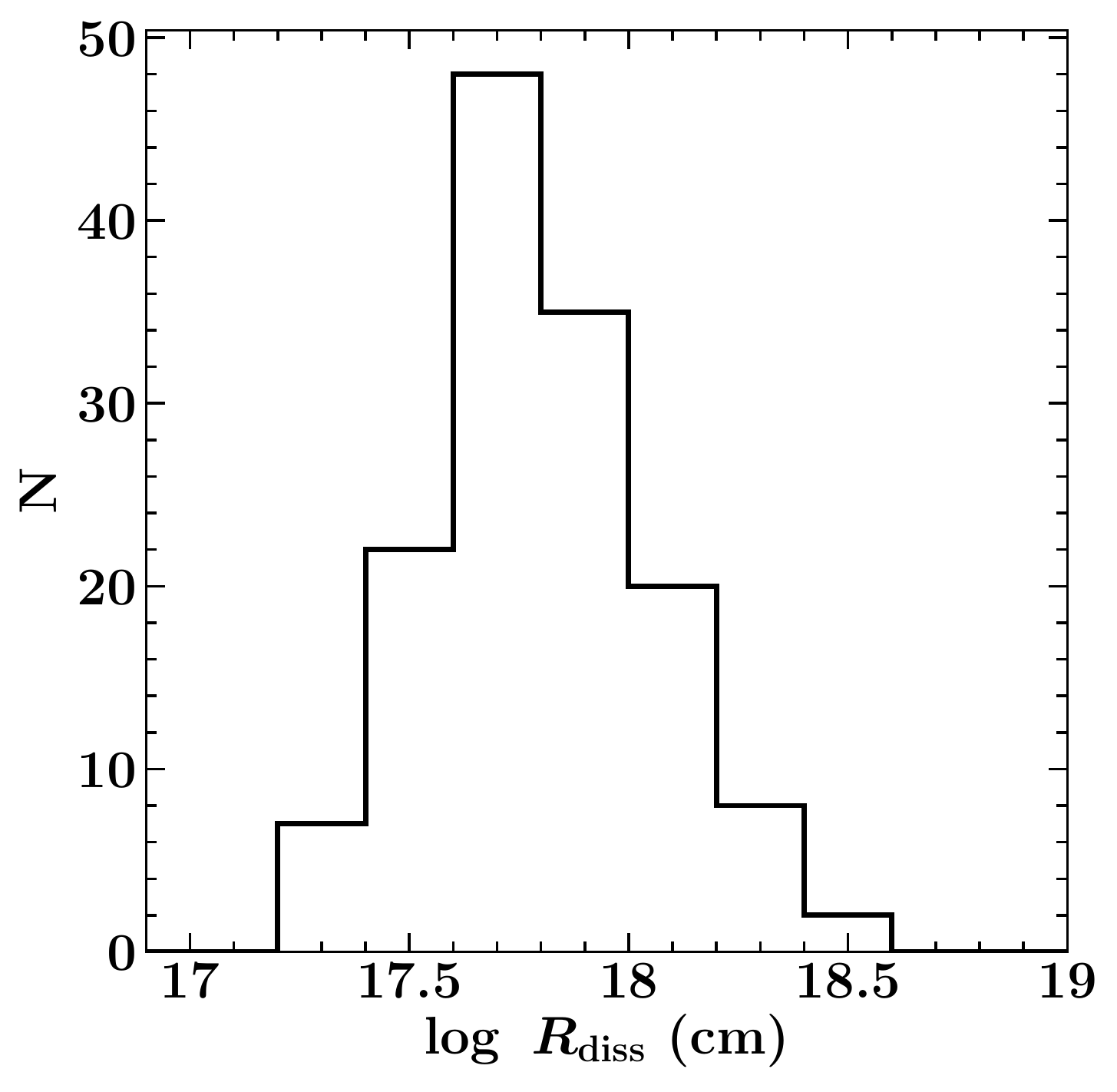}{0.24\textwidth}{(f)}
          }
\gridline{\fig{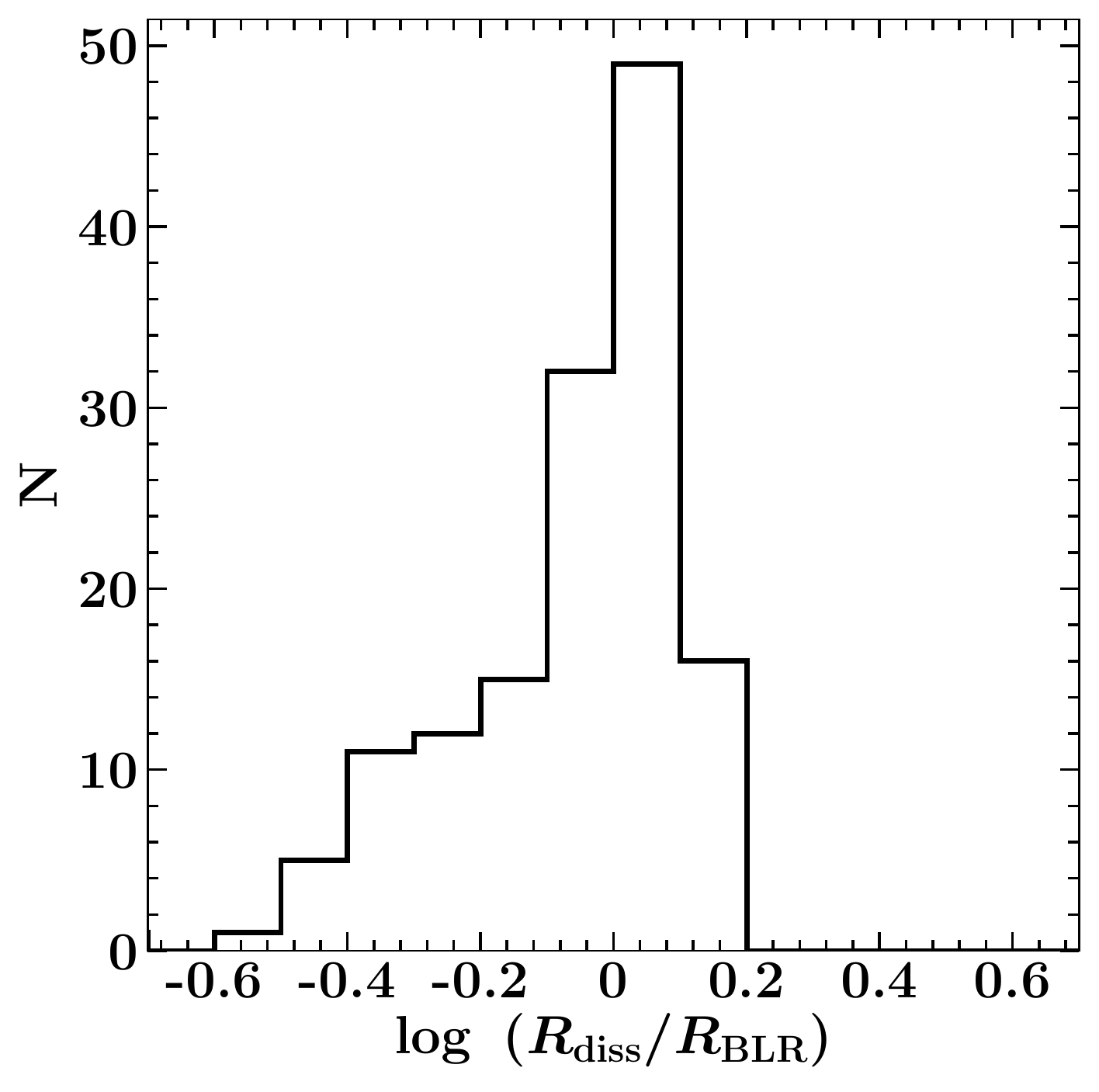}{0.24\textwidth}{(g)}
          \fig{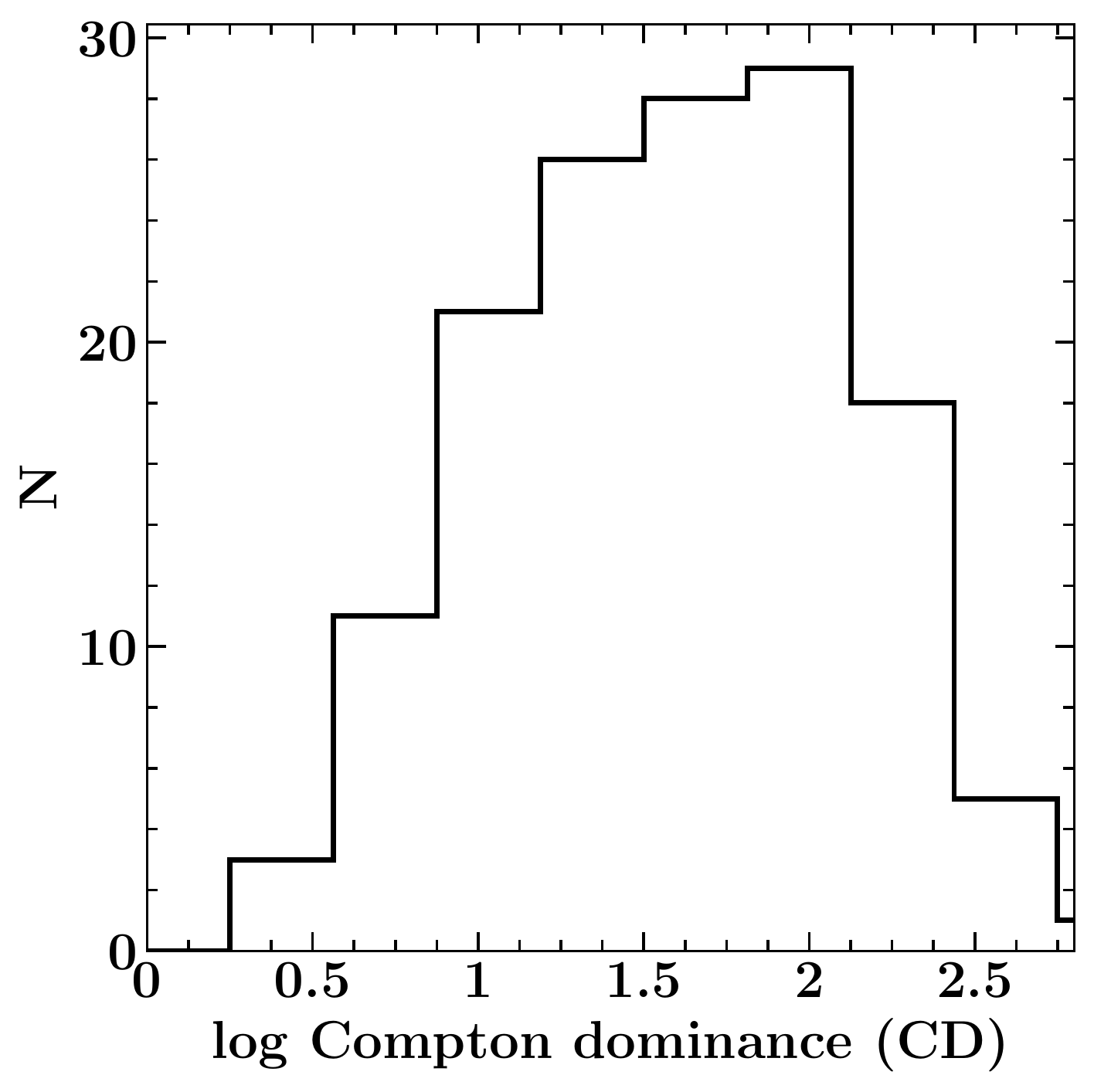}{0.24\textwidth}{(h)}
          \fig{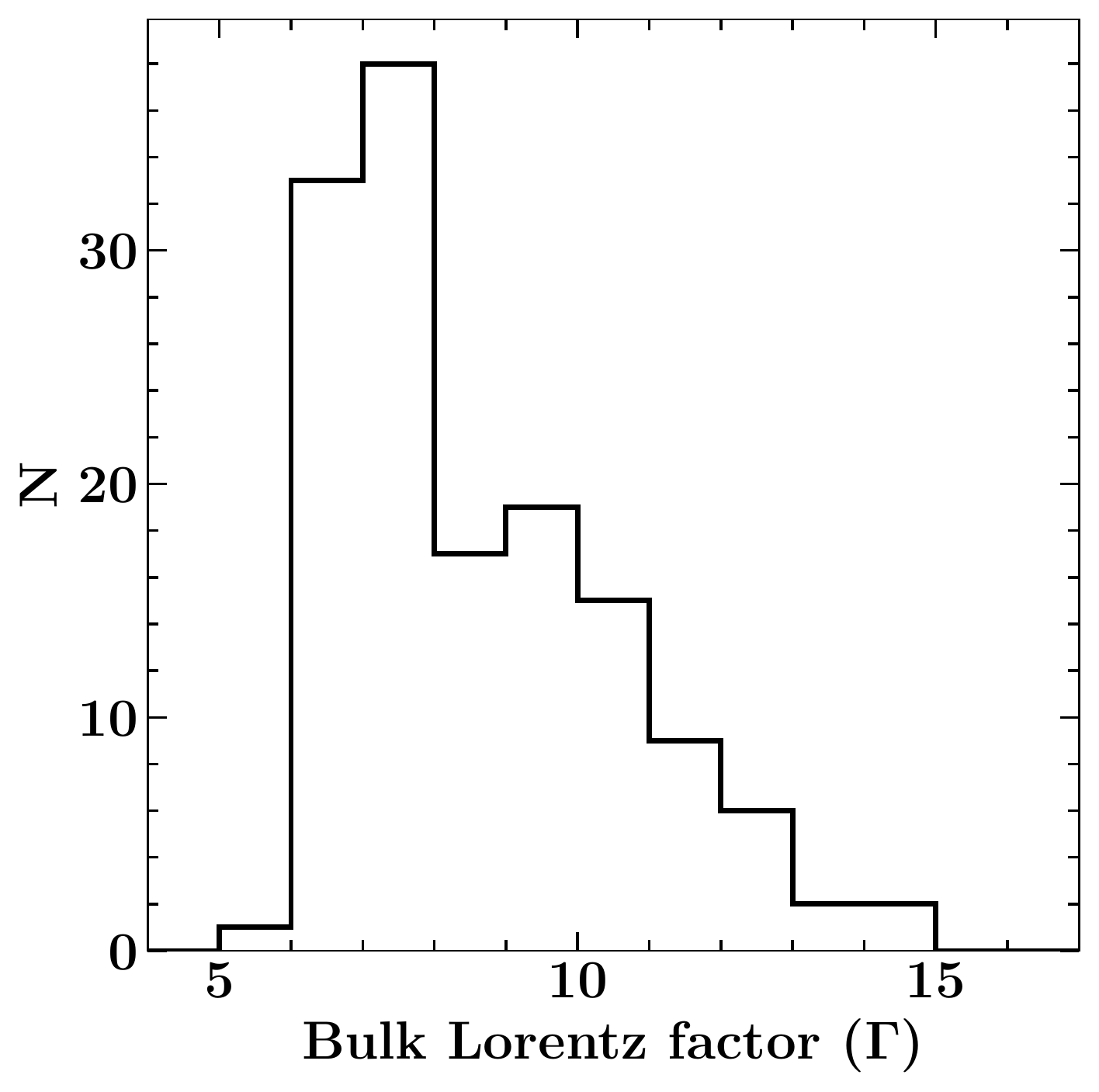}{0.24\textwidth}{(i)}
          \fig{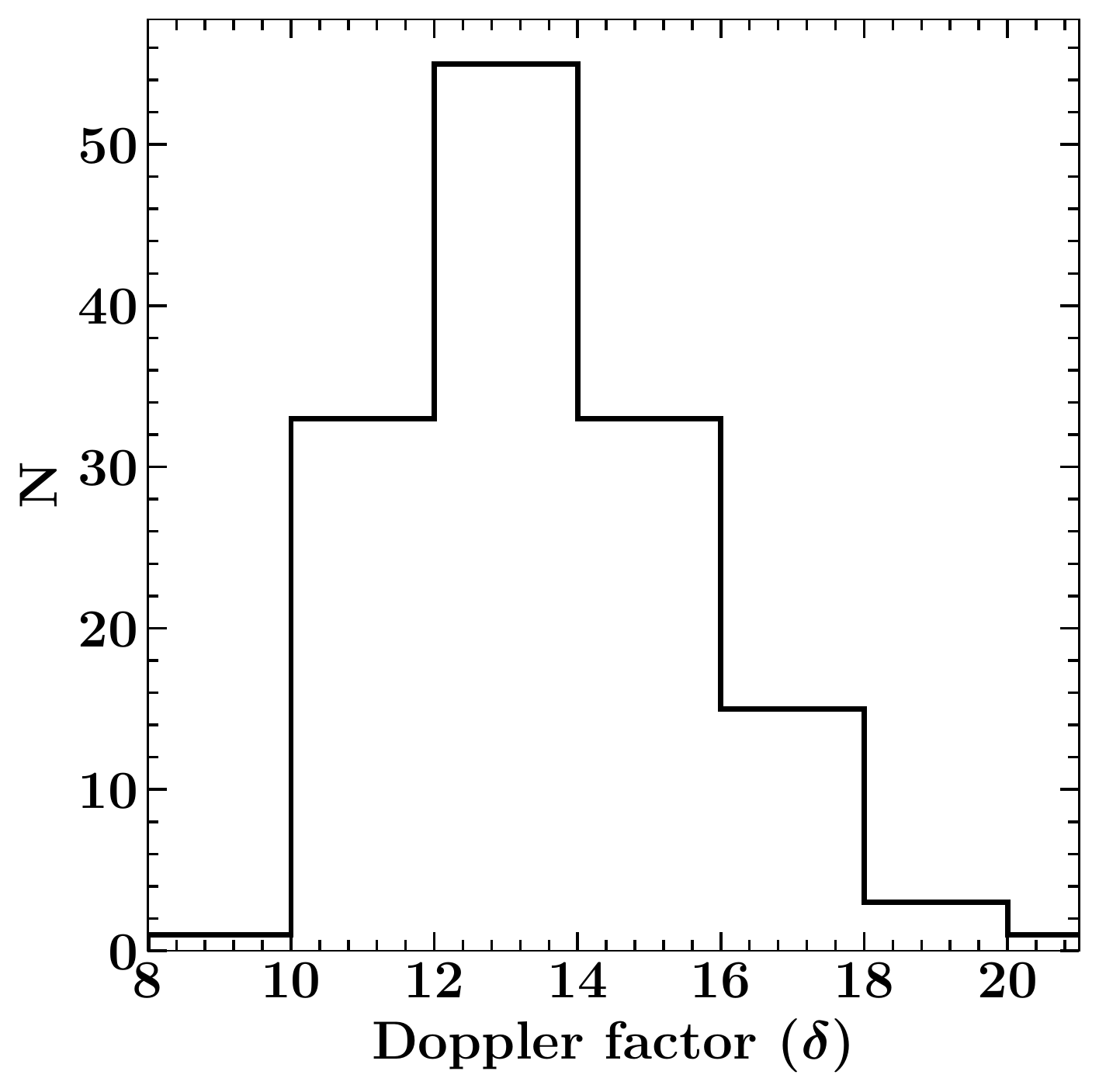}{0.24\textwidth}{(j)}
          }
\caption{Histograms of the SED parameters for the high-redshift blazars.\label{fig:histo}}
\end{figure*}

\begin{figure}[t!]
\includegraphics[scale=0.5]{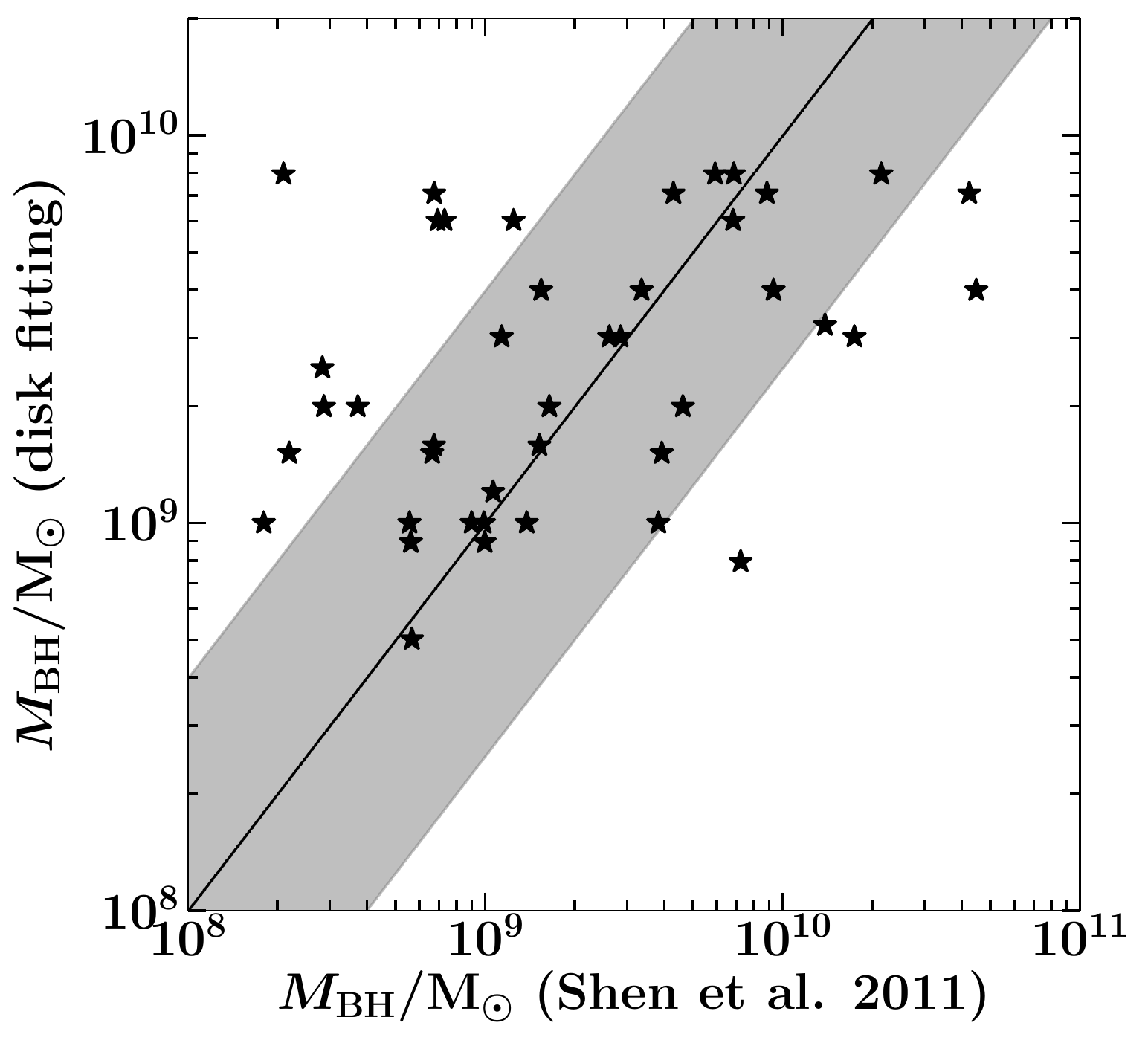}
\caption{A comparison of \mbh~values reported by \citet[][]{2011ApJS..194...45S} using C{\sevenrm IV} emission line with that derived using disk modeling method in this paper. The shaded area demonstrate an uncertainty factor of 4 associated with the virial technique.} \label{fig:bh_comp}
\end{figure}

We compute \mbh~and \ld~for all sources by reproducing their IR-optical emission with a standard \citet[][]{1973A&A....24..337S} accretion disk model. This approach is similar to that adopted in various recent studies \citep[e.g.,][]{2015MNRAS.450L..34G,2016MNRAS.462.1542S}. As discussed in Section~\ref{sec:model}, we considered only data points redder than the rest-frame frequency of the Lyman-$\alpha$ line. This is because the bluer data points may not reveal the true flux level of the disk due to absorption by the intervening Lyman-$\alpha$ clouds whose nature is uncertain. In the top panel of Figure~\ref{fig:histo}, we show the estimated \mbh~and \ld~values.  For the sake of completeness, we have compared \mbh~values derived from the disk fitting approach with C{\sevenrm IV} emission line based measurements for 45 blazars also studied by \citet[][]{2011ApJS..194...45S}.  As can be seen in Figure~\ref{fig:bh_comp}, a  majority of sources have comparable \mbh~values within the uncertainties associated with the virial technique. There is a large spread in \mbh~values derived from the virial method, likely due to complexity involved with the C{\sevenrm IV} emission lines \citep[e.g., narrow absorption troughs, cf.][]{2014ApJS..215...12C}.

The top panels of Figure~\ref{fig:histo} demonstrate the high-redshift sources host powerful central engines. The mean luminosity of the accretion disk is found to be $\langle \log~L_{\rm disk} \rangle=46.7$ (in units of \lum) for sources present in our sample. Moreover, the  high-redshift blazars are powered by massive black holes with $\langle \log~M_{\rm BH,~M{\odot}} \rangle=9.5$. These numbers are on the higher side with respect to the low-redshift blazar population \citep[][]{2017ApJ...851...33P,2019ApJ...881..154P}. This observation is likely due to a selection effect since only the most powerful objects are expected to be detected at high redshifts. However, this is a favorable bias as it allows us to identify and study the most massive black holes at the beginning of the Universe.

\subsection{Other SED Parameters}
In Figure~\ref{fig:histo}, we show the distributions of various SED parameters derived from the leptonic modeling and also overplot the same computed for low-redshift blazars for a comparison.

{\it Particle energy distribution:} The distribution of the low-energy index of the broken power law spectrum peaks at $\langle p \rangle=1.8$. Interestingly, the break Lorentz factor or \gm$_{b}$ for the high-redshift blazars has an average value $\langle \log~\gamma_{\rm b} \rangle=1.8$ with  a narrow dispersion (Table~\ref{tab:stat}). Since \gm$_{\rm b}$ indicates the SED peak locations, the derived results suggest the SED peaks of the high-redshift objects to lie at low frequencies. These results support the idea of the high-redshift blazars to be MeV-peaked and thus brighter in the hard X-ray band \citep[e.g.,][]{2010MNRAS.405..387G}.

{\it Magnetic Field and the Dissipation Distance:} According to our analysis, the average magnetic field strength in the high-redshift sources is $\langle B \rangle=1.0$ G (Figure~\ref{fig:histo}, panel (c)). Considering the distance of the emission region from the central black hole in absolute units, the mean is $\langle \log~R_{\rm diss,~cm} \rangle=17.8$. When normalized in $R_{\rm BLR}$ units, we noticed that a majority of the high-redshift blazars have a dissipation region located within the BLR (Figure~\ref{fig:histo}, panel (g)). This is because, in our model, the size of the BLR and dusty torus are a function of \ld~\citep[see also][]{2009MNRAS.397..985G} which is found to be larger, hence a bigger BLR, for the high-redshift sources.

{\it Compton Dominance:} The SEDs of the high-redshift blazars are found to be Compton dominated (Table~\ref{tab:stat}). This can be understood in terms of a relatively enhanced X-ray emission noticed in the high-redshift sources with respect to their radio emission \citep[see, e.g.,][]{2011ApJ...738...53S,2013ApJ...763..109W,2019MNRAS.482.2016Z}. Since the X-ray and radio fluxes are used to constrain the inverse Compton and synchrotron spectra, respectively, a larger Compton dominance is expected. In addition to that, Compton dominance is reported to be positively correlated with \ld~\citep[][]{2017ApJ...851...33P}. Therefore, the observation of Compton dominated SEDs in the high-redshift blazars can be understood since they have luminous accretion disks (Figure~\ref{fig:histo}).

\begin{table}
\caption{Various jet powers derived from the SED modeling.\label{tab:jet}}
\begin{center}
\begin{tabular}{lccccc}
\hline
Name &  $P_{\rm ele}$ & $P_{\rm mag}$ & $P_{\rm rad}$ & $P_{\rm kin}$ & $P_{\rm jet}$\\
~[1] & [2] & [3]  & [4] & [5] & [6]  \\ 
\hline
J000108+191434	&44.98	& 45.13	& 45.59	& 47.38	& 47.38\\
J000657+141546	&45.10	   & 45.07	    & 46.47	& 47.67	& 47.68\\
J001708+813508	&45.67	   & 46.51	   & 47.29	    & 48.18	& 48.19\\
J012100-280623	&44.61	   & 45.45	   & 46.03	& 47.03	& 47.04\\
J012201+031002	&45.48	& 44.79	& 46.65	& 48.12	& 48.12\\

\hline
\end{tabular}
\end{center}
\tablecomments{The column contents are as follows: Col.[1] name the source; Col.[2], [3], [4], [5], and [6]: log-scale electron, magnetic, radiative, kinetic, and total jet power, respectively. Note that $P_{\rm jet}$ = $P_{\rm ele}$ + $P_{\rm mag}$ + $P_{\rm kin}$.\\
(This table is available in its entirety in a machine-readable form in the online journal. A portion is shown here for guidance regarding its form and content.)}
\end{table}

{\it Jet Velocity:} The derived bulk Lorentz factor and Doppler factor for the high-redshift blazar population (Figure~\ref{fig:histo}, panel (i) and (j)), is $\langle \Gamma \rangle=7$ and $\langle \delta \rangle=12.3$, which are relatively smaller compared to that determined for other blazars located at $z<3$ \citep[cf.][]{2014Natur.515..376G}. In fact, \citet[][]{2011MNRAS.416..216V} proposed a decrease in $\Gamma$ as a likely factor to explain the deficiency of the parent population members of blazars at high redshifts. This is because, for each blazar with the jet Lorentz factor $\Gamma$, there are 2$\Gamma^2$ sources expected to be present in the same redshift bin and hence a low value of $\Gamma$ indicates fewer misaligned radio-loud quasars. Though model dependent, our findings provide crucial insights about the blazar evolution scenario and they are consistent not only with other studies where a low $\Gamma$ was estimated from the SED modeling \citep[][]{2018ApJ...856..105A} but also that inferred from radio studies \citep[e.g.,][]{2020NatCo..11..143A}. However, we cannot make a strong claim due to lack of $>$10 keV data for most of the sources. Observations in the hard X-ray band, e.g., with \nustar, are crucial to better constrain $\Gamma$ as shown in recent studies \citep[see, e.g.,][]{2013ApJ...777..147S,2018ApJ...856..105A}.

\begin{figure}
\includegraphics[scale=0.6]{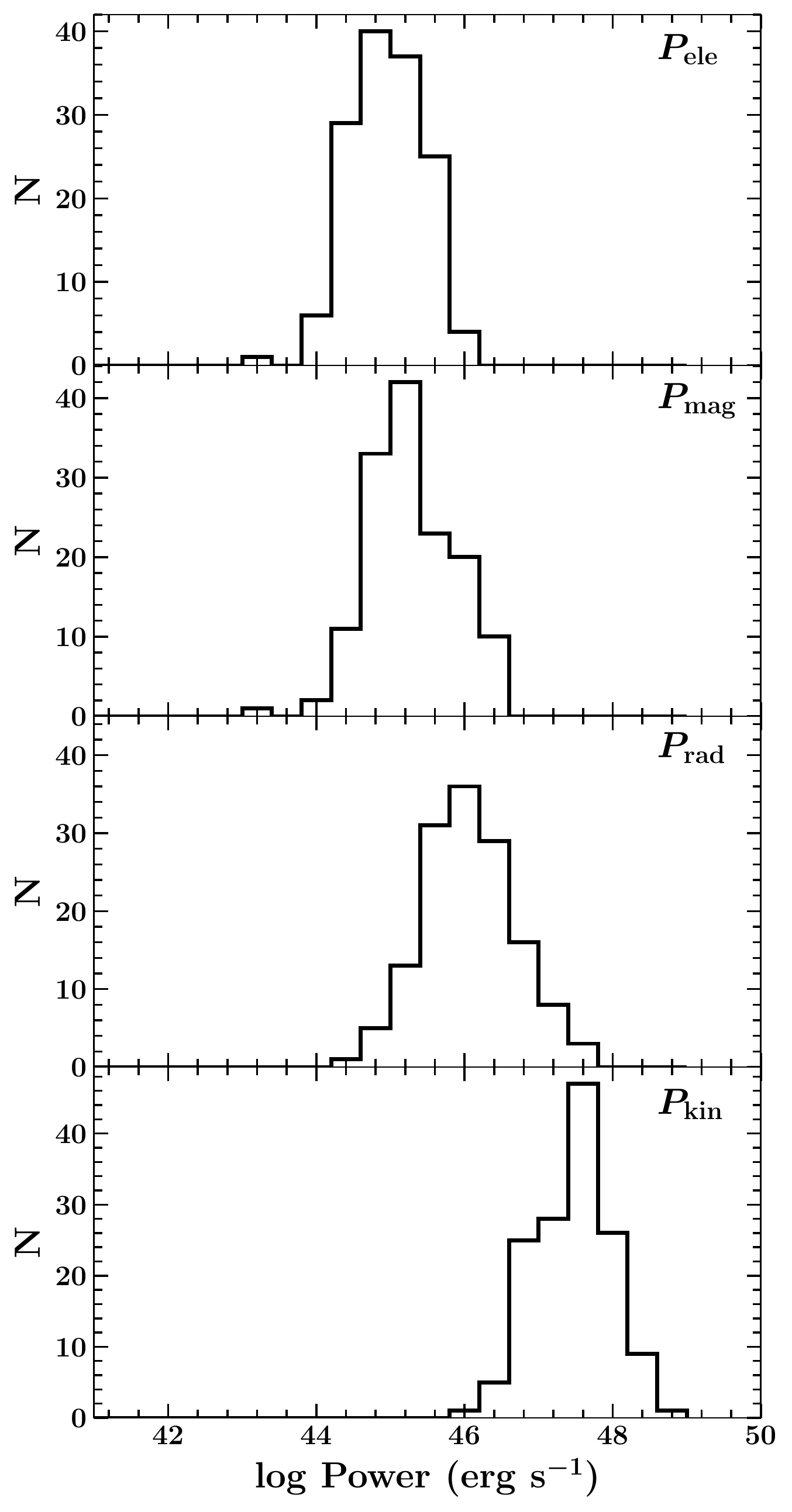}
\caption{Distributions of various jet powers, as labeled. Other information are same as in Figure \ref{fig:sed}. The statistics of the parameters are provided in Table~\ref{tab:stat}. \label{fig:jet}}
\end{figure}

\begin{figure}[t!]
\includegraphics[scale=0.55]{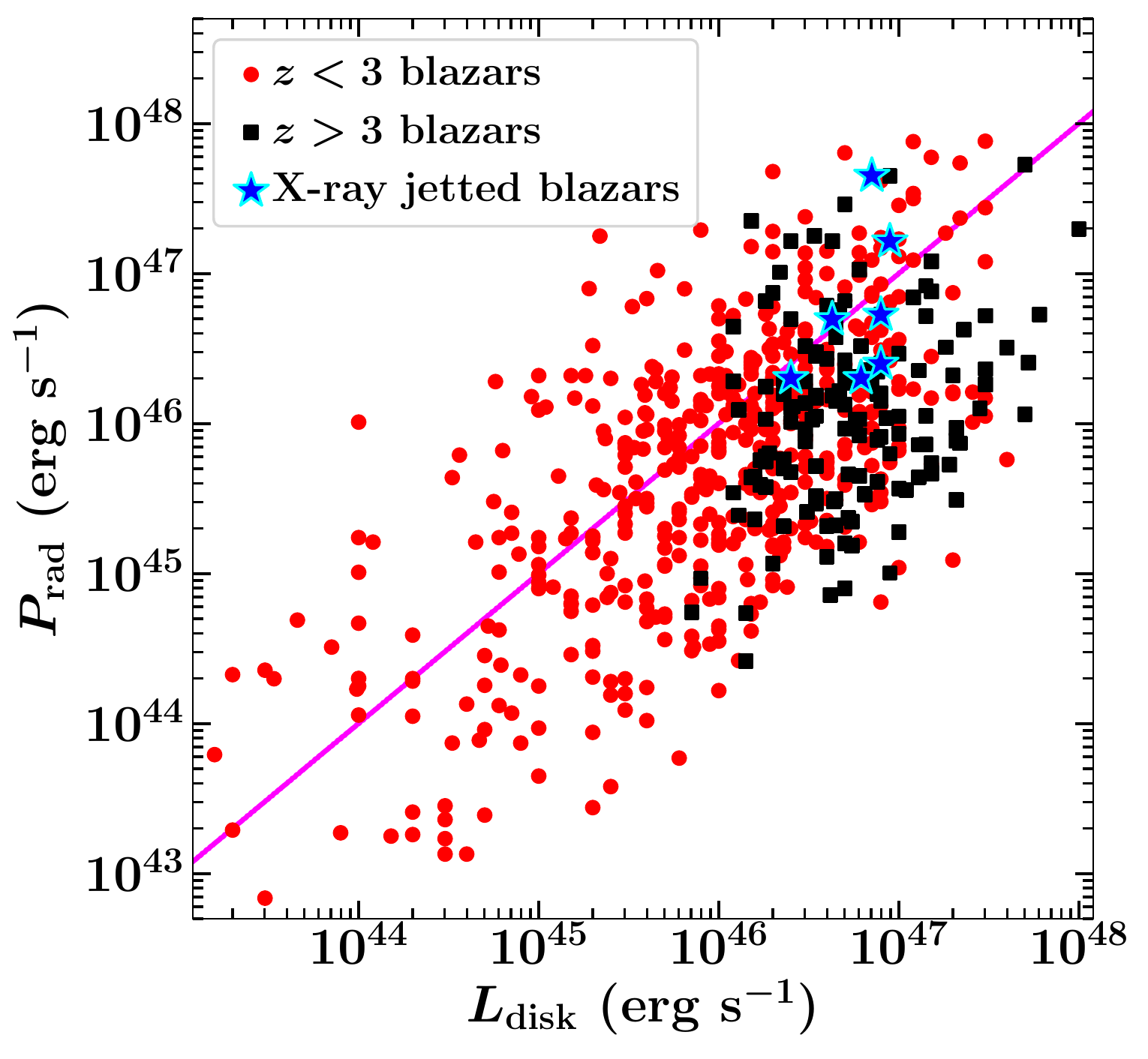}
\includegraphics[scale=0.55]{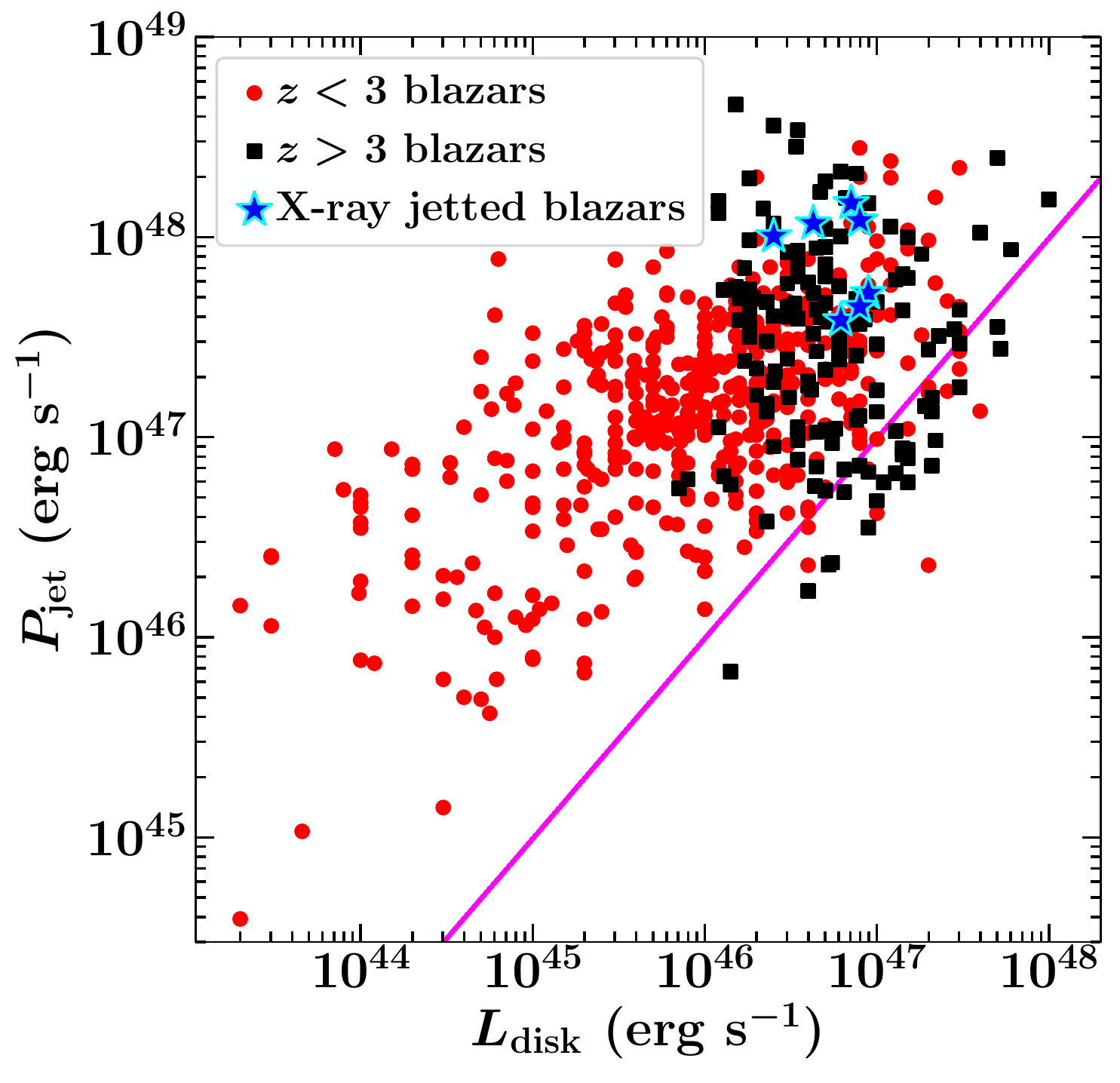}
\caption{The radiative jet power ($P_{\rm rad}$) and total jet power ($P_{\rm j}=P_{\rm p}+P_{\rm e}+P_{\rm m}$) as a function of the $L_{\rm d}$ are shown in the top and bottom panels, respectively. The high-redshift and $z<3$ blazars are displayed with black squares and red circles, respectively. We also plot 7 X-ray jetted blazars with blue stars. The pink line corresponds to the one-to-one correlation of the plotted quantities.} \label{fig:jet-disk}
\end{figure}

\begin{figure*}[t!]
\gridline{\fig{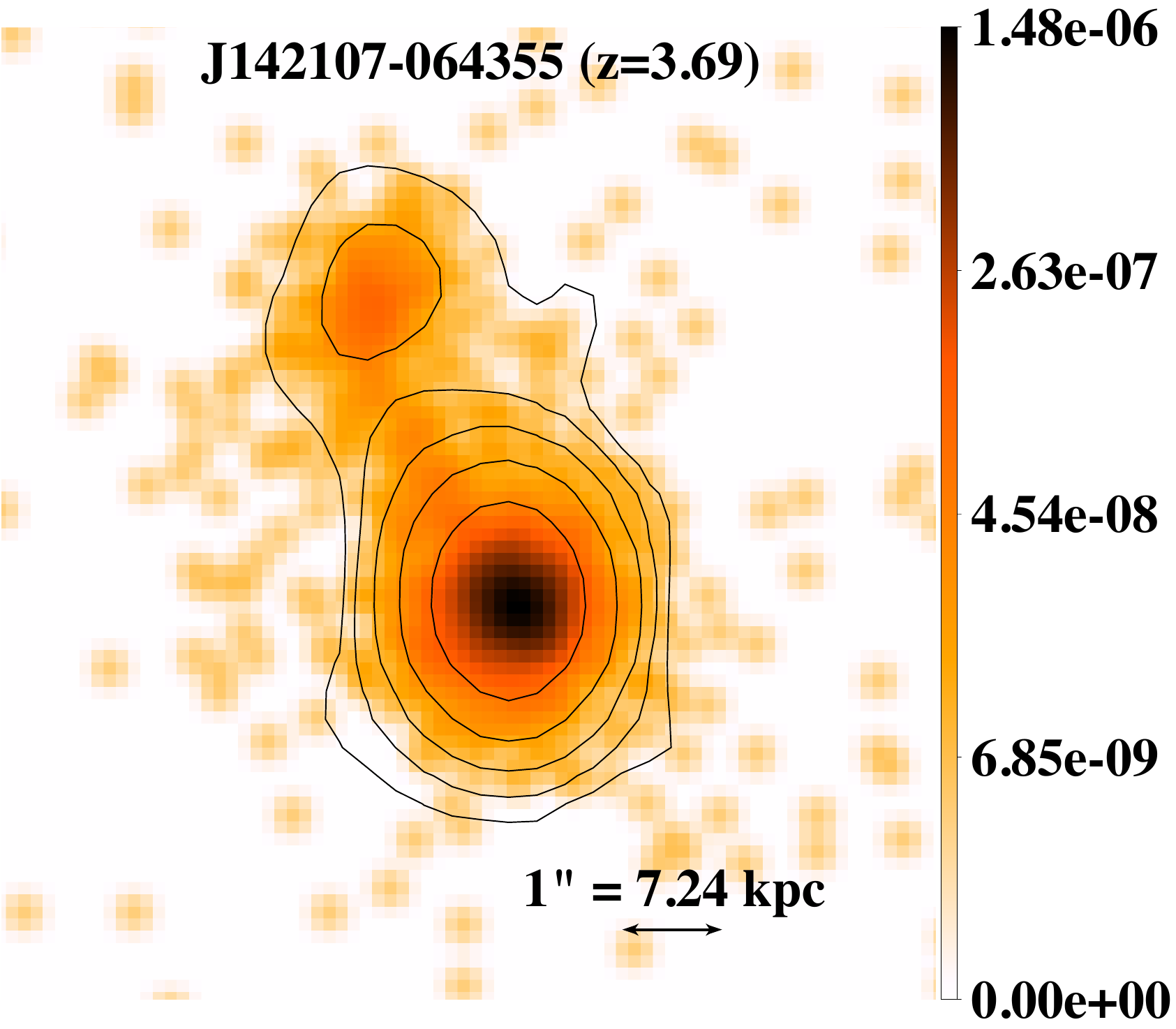}{0.3\textwidth}{(a)}
          \fig{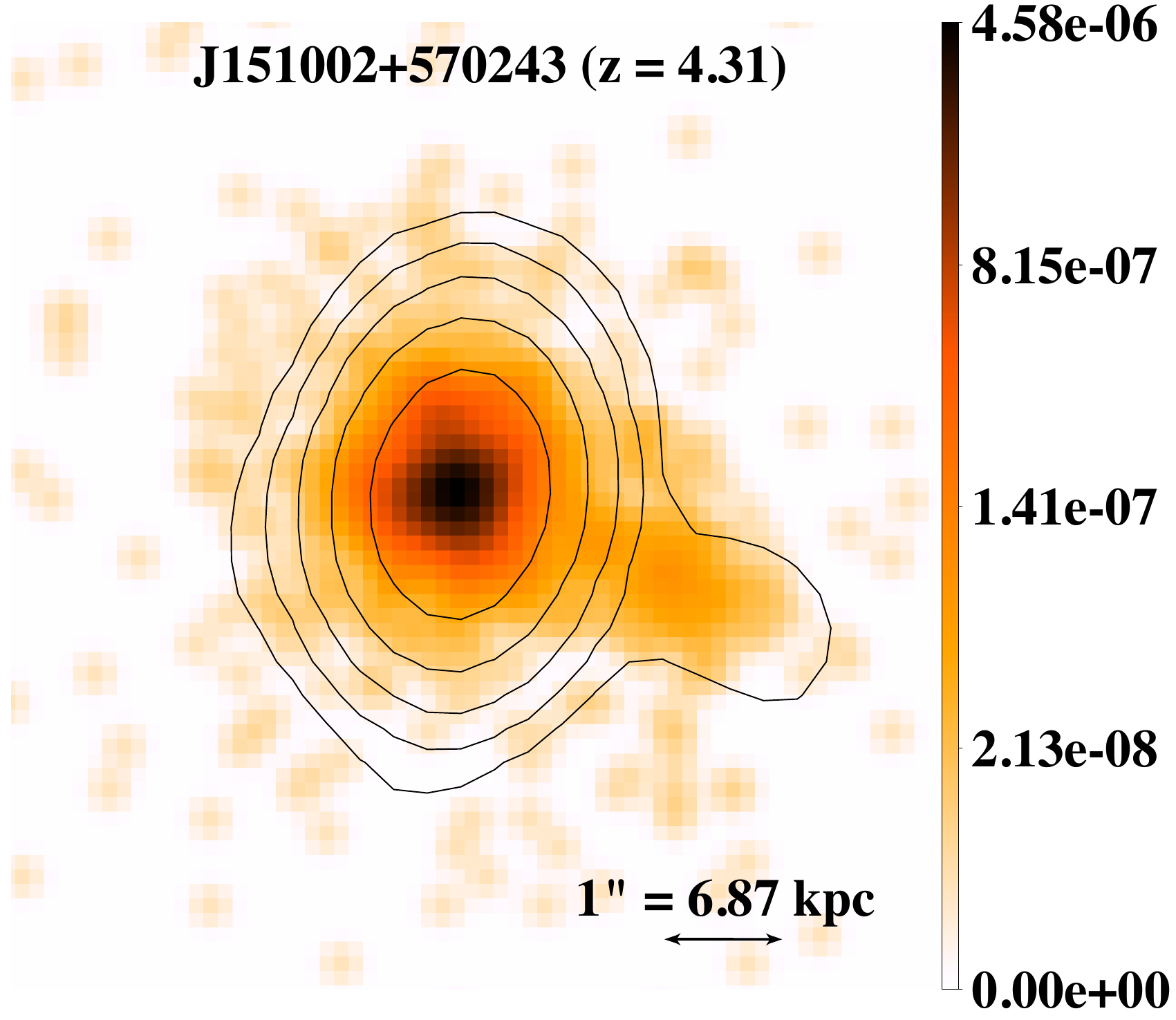}{0.3\textwidth}{(b)}
          \fig{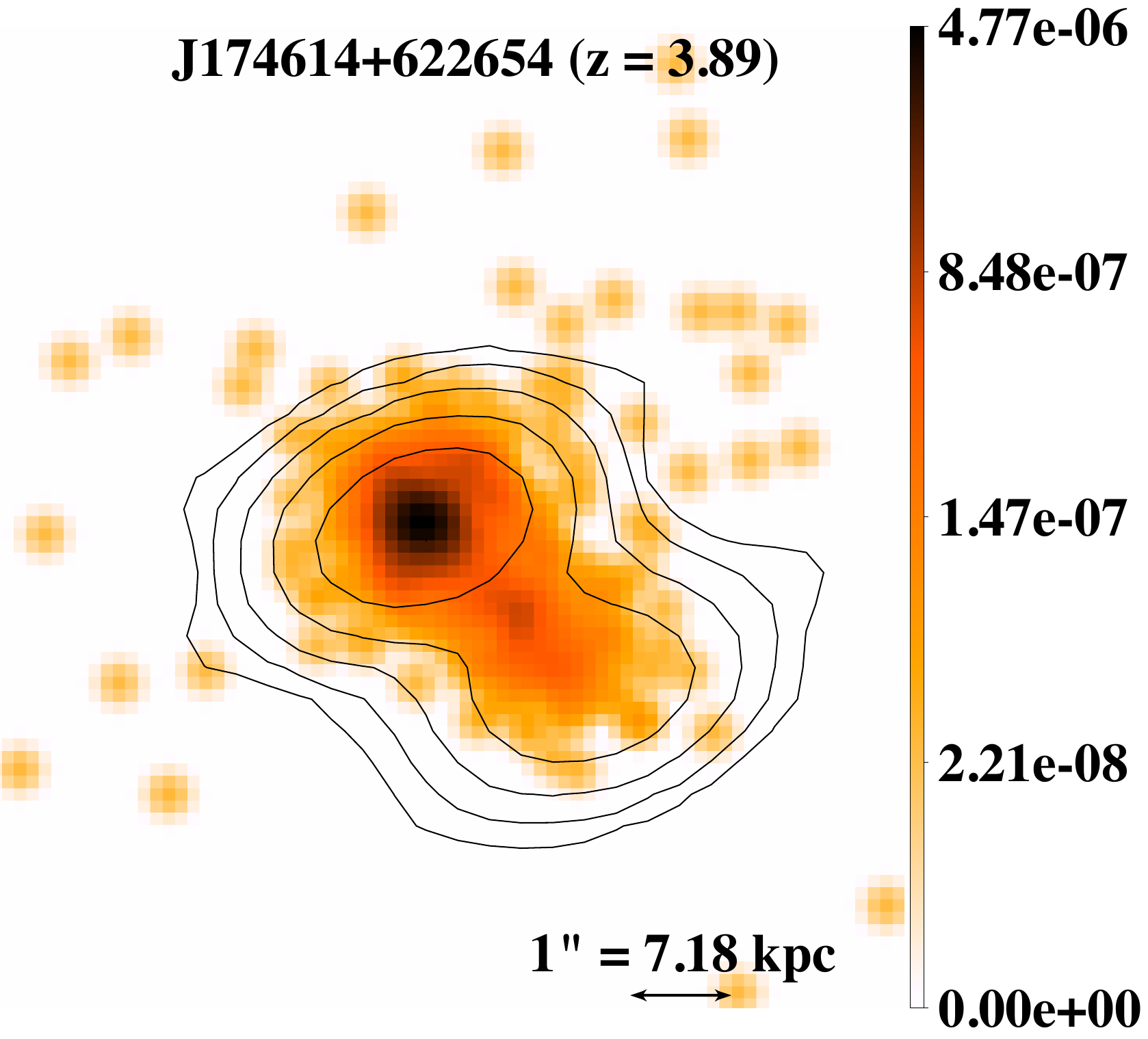}{0.3\textwidth}{(c)}
          }
\gridline{\fig{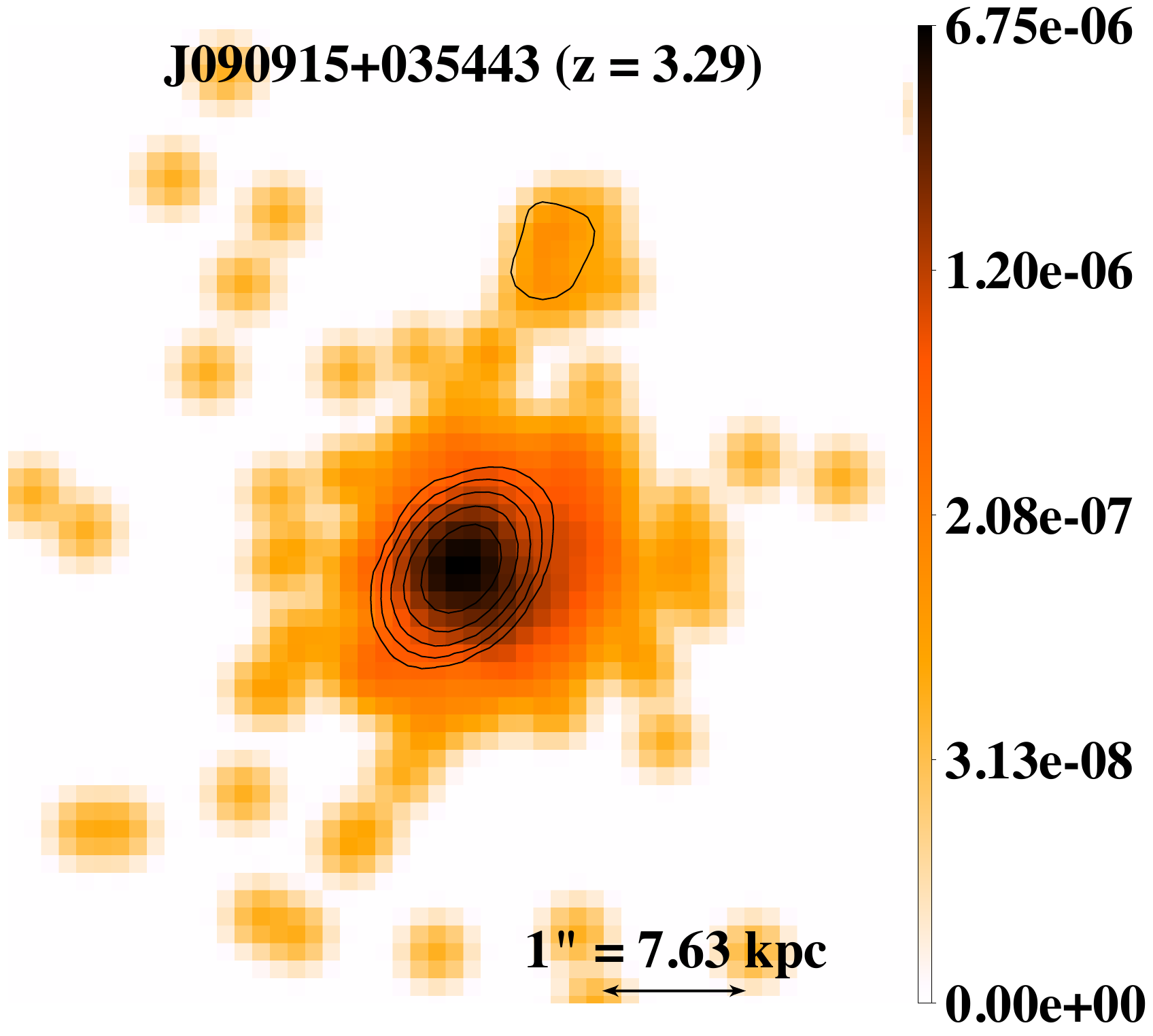}{0.24\textwidth}{(d)}
          \fig{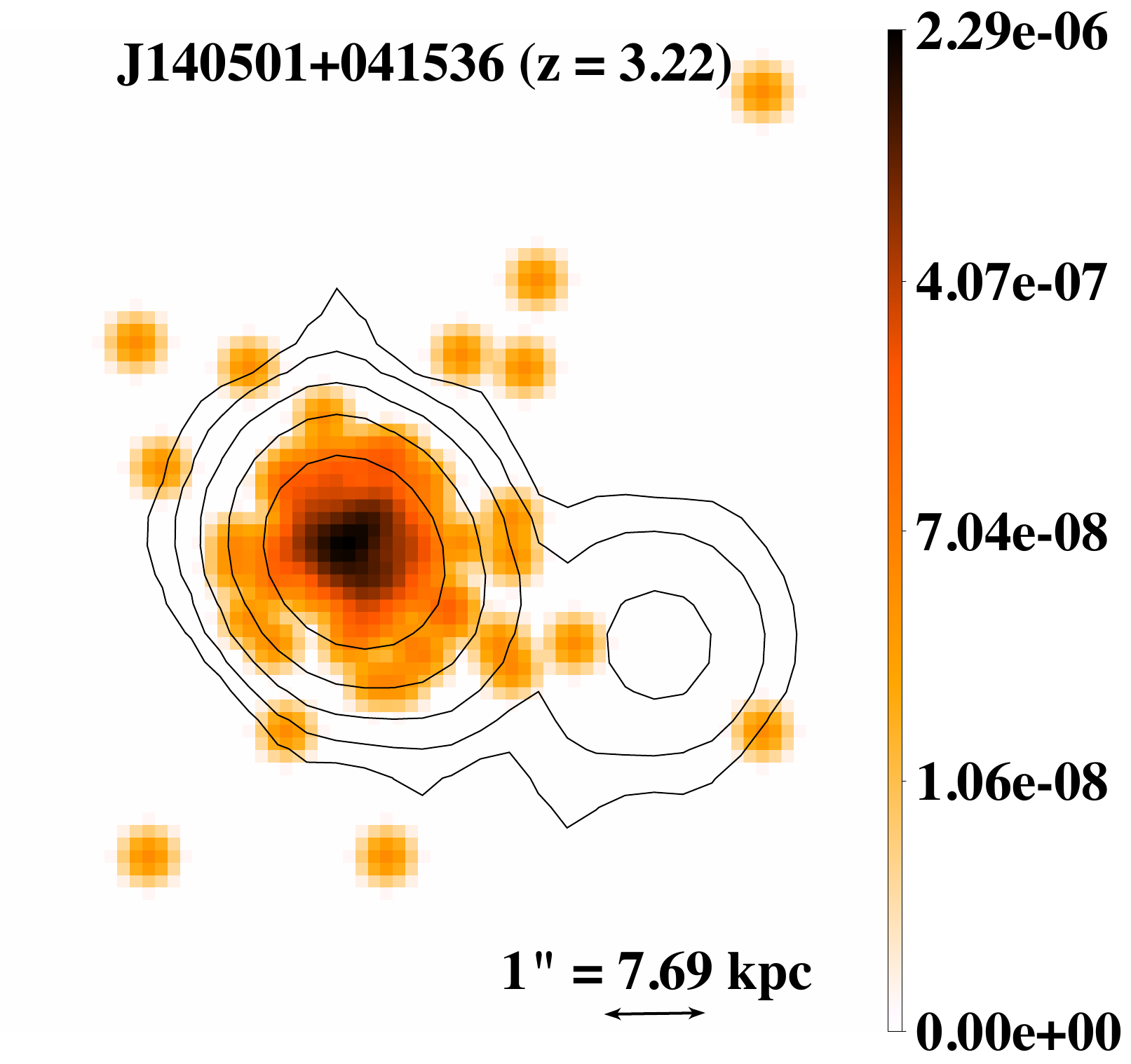}{0.24\textwidth}{(e)}
          \fig{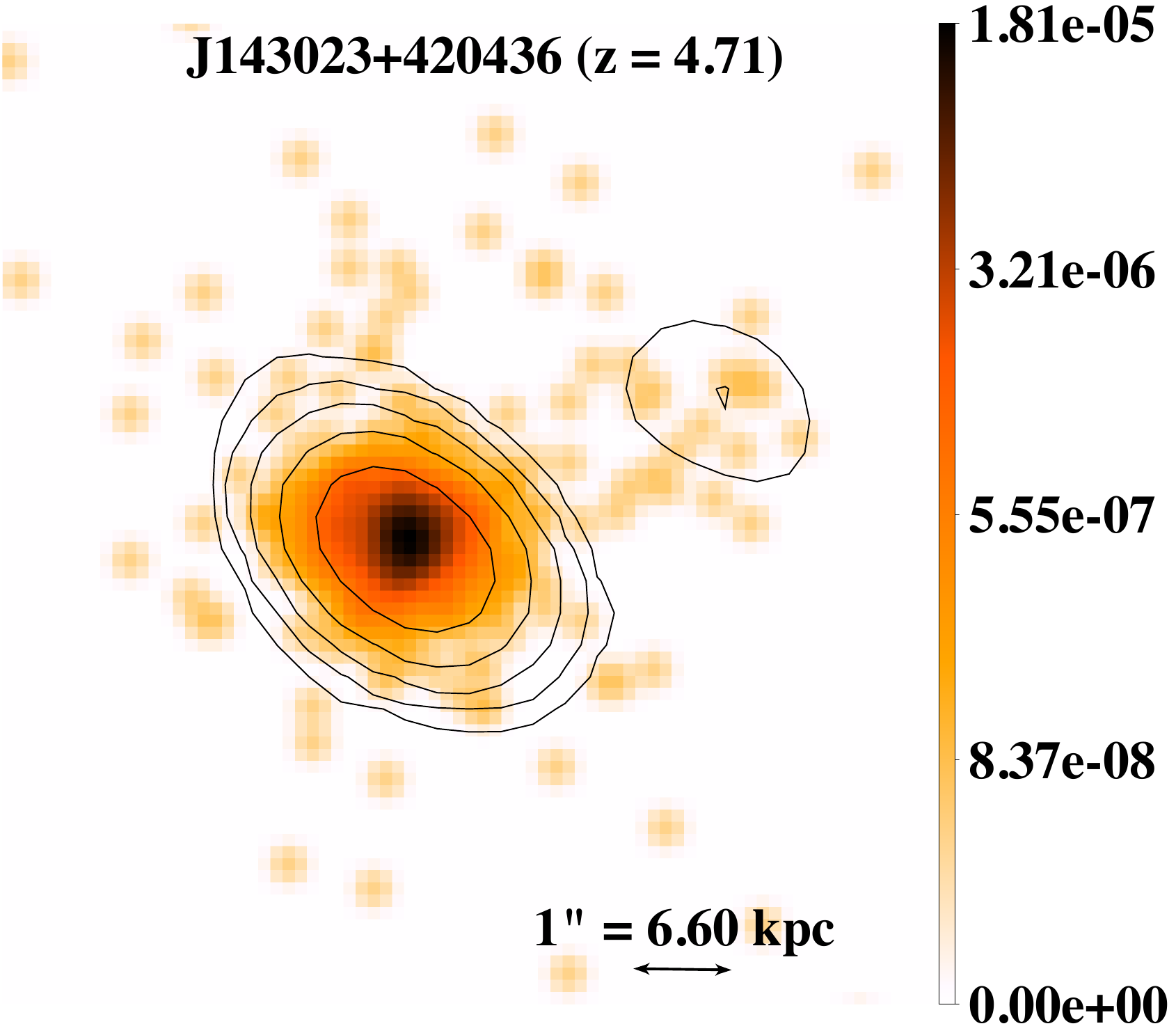}{0.24\textwidth}{(f)}
          \fig{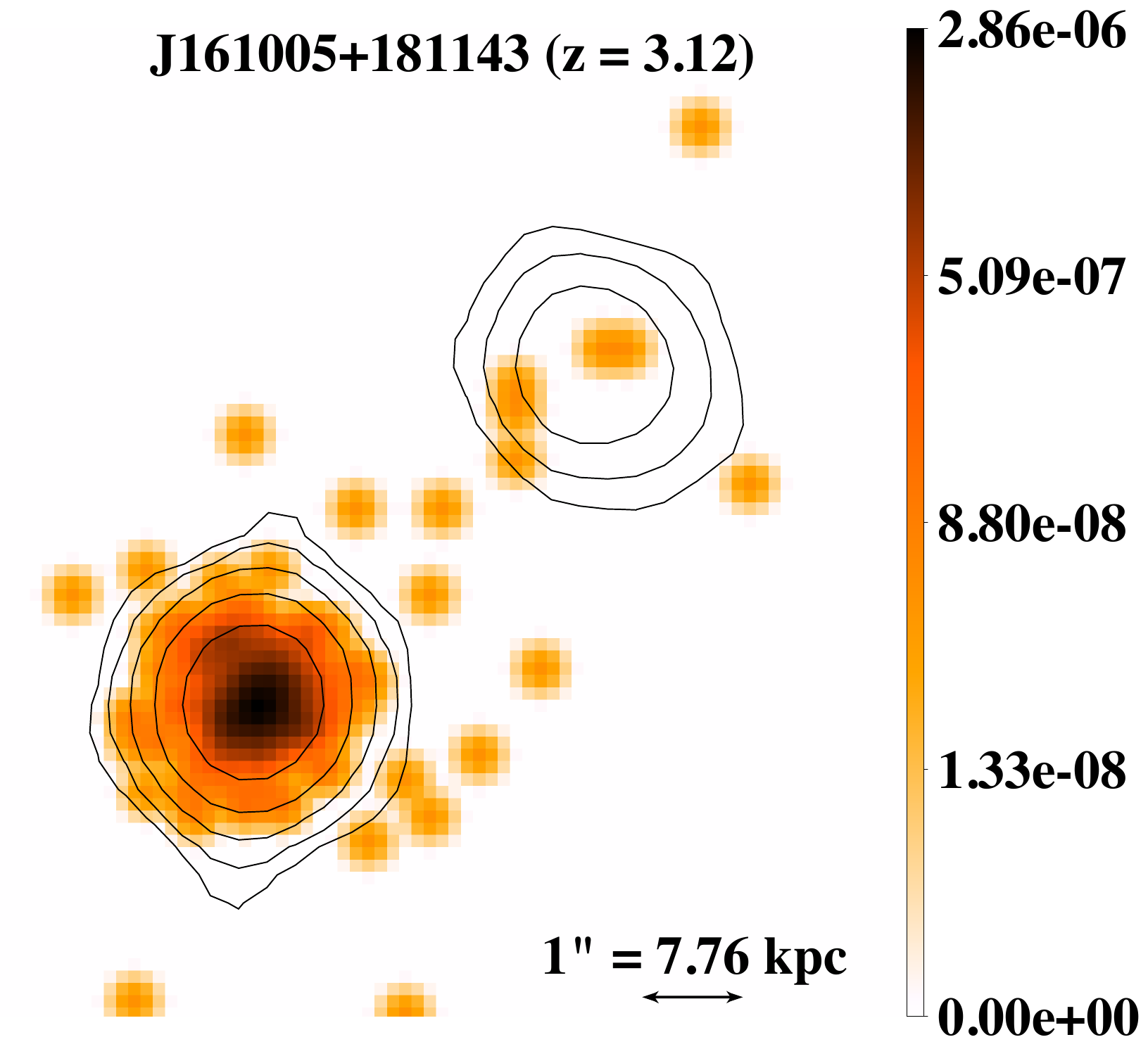}{0.24\textwidth}{(g)}
          }
\caption{The 0.5$-$7 keV \chandra~images of the high-redshift blazars that exhibit traces of extended X-ray emission. The image scales are labeled. Top panels belong to sources with the most prominent X-ray jets, whereas, bottom ones are with relatively weaker emission. Overplotted contours are from VLA observations in six equispaced logarithmic intervals and start from 5 times the off-source rms. The X-ray data are binned by 1/4 of the native 0$^{\prime\prime}$.492 pixel giving an effective resolution of 0.123$^{\prime\prime}$. The images are Gaussian smoothed with kernel radius of 3 pixels. The color bars are in units of \phflux.} \label{fig:x-ray_jet}
\end{figure*}

{\it Jet Powers:} The jet powers derived from the SED modeling can be found in Table~\ref{tab:jet} and we plot their distributions in Figure~\ref{fig:jet}. On average, the high-redshift sources have powerful jets, especially the proton and radiative jet powers as can be seen in Table~\ref{tab:stat}. On comparing the jet powers with the respective accretion luminosities (Figure~\ref{fig:jet-disk}), we find that the high-redshift blazars follow the accretion-jet connection known for other, relatively nearby objects \citep[e.g.,][]{2014Natur.515..376G}. We quantify the correlation by determining the partial Spearmann's correlation coefficient \citep[$\rho_{\rm s}$;][]{1992A&A...256..399P} and probability of no-correlation (PNC) which takes into account the common redshift dependence. The derived values are $\rho_{\rm s}=0.23\pm0.07$, PNC $<$10$^{-10}$ and $\rho_{\rm s}=0.58\pm0.05$, PNC $<$10$^{-10}$ for \ld~versus  $P_{\rm rad}$ and \ld~versus  $P_{\rm jet}$ correlations, respectively.

 Interestingly, as can be seen in the top panel of Figure~\ref{fig:jet-disk}, a major fraction of the high-redshift blazar population lies below the one-to-one correlation line, indicating their accretion power to be larger than their radiative jet luminosity. Considering the total jet power versus \ld (Figure~\ref{fig:jet-disk}, bottom panel), most of the sources do exhibit jet powers that exceed their accretion luminosities though about a quarter of them have lower jet powers. Keeping in mind the fact that the presence of pairs in the jet can reduce the total jet power by a factor of a few \citep[e.g.,][]{2017MNRAS.465.3506P}, we conclude that \ld~in the high-redshift blazars is comparable to their total jet powers.  Additionally, we caution that the results derived in this work are mainly driven by the soft X-ray observations. In order to better estimate the SED parameters and jet powers, observations in the hard X-ray band are necessary. This is because \nustar~data permit us to put tighter constraints on the low-energy slope of the particle energy distribution and also, along with the soft X-ray measurements, the minimum energy of the emitting electron population and the bulk Lorentz factor. These parameters are crucial to accurately compute the jet powers. Looking into the future, observations from the next generation all-sky MeV missions, e.g. All-sky Medium Energy Gamma-ray Observatory \citep[AMEGO, energy coverage 200 keV to 10 GeV;][]{2019BAAS...51g.245M}, will allow us to cover the broad range of the inverse Compton emission including the high-energy SED peak \citep[e.g.,][]{2019arXiv190306106P}, leading to an unprecedented measurement of the physical properties of the high-redshift blazars. 

It can also be noticed in Figure~\ref{fig:jet-disk} that both \ld~and jet power appear to saturate around 10$^{48}$ \lum. This observation might be connected to the upper limit of the black hole mass that can be achieved via accretion \citep[$\sim$a few times 10$^{10}$ \Msun, see, e.g.,][]{2016ApJ...828..110I,2016MNRAS.456L.109K}, and hence, to the maximum accretion rate in Eddington units. In other words, the average jet power and \ld~appear to saturate around the maximum possible Eddington luminosity.

\section{Extended X-ray Jets}\label{sec:x-ray_jets}
There are seven high-redshift blazars that have exhibited traces of extended X-ray emission in their \chandra~observations. We show 0.5$-$7 keV \chandra~images of these sources in Figure~\ref{fig:x-ray_jet} and overplot the VLA radio contours to look for radio counterparts of the X-ray jets. Note that the presence of X-ray jets in these objects has already been reported in various previous works \citep[see, e.g.,][]{2003ApJ...598L..15S,2004ApJ...600L..23C,2006ApJ...650..679C,2012ApJ...756L..20C,2016ApJ...833..123M,2019AN....340...30S}. However, instead of focusing on the properties of the extended X-ray emission as done in those works, we explore the properties of the blazar core with the motivation to search for any possible pattern in the physical properties which may reveal the origin of kpc-scale X-ray jets.

\begin{figure*}[t!]
\includegraphics[scale=0.4]{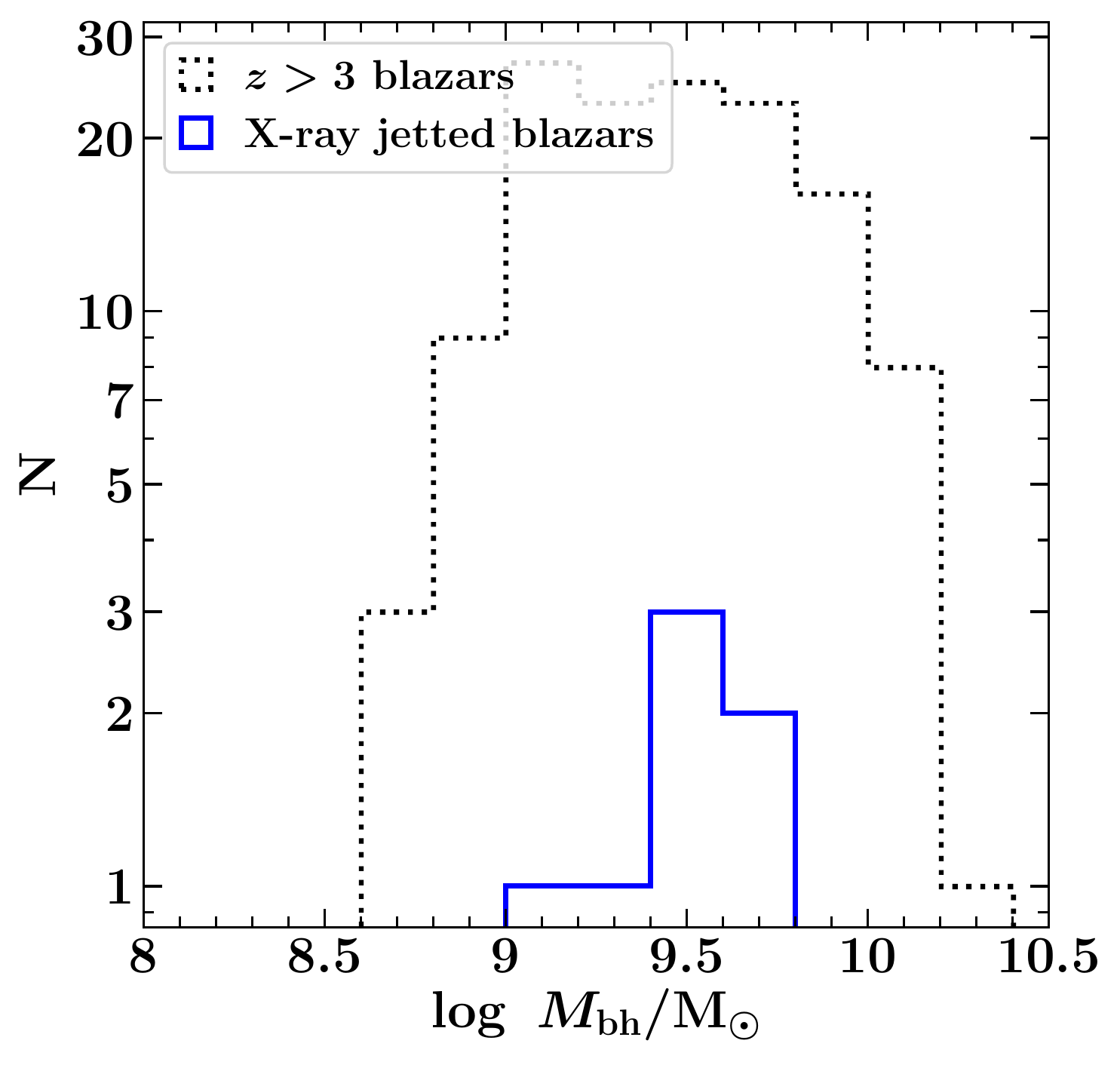}
\includegraphics[scale=0.4]{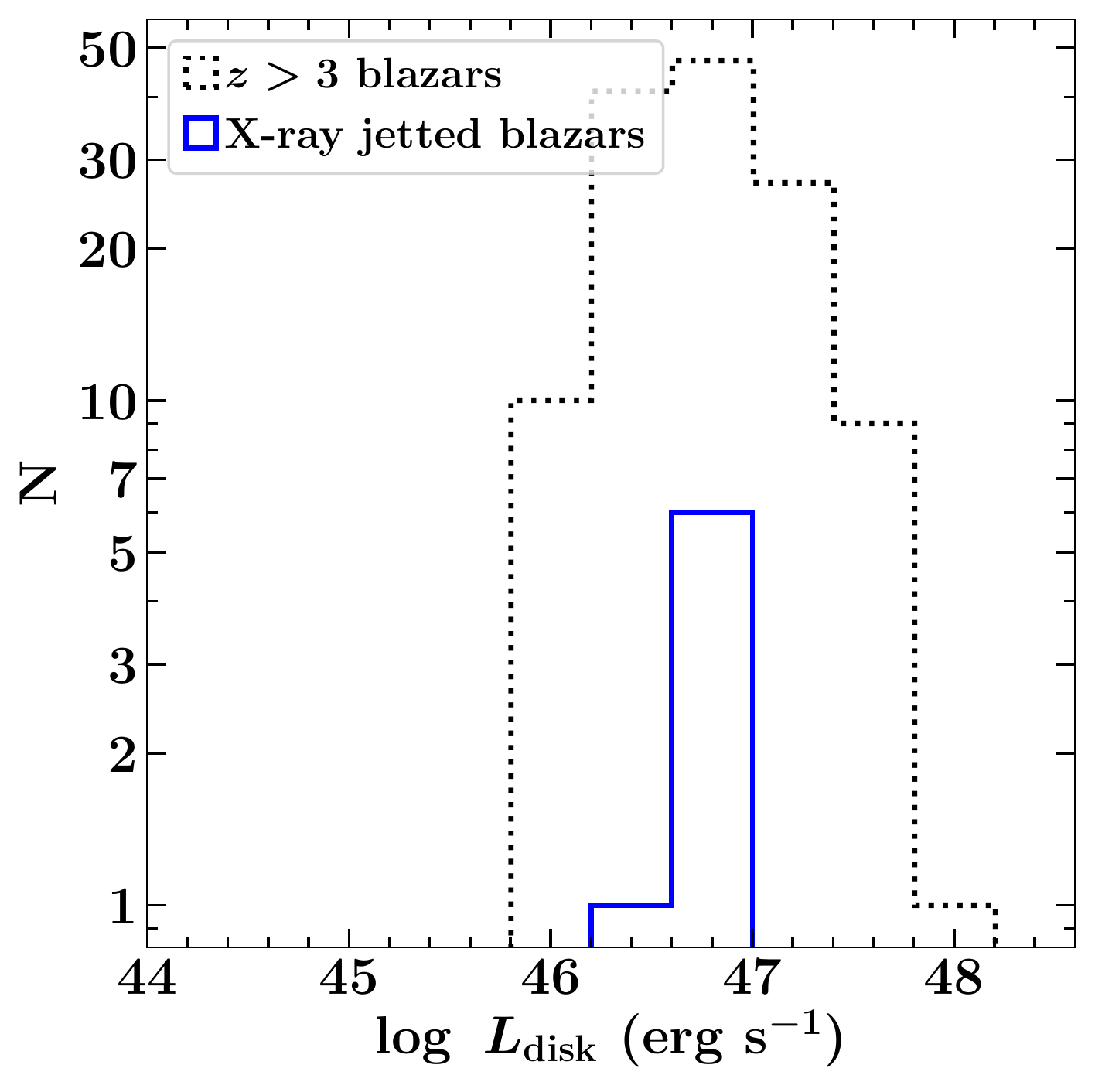}
\includegraphics[scale=0.4]{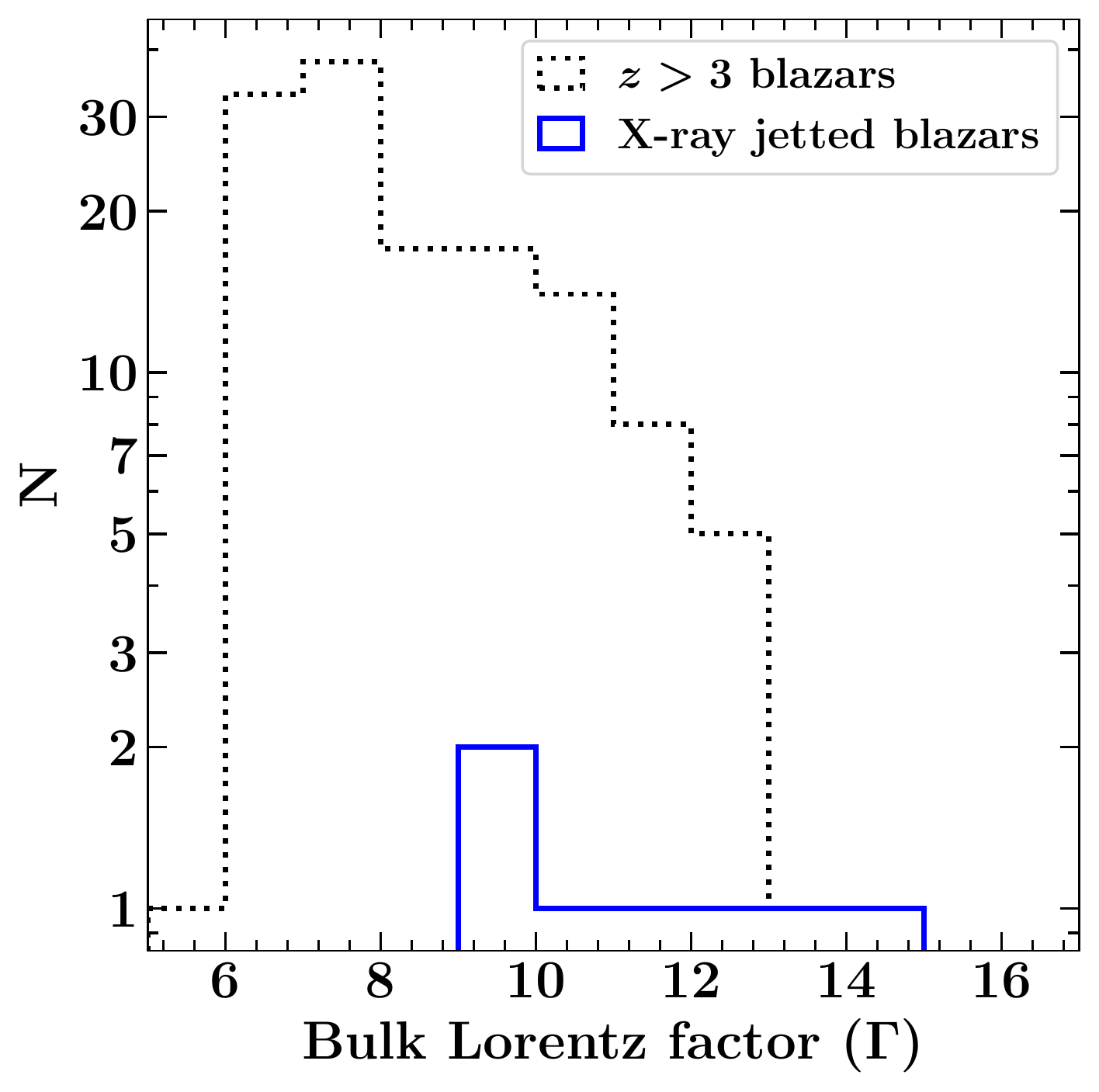}
\caption{The histograms of \mbh~(left), \ld~(middle), and $\Gamma$ (right) for X-ray jetted sources (blue solid) and other high-redshift blazars (black dotted). See the text for details.} \label{fig:sed_jet}
\end{figure*}

Figure~\ref{fig:sed_jet} shows the distribution of the central engine parameters (\mbh~and \ld) and the bulk Lorentz factor of the jet for X-ray jetted objects and other high-redshift blazars. Both \mbh~and \ld~have similar average values for two populations. In the \ld~versus jet power diagram, these objects tend to lie in the regime with higher jet powers (see Figure~\ref{fig:jet-disk}). Interestingly, we noticed a relatively higher $\Gamma$ in X-ray jetted blazars compared to other $z>3$ sources (Figure~\ref{fig:sed_jet}, right panel). This observation suggests faster moving jets in objects showing extended X-ray emission. Interestingly, recent VLBA observations of the most distant X-ray jetted blazar, NVSS J143023+420436 ($z=4.71$), also revealed a rapidly moving jet with $\Gamma=14.6\pm3.8$ \citep[][]{2019arXiv191212597Z}, similar to $\Gamma=14$ found by us via SED modeling. Therefore, it appears that plasma in X-ray jetted blazars remains highly relativistic at parsec scale distances or even further down the jet. However, the sample of the known extended X-ray jets in the parent sample of the high-redshift objects is small and therefore a strong claim cannot be made. One needs also to consider relatively nearby (i.e., $z<3$) X-ray jets to increase the sample size and ascertain the findings reported here.

\begin{figure}
\hbox{\hspace{0cm}
\includegraphics[width=\linewidth]{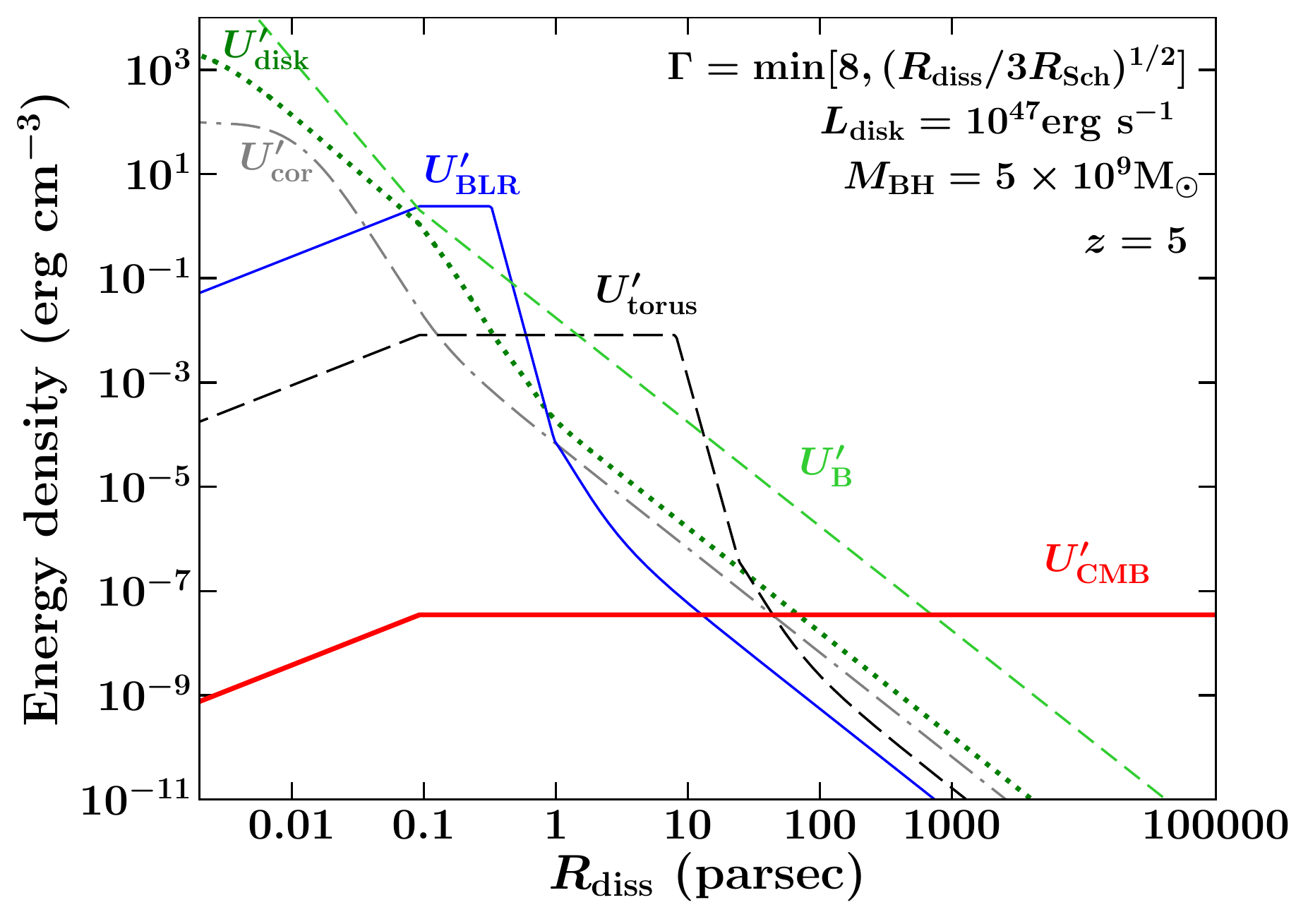}
}
\caption{This plot shows the variation of the comoving-frame energy densities of different AGN components as a function of the distance from the central black hole. For a comparison, we also plot the CMB energy density as seen in the moving plasma frame. We adopt a \mbh$=5\times10^9$\Msun, \ld$=10^{47}$\lum and a bulk Lorentz factor $\Gamma=8$. Magnetic energy density ($U'_{\rm B}$) is derived by considering the Poynting jet power as 10\% of \ld. Note that CMB energy density becomes dominant only after a kpc distance from the black hole.} \label{fig:ene_den}
\end{figure}

The X-ray emission in the high-redshift, radio-loud quasars is found to be significantly enhanced compared to low-redshift sources with matched properties in other wavebands \citep[][]{2013ApJ...763..109W}. One of the possible explanations put forward is due to interaction of the jet electrons with the Cosmic Microwave Background (CMB) photons whose energy density has a strong redshift dependence: $U_{\rm CMB} \propto(1+z)^4$. Considering the fact that the radio-loudest quasars usually belong to blazar population, it may be instructive to use the high-redshift blazars to study this problem. In Figure~\ref{fig:ene_den}, we show the variations of the energy densities of various AGN components, e.g., BLR/torus, as a function of the distance from the central black hole, as seen in the comoving frame of the jet plasma at $z=5$ \citep[][]{2009MNRAS.397..985G}. We assume \mbh$=5\times10^9$ \Msun~and \ld$=10^{47}$ \lum~and $\Gamma=8$. 
According to this diagram, CMB energy density becomes dominant over other AGN components only after a kpc from the black hole and even larger if the disk is more luminous. Therefore, if the observed X-ray enhancement is due to inverse Compton scattering of CMB photons (IC-CMB), the emission region is expected to be located far ($>$1 kpc) from the central engine which is rather unconvincing due to rapid flux variability observed from blazars. In fact, CMB energy density is comparatively small and may not be able to explain the bright X-ray emission which would be dominated by emission regions located closer to the central black hole due to strong BLR/torus photon field. An alternative possibility to explain the enhanced X-ray brightness could be due to the shift of the SED peaks to lower frequencies as the redshift increases, thereby making the blazar more luminous in the X-ray band. Due to synchrotron self absorption, however, this hypothesis cannot be tested at GHz frequencies where the peak of the synchrotron emission is located. In addition to that, the presence of multiple emission regions cannot be excluded with a fraction of the observed X-ray emission being originated via the IC-CMB mechanism. Even in this case, the observed X-ray radiation will be dominated by that produced within the central few parsecs from the black hole. Therefore, a pure IC-CMB model is not supported by the observations \citep[see also][]{2019MNRAS.482.2016Z}.

\section{Summary}\label{sec:summary}
We have carried out a broadband study of 142 high-redshift ($z>3$), radio-loud quasars that exhibit blazar like characteristics, including 9 \gm-ray detected and 15 with hard X-ray observations with \nustar. Below we summarize our main findings.

\begin{enumerate}
\item The members of the high-redshift blazar population are faint \gm-ray emitters with steep spectra, as revealed by the stacking analysis.
\item In the X-ray band, these objects have been selected in order to have flat ($\Gamma_{\rm X}$) spectra and are luminous.
\item High-redshift blazars present in our sample host massive black holes ($>10^9$ \Msun) and luminous accretion disks ($>10^{46}$ \lum) at their centers.
\item Based on a simple one-zone leptonic emission modeling, we have found that the high-redshift objects are MeV peaked and have Compton dominated SEDs, thus indicating that a major fraction of their bolometric output is radiated in the form of high-energy X- to \gm-ray emission. Furthermore, a rather low value of the bulk Lorentz factor based on available data can possibly explain the identification of fewer number of their parent population. However, a strong claim cannot be made due to lack of hard X-ray observations, e.g., with \nustar, which are necessary to accurately constrain $\Gamma$.
\item The known accretion-jet connection noticed in the low-redshift blazars is also followed by the high-redshift ones. There are indications that both jet power and accretion luminosity have a maximum at $\sim$10$^{48}$ \lum.
\item A small fraction of our sample that have available \chandra~observations (7 out of 54), exhibits extended X-ray jets. These sources tend to have higher total jet powers with respect to other $z>3$ blazars and more importantly, have faster moving jets, though the results are model dependent. Further investigation considering a larger sample of X-ray jetted AGNs is needed to confirm this finding.
\item The observed X-ray enhancement of the high-redshift sources cannot be explained with a pure IC-CMB model. Among a few alternative possibilities, one could be presence of multiple emission regions with those located at hundreds of parsecs far away from the central black hole may contribute via IC-CMB mechanism, though the overall emission may be dominated by those lying within the central parsec region of the AGN. A shift of the high-energy SED peak to lower frequencies (i.e., towards X-rays) as the redshift increases, could be another possible explanation.

\end{enumerate}

\acknowledgments
We are grateful to the journal referee for a constructive criticism. VSP and MA acknowledge funding under NASA contracts 80NSSC18K0580 and NNX17AC35G. This work was supported by the Initiative and Networking Fund of the Helmholtz Association. H-MC acknowledges support by the National Natural Science Foundation of China (Grant No. U1731103). AK is thankful to N. {\'A}lvarez Crespo, E. J. Marchesini, H. A. Pe{\~n}a-Herazo, and F. Massaro for a useful discussion on the optical spectral analysis. 

The \textit{Fermi} LAT Collaboration acknowledges generous ongoing support from a number of agencies and institutes that have supported both the development and the operation of the LAT as well as scientific data analysis. These include the National Aeronautics and Space Administration and the Department of Energy in the United States, the Commissariat \`a l'Energie Atomique and the Centre National de la Recherche Scientifique / Institut National de Physique Nucl\'eaire et de Physique des Particules in France, the Agenzia Spaziale Italiana and the Istituto Nazionale di Fisica Nucleare in Italy, the Ministry of Education, Culture, Sports, Science and Technology (MEXT), High Energy Accelerator Research Organization (KEK) and Japan Aerospace Exploration Agency (JAXA) in Japan, and the K.~A.~Wallenberg Foundation, the Swedish Research Council and the Swedish National Space Board in Sweden. Additional support for science analysis during the operations phase is gratefully acknowledged from the Istituto Nazionale di Astrofisica in Italy and the Centre National d'\'Etudes Spatiales in France. This work performed in part under DOE Contract DE- AC02-76SF00515.

This research has made use of data obtained through the High Energy Astrophysics Science Archive Research Center Online Service, provided by the NASA/Goddard Space Flight Center. This research has made use of the NASA/IPAC Extragalactic Database (NED), which is operated by the Jet Propulsion Laboratory, California Institute of Technology, under contract with the National Aeronautics and Space Administration. Part of this work is based on archival data, software or online services provided by the ASI Data Center (ASDC). 

This work made use of data from the NuSTAR mission, a project led by the California Institute of Technology, managed by the Jet Propulsion Laboratory, and funded by the National Aeronautics and Space Administration. We thank the NuSTAR Operations, Software, and Calibration teams for support with the execution and analysis of these observations. This research has made use of the NuSTAR Data Analysis Software (NuSTARDAS) jointly developed by the ASI Science Data Center (ASDC, Italy) and the California Institute of Technology (USA). This research has made use of the XRT Data Analysis Software (XRTDAS). This work made use of data supplied by the UK Swift Science Data Centre at the University of Leicester.

This work is based on observations obtained with XMM-Newton, an ESA science mission with instruments and contributions directly funded by ESA Member States and NASA. The scientific results reported in this article are based on data obtained from the Chandra Data Archive. This research has made use of software provided by the Chandra X-ray Center (CXC) in the application packages CIAO, ChIPS, and Sherpa. This publication makes use of data products from the Widefield Infrared Survey Explorer, which is a joint project of the University of California, Los Angeles, and the Jet Propulsion Laboratory/California Institute of Technology, funded by NASA.

This research made use of Astropy,\footnote{http://www.astropy.org} a community-developed core Python package for Astronomy \citep{2013A&A...558A..33A,2018AJ....156..123A}.

\vspace{5mm}
\facilities{\fermi-LAT, \swift, \xmm, \nustar, \chandra}

\software{CIAO (v.4.9), SAS (v.15.0.0), XSPEC \citep[v 12.10.1;][]{1996ASPC..101...17A}, Astropy \citep[][]{2013A&A...558A..33A,2018AJ....156..123A},  Swift-XRT data product generator \citep[][]{2009MNRAS.397.1177E}, fermiPy \citep[][]{2017arXiv170709551W}}



\appendix

\section{Uncertainty Measurement in the Disk Fitting Technique}\label{sec:disk_uncer}
 \begin{figure*}[t!]
\hbox{\hspace{0cm}
\includegraphics[scale=0.6]{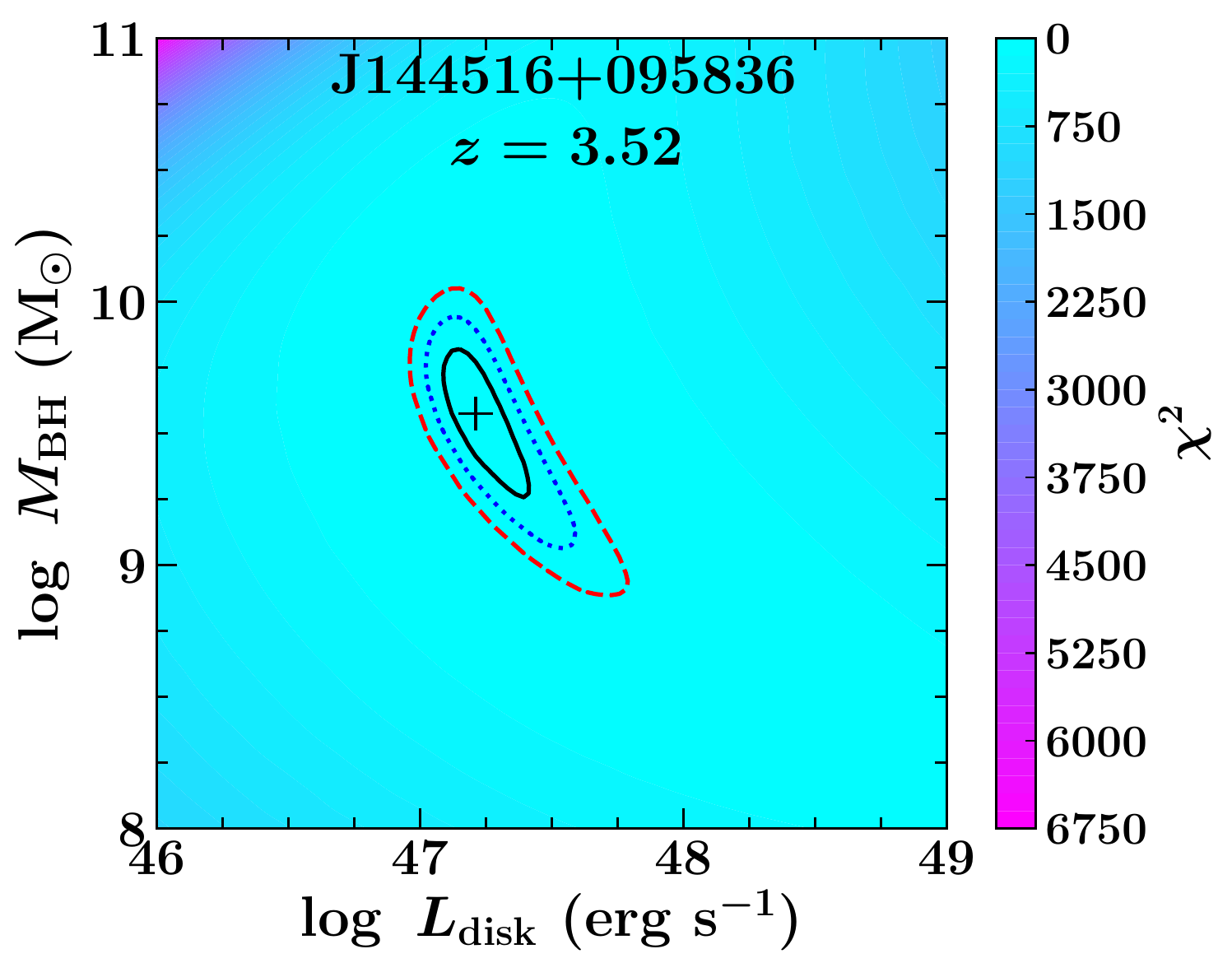}
\includegraphics[scale=0.6]{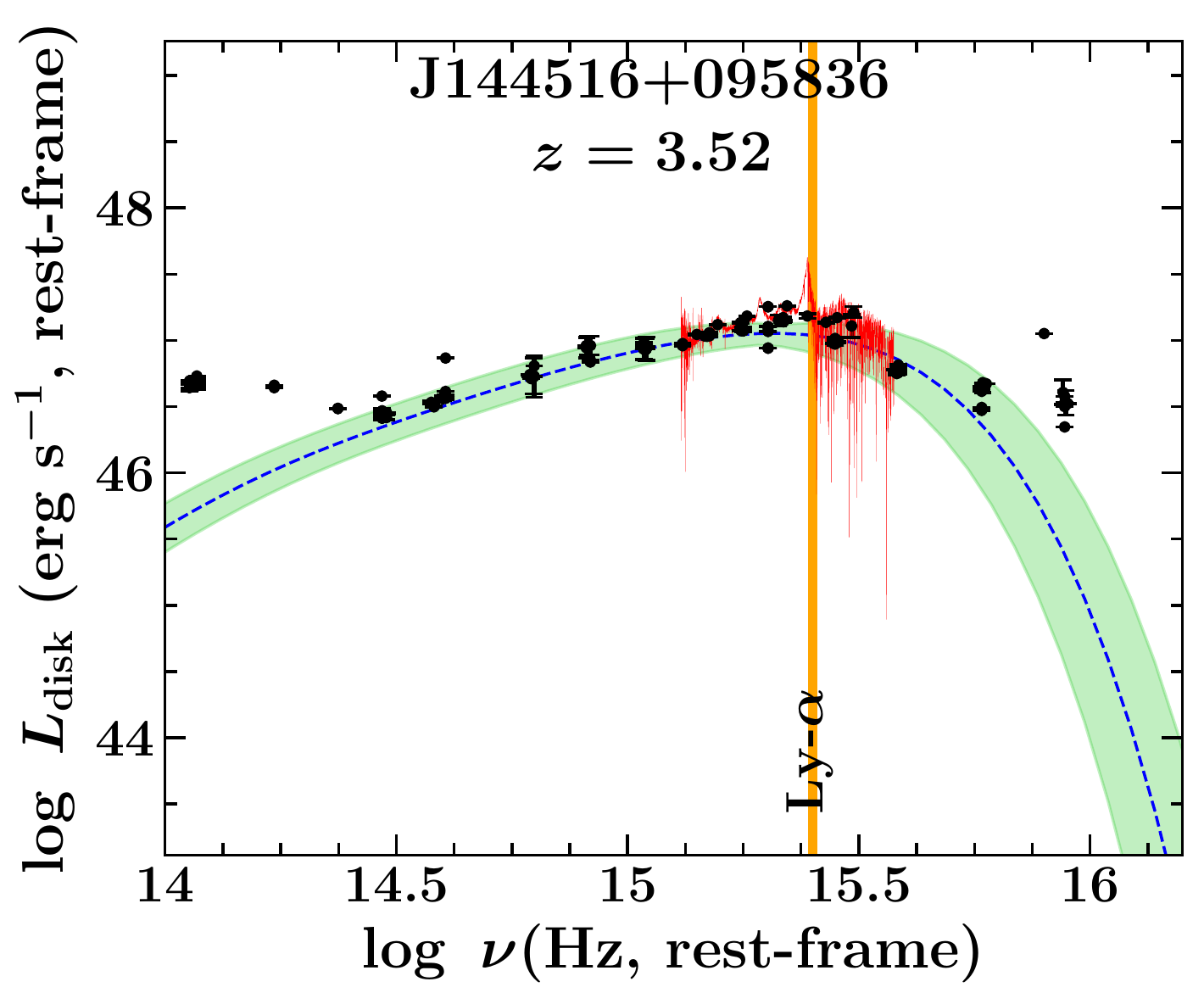}
}
\caption{Left: A $\chi^2$ grid map of \mbh~and \ld~for NVSS J144516+095836. The `+' sign denotes the minimum of the $\chi^2$ surface corresponding to the best-fitted \mbh~and \ld~values. Confidence contours are at 1$\sigma$ (black solid), 2$\sigma$ (blue dotted), and 3$\sigma$ (red dashed) levels. Right: The IR-UV SED of the same object. Black data points are archival observations from SSDC and the SDSS spectrum (not used in the fit) is shown with the red line. Blue dashed line refers to the best-fitted accretion disk model corresponding to the \mbh~and \ld~values derived from the grid scan, as shown in the left panel. The green shaded area denotes the 1$\sigma$ uncertainty in the fitted model.} \label{fig:disk_app}
\end{figure*}

As with any fitting method, the accuracy of the \mbh~and \ld~computed from the accretion disk modeling technique depends on the IR-UV data coverage. For a good quality spectrum, both numbers can be constrained within a factor of two. To demonstrate this, we performed a simple $\chi^2$ test for the blazar NVSS J144516+095836 ($z=3.52$) which has a good quality IR-optical data available. We generated a library of accretion disk spectrum for a large range of [\mbh, \ld] pairs, e.g., [10$^7$ \Msun, 10$^{45}$ \lum], [10$^7$ \Msun, 10$^{45.1}$ \lum] ... [10$^{10}$ \Msun, 10$^{50}$ \lum] and so on. We, then compared the disk spectrum generated for each [\mbh, \ld] pair with the data to derive $\chi^2$, thus effectively generating a $\chi^2$ grid. The global minimum of the generated grid and  1$\sigma$, 2$\sigma$, and 3$\sigma$ confidence levels were determined by fitting a cubic spline function (Figure~\ref{fig:disk_app}). This exercise led to the best-fitted, log-scale \mbh~(in \Msun) and \ld~(in \lum) as 9.60$\pm$0.14 and 47.21$\pm$0.08, respectively, which is very close to \mbh~(in \Msun)= 9.48 and \ld~(in \lum)= 47.26 used in the paper. These results suggest a typical uncertainty of a factor of $\lesssim$2 associated with the disk modeling approach. Note that for objects with poorer data coverage, the reliability of disk fitting technique also depends on the additional piece of information, i.e., range of \ld~from broad line luminosities, introducing another factor of uncertainty, $\sim$0.3 dex, in \ld~and \mbh~measurement.

\section{Stacking Analysis from 100 MeV}\label{sec:stacking}
In Section~\ref{subsec:gamma}, we presented the results derived from the stacking analysis with the minimum energy set as $E_{\rm min}=300$ MeV. Here we explain the reasons behind the adopted choice of $E_{\rm min}$, instead of considering 100 MeV which is conventionally used in the standard \fermi-LAT data analysis.

\begin{figure*}[t!]
\hbox{\hspace{0cm}
\includegraphics[scale=0.32]{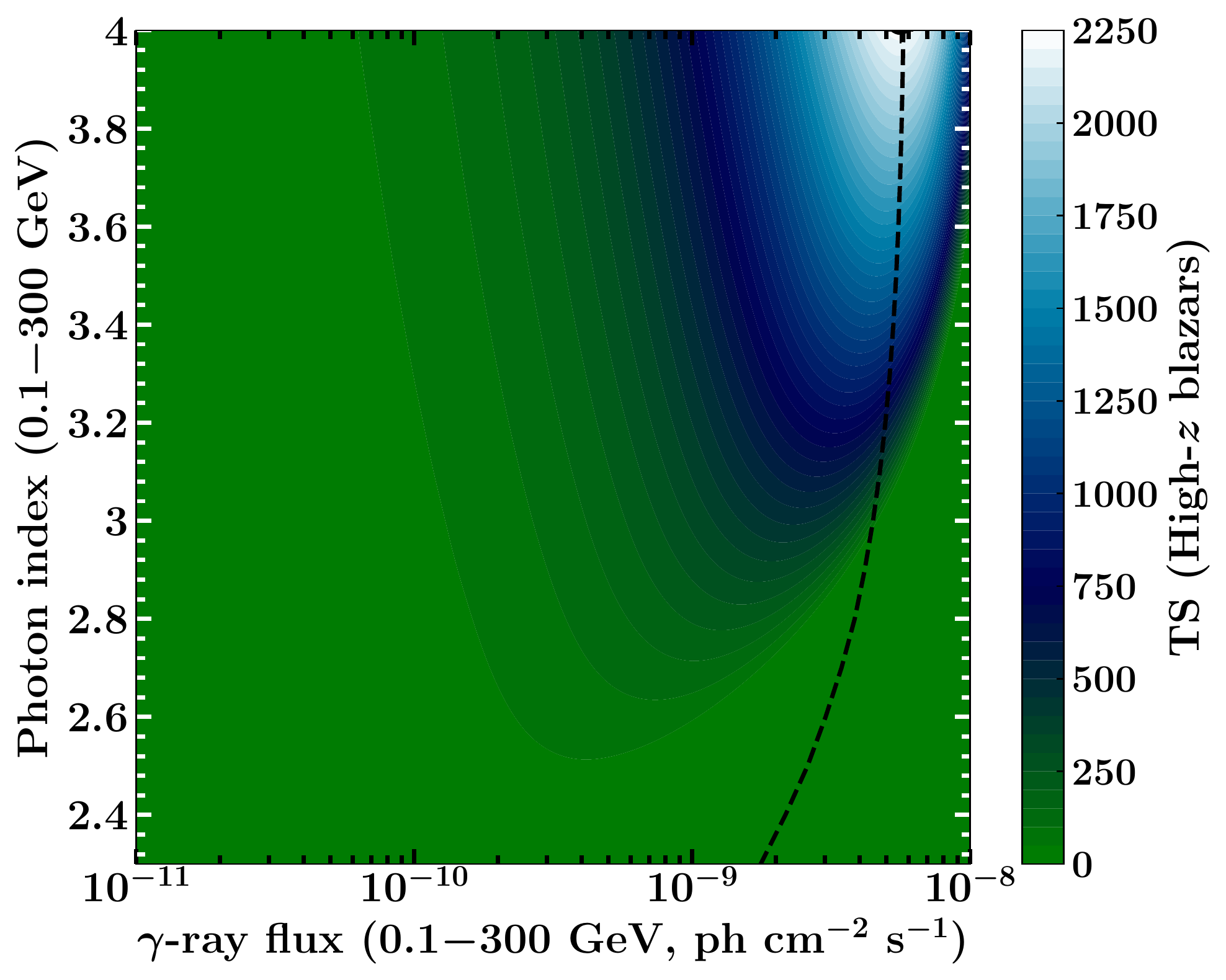}
\includegraphics[scale=0.3]{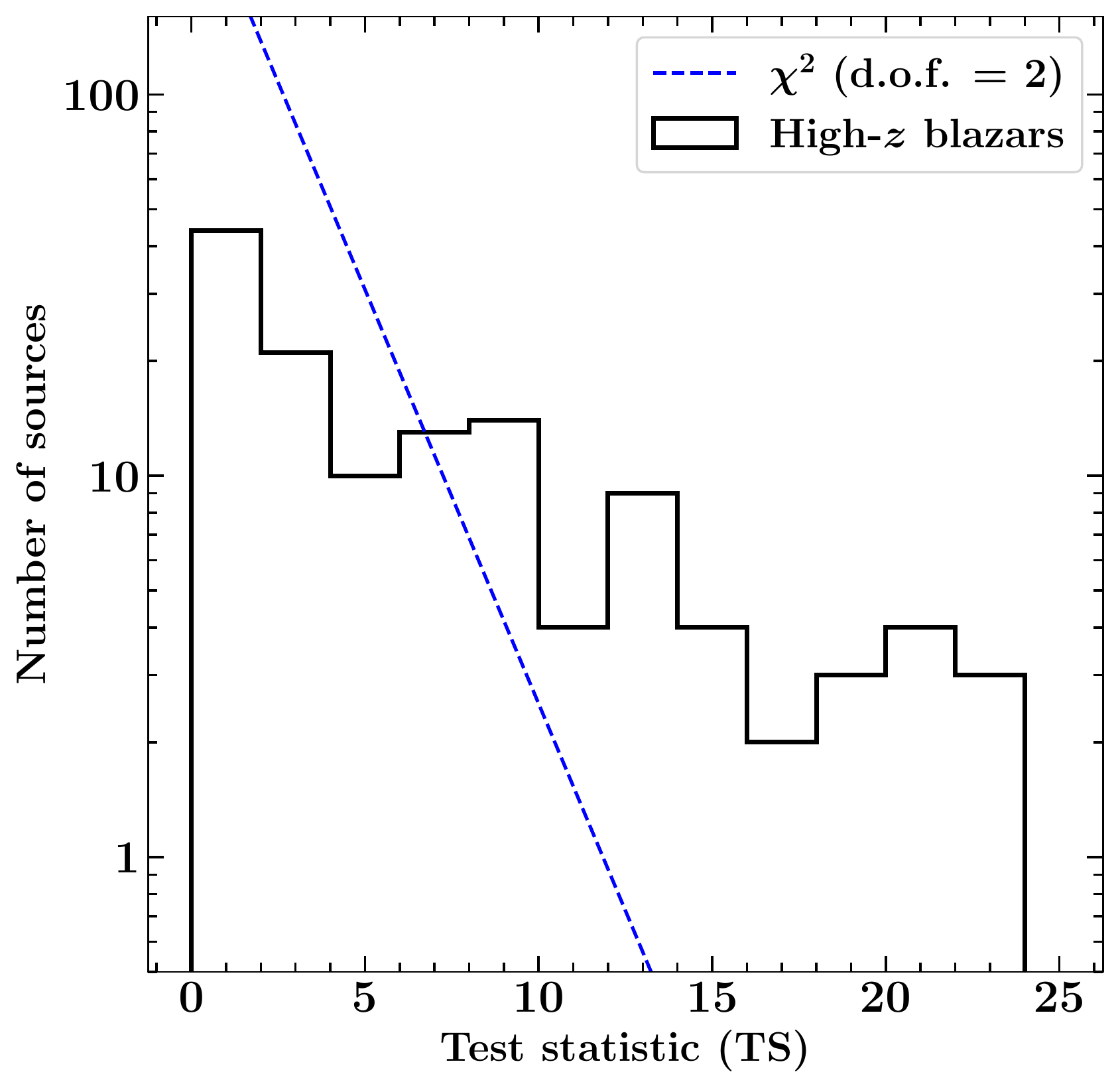}
\includegraphics[scale=0.32]{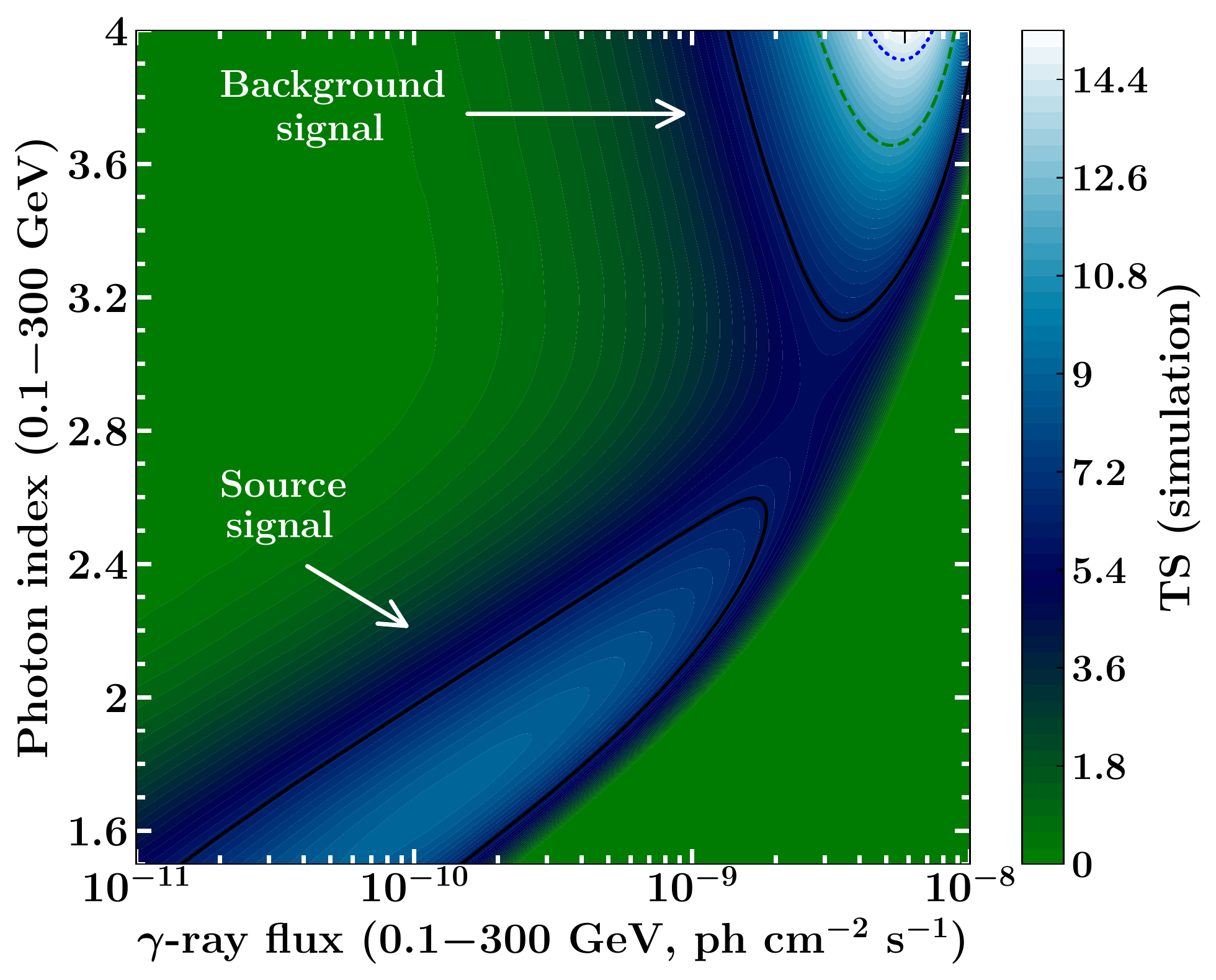}
}
\caption{Left: The stacked TS profile of 133 \gm-ray undetected sources when $E_{\rm min}$ for the analysis is set as 100 MeV. The black dahsed line shows the \fermi-LAT sensitivity limit for the time period covered in this work. Middle: The distributions of the TS for \gm-ray undetected sources. The blue dashed line shows the $\chi^2$ distribution for 2 degrees of freedom corresponding to the null hypothesis of no source, i.e., random fluctuations. Right: The significance profile of a simulated \gm-ray point object with a hard and faint \gm-ray spectrum. Note the bright and soft background emission clearly distinguishable from the point-source signal. We have masked the negative TS values to highlight the positive signal. See the text for details.} \label{fig:stacking_app}
\end{figure*}

The left panel of Figure~\ref{fig:stacking_app} shows the combined significance profile of 133 \fermi-LAT undetected blazars when the analysis was carried out using $E_{\rm min}=100$ MeV. A bright and extremely soft \gm-ray emission can be noticed. However, this emission may not have originated from the high-redshift blazars. This is due to three reasons: (i) none of the known \gm-ray blazars, including the high-redshift ones, exhibit such a steep \gm-ray spectrum in 0.1$-$300 GeV energy range, (ii) a comparison with the \fermi-LAT sensitivity limit for the period covered in this work suggests that individual objects with such a soft spectrum should have already been detected (see Figure~\ref{fig:stacking_app}, left panel), and (ii) even after assuming that all 133 sources have the same photon flux and index, the combined TS cannot reach a value as large as TS = 2250. This is demonstrated in the middle panel of Figure~\ref{fig:stacking_app} where we show the TS distributions for the considered high-redshift blazars and compare with a $\chi^2$ distribution with 2 degrees of freedom representing the null hypothesis. This plot also explains that the derived \gm-ray signal (Figure~\ref{fig:stacking}) is not due to random background fluctuations and belongs to the real blazar population. Therefore, we conclude that the soft and bright emission observed in the stacked profile is most likely due to isotropic background embedded in the \fermi-LAT data. This is further confirmed with the simulation of a hard spectrum blazar assuming its 0.1$-$300 GeV photon index $\Gamma_{\rm 0.1-300~GeV}=1.7$, a photon flux $F_{\rm 0.1-300~GeV}=10^{-10}$ \phflux, and a faint signal TS = 8. The significance profile for this simulated blazar can be seen in the right panel of Figure~\ref{fig:stacking_app}. Due to the input assumption of the hard spectrum, we are able to disentangle the soft background emission as can be seen in this plot.

To remove the observed background emission from the stacking, we carried out a number of tests and simulations, e.g., by changing $z_{\rm max}$ or $E_{\rm min}$ thresholds. It was noticed that only after increasing the minimum energy from 100 MeV to 300 MeV, we are able to get rid of the background. This is expected since a soft emission is brightest at the lowest energies. As can be seen in the right panel of Figure~\ref{fig:stacking}, when considering $E_{\rm min}=300$ MeV, the background is completely removed from the stacking. Therefore, we repeated the whole exercise with $E_{\rm min}=300$ MeV.

\begin{figure*}
\centering
\hbox{
\includegraphics[width=\linewidth]{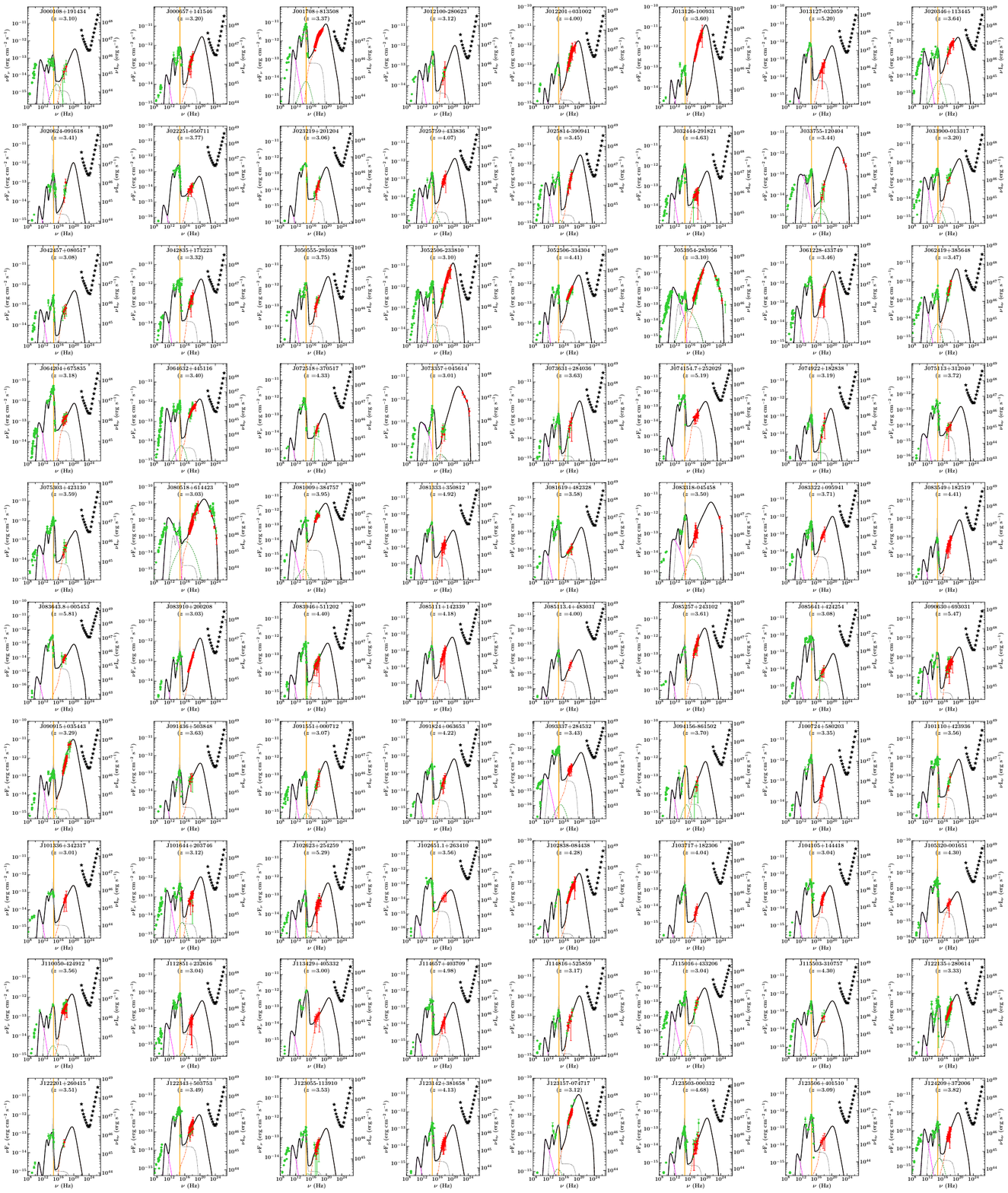}
}
\caption{Modeled SEDs of the high-redshift blazars.}\label{fig_all_SED1}
\end{figure*}
\begin{figure*}
\centering
\hbox{
\includegraphics[width=\linewidth]{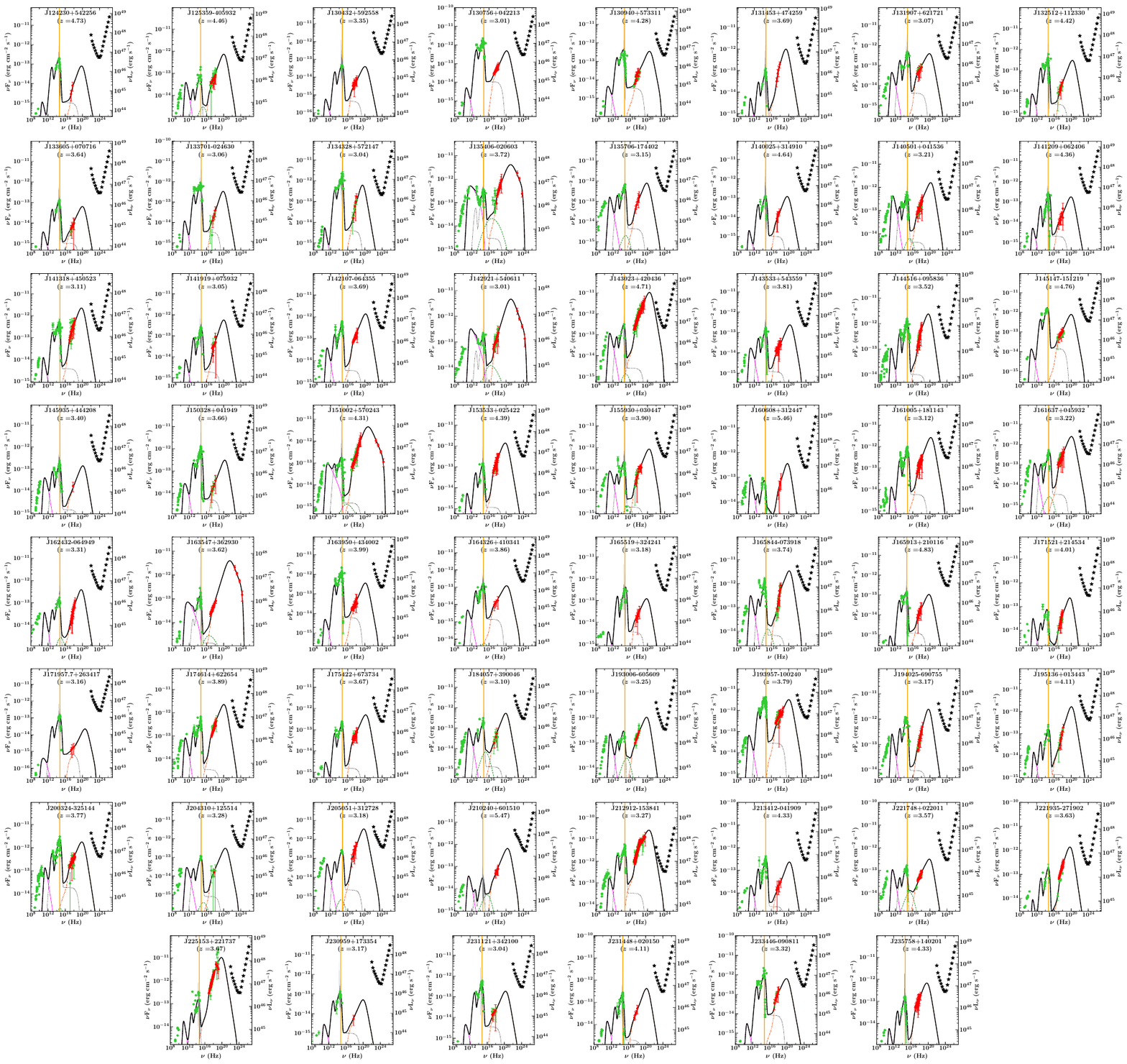}
}
\caption{Modeled SEDs of the high-redshift blazars.}\label{fig_all_SED2}
\end{figure*}

\begin{figure*}
\centering
\hbox{
\includegraphics[width=\linewidth]{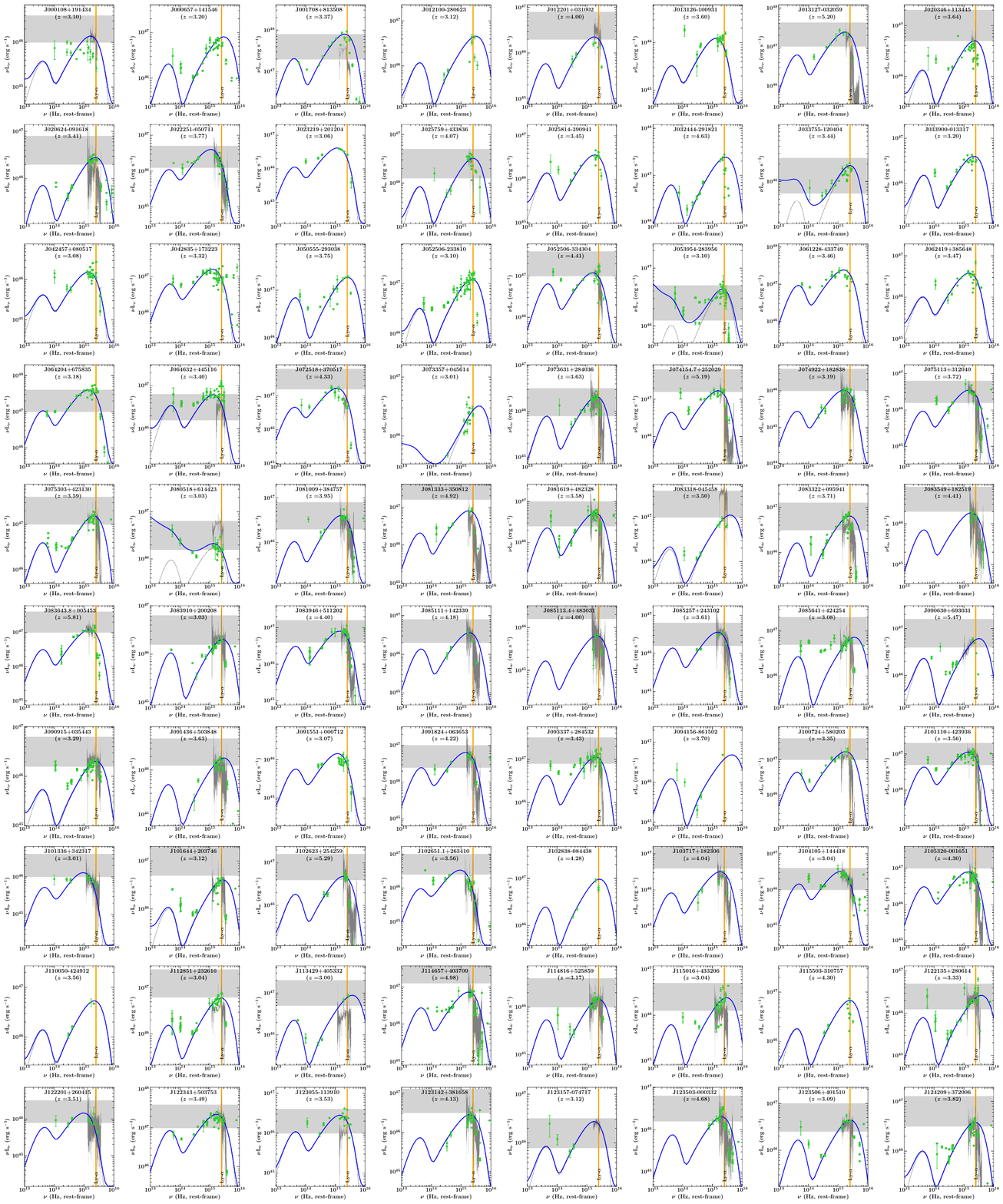}
}
\caption{Modeled IR-UV SEDs of the high-redshift blazars.}\label{fig_all_disk1}
\end{figure*}
\begin{figure*}
\centering
\hbox{
\includegraphics[width=\linewidth]{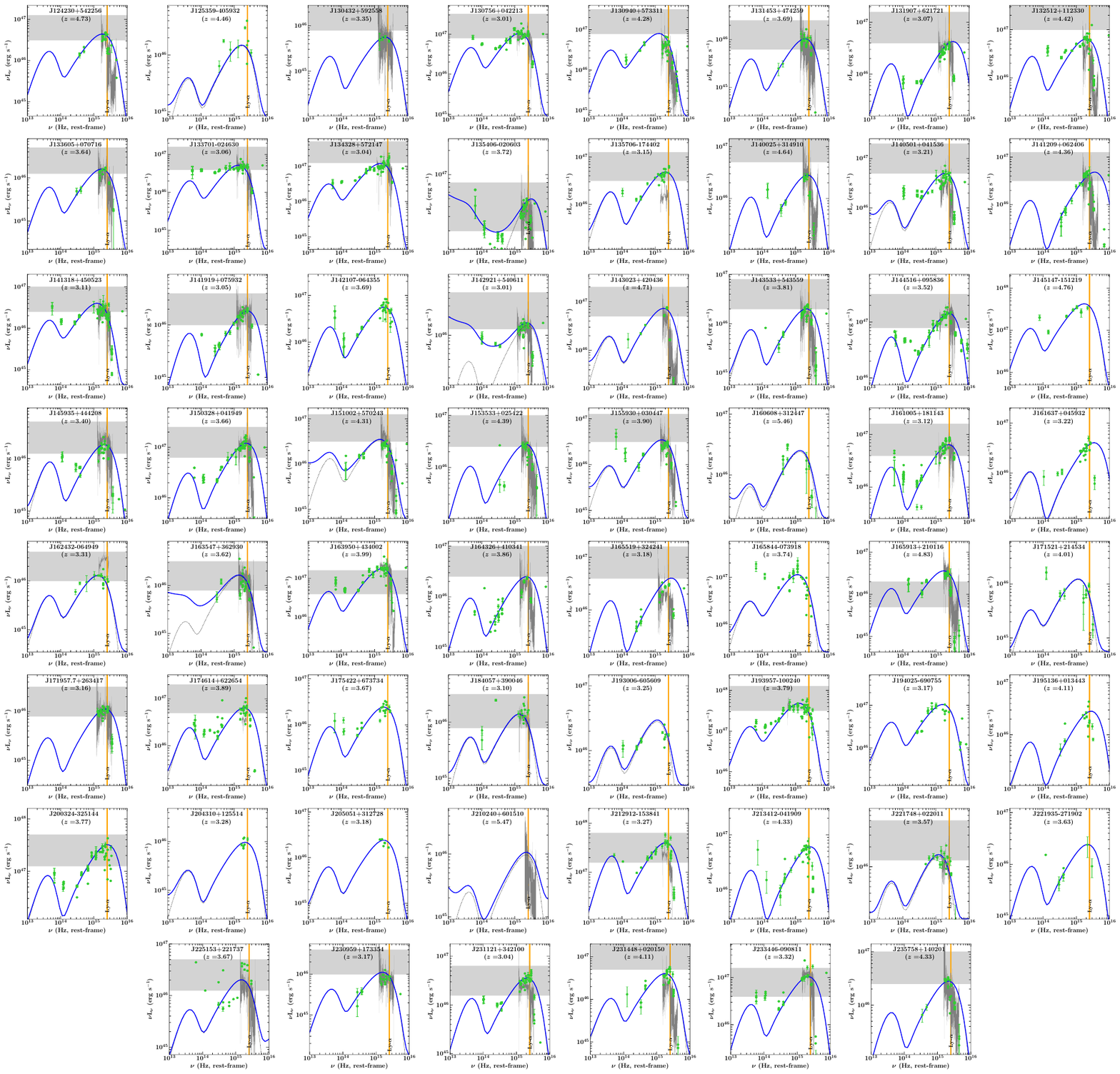}
}
\caption{Modeled IR-UV SEDs of the high-redshift blazars.}\label{fig_all_disk2}
\end{figure*}

\bibliography{Master}{}
\bibliographystyle{aasjournal}

\end{document}